\documentclass{report}

\usepackage{a4}
\usepackage[utf8]{inputenc}
\usepackage{graphicx,caption,subcaption}
\usepackage{verbatim}
\usepackage[total={6in, 9in}, top=1in,left=1.1in]{geometry} 
\usepackage{moreverb}
\usepackage{psfrag}
\usepackage{amsmath,amssymb,amsfonts,mathrsfs,mathcomp}
\usepackage{upquote}
\usepackage{fancyhdr}

\newcommand{\changefont}{\fontsize{10}{11}\selectfont}
\newcommand{\HRule}{\rule{\linewidth}{0.5mm}}
\usepackage{imakeidx}
\makeindex

\usepackage{upquote} %To make straight quotes work in pdf
\usepackage[pagebackref=false,colorlinks,linkcolor=blue,citecolor=blue,urlcolor=blue]{hyperref}

\usepackage{xcolor}

\usepackage[nameinlink]{cleveref}
\crefformat{equation}{(#2#1#3)}

\newcommand{\tx}[1]{{\textrm{#1}}}
\usepackage{mhchem}
\usepackage{amsmath}
\usepackage{tikz}

\usepackage{dirtree}
\crefname{lstlisting}{listing}{listings}
\Crefname{lstlisting}{Listing}{Listings}

\usepackage{fancyvrb}

\usepackage{listings}

\usepackage{accsupp}
\newcommand{\noncopynumber}[1]{%
    \BeginAccSupp{method=escape,ActualText={}}%
    #1%
    \EndAccSupp{}%
}

\colorlet{mygray}{black!4}
\definecolor{mygreen}{rgb}{0,0.4,0}
\colorlet{myred}{red!80!blue}

\lstdefinestyle{cpp}{
	language=C++,
	basicstyle=\ttfamily\footnotesize,
	keepspaces=true,
  	columns=fixed,%fullflexible
	fontadjust=true,
	basewidth=0.5em,
	backgroundcolor=\color{mygray},
	tabsize=4,
	captionpos=b,
	frame=single,
	numbers=left,
	numberstyle=\tiny\noncopynumber,
	numbersep=5pt,
	breaklines=true,
	showstringspaces=false,
	keywordstyle=\color{blue},
	commentstyle=\color{mygreen},
	stringstyle=\color{myred}
}

\lstdefinestyle{OpenFOAMDict}{
	language=C++,
	basicstyle=\ttfamily\footnotesize,
	keepspaces=true,
  	columns=fixed,%fullflexible
	fontadjust=true,
	basewidth=0.5em,
	backgroundcolor=\color{mygray},
	tabsize=4,
	captionpos=b,
	frame=single,
	numbers=left,
	numberstyle=\tiny\noncopynumber,
	numbersep=5pt,
	breaklines=true,
	showstringspaces=false,
	commentstyle=\color{mygreen},
}

\lstdefinestyle{python}{
	language=Python,
	basicstyle=\ttfamily\footnotesize,
	keepspaces=true,
  	columns=fixed,%fullflexible
	fontadjust=true,
	basewidth=0.5em,
	backgroundcolor=\color{mygray},
	tabsize=4,
	captionpos=b,
	frame=single,
	numbers=left,
	numberstyle=\tiny\noncopynumber,
	numbersep=5pt,
	breaklines=true,
	showstringspaces=false,
	keywordstyle=\color{blue},
	commentstyle=\color{mygreen},
	stringstyle=\color{myred}
}

\makeatletter
\newcommand*{\textalltt}{}
\DeclareRobustCommand*{\textalltt}{%
	\begingroup
	\let\do\@makeother
	\dospecials
	\catcode`\\=\z@
	\catcode`\{=\@ne
	\catcode`\}=\tw@
	\verbatim@font\@noligs
	\@vobeyspaces
	\frenchspacing
	\@textalltt
}
\newcommand*{\@textalltt}[1]{%
	#1%
	\endgroup
}
\makeatother
\makeatletter
\let\orig@lstnumber=\thelstnumber
\newcommand\lstsetnumber[1]{\gdef\thelstnumber{#1}}
\newcommand\lstresetnumber{\global\let\thelstnumber=\orig@lstnumber}
\makeatother

\usepackage{siunitx}
\usepackage{nomencl}
\usepackage{ifthen}
\renewcommand\nomgroup[1]{%
\ifthenelse{\equal{#1}{A}}{%
\vspace{0.5cm}%
\item[\textbf{Acronyms}]}{%              A - Acronyms
\ifthenelse{\equal{#1}{E}}{%
\vspace{0.5cm}%
\item[{\textbf{English symbols}}]}{%           E - English
\ifthenelse{\equal{#1}{G}}{%
\vspace{0.5cm}
\item[{\textbf{Greek symbols}}]}{%           G - Greek
\ifthenelse{\equal{#1}{S}}{%
\vspace{0.5cm}%	
\item[\textbf{Superscripts}]}{%            S - Superscripts
\ifthenelse{\equal{#1}{U}}{%
\vspace{0.5cm}%	
\item[\textbf{Subscripts}]}{%              U - Subscripts
\ifthenelse{\equal{#1}{X}}{%
\vspace{0.5cm}%	
\item[\textbf{Other symbols}]}{%           X - Other Symbols
{}}}}}}}}

\newcommand{\nomunit}[1]{\renewcommand{\nomentryend}{\dotfill#1}}
\makenomenclature
\begin{document}

% ----------------------- Front Matter ----------------------- %
\begin{titlepage}
\begin{center}

%%%Some main details, as new commands, to avoid making errors:
%Title of project:
\newcommand{\myTitle}{Implementation of Analytical Jacobian and Chemical Explosive Mode Analysis (CEMA) in OpenFOAM}
%Year of course:
\newcommand{\thisYear}{2021}
%%%

% {\small Cite as: LastName, F.: \myTitle. In Proceedings of CFD with OpenSource Software, \thisYear, Edited by Nilsson. H., \textalltt{http://dx.doi.org/10.17196/OS_CFD#YEAR_\thisYear}}

{\small Cite as: Gadalla, M.: \myTitle. In Proceedings of CFD with OpenSource Software, \thisYear, Edited by Nilsson, H.,
\href{http://dx.doi.org/10.17196/OS_CFD#YEAR_\thisYear}
{\textalltt{http://dx.doi.org/10.17196/OS\_CFD\#YEAR\_\thisYear}}}

~\\[0.6cm]
\textsc{\LARGE CFD with OpenSource software}\\[0.6cm]	
\textsc{A course at Chalmers University of Technology}\\
\textsc{Taught by H{\aa}kan Nilsson}\\[0.6cm]	

\HRule \\[0.5cm]
{ \huge \bfseries \myTitle}\\[0.2cm]
\HRule \\[1cm]

\begin{minipage}{0.35\textwidth}
Developed for OpenFOAM-v2006\\
Requires: PyJac
\end{minipage}\\[1.0cm]

\begin{minipage}{0.4\textwidth}
\begin{flushleft} \large
\emph{Author:}\\
    Mahmoud \textsc{Gadalla}\\
    Aalto University,\\ School of Engineering.\\
    mahmoud.gadalla@aalto.fi\\
    gadalla.mah@gmail.com\\
\end{flushleft}
\end{minipage}
\begin{minipage}{0.46\textwidth}
\begin{flushright} \large
\emph{Peer reviewed by:} \\
% Konstantinos~\textsc{Missios}, M.~Sc., Roskilde U.\\
% Islam~\textsc{Kabil}, M.~Sc., Connecticut U.\\
% Saeed~\textsc{Salehi}, Ph.~D, Chalmers U.\\
% \mbox{Ville~\textsc{Vuorinen}, Prof., Aalto U.}
Konstantinos~\textsc{Missios}, Roskilde Univ.\\
Islam~\textsc{Kabil}, Connecticut Univ.\\
Saeed~\textsc{Salehi}, Chalmers Univ.\\
Ville~\textsc{Vuorinen}, Aalto Univ.
\end{flushright}
\end{minipage}

~\\[1cm]

Library on Github: \url{https://github.com/Aalto-CFD/CEMAFoam}
~\\[1cm]

Licensed under CC-BY-NC-SA, \verb+https://creativecommons.org/licenses/+

\vfill

{Disclaimer: This is a student project work, done as part of a course where OpenFOAM and some other OpenSource software are introduced to the students. Any reader should be aware that it might not be free of errors. Still, it might be useful for someone who would like learn some details similar to the ones presented in the report and in the accompanying files. The material has gone through a review process. The role of the reviewer is to go through the tutorial and make sure that it works, that it is possible to follow, and to some extent correct the writing. The reviewer has no responsibility for the contents.}\\[2cm]

% {\large \today} % February 12, 2022
{\large January 18, 2022}

\end{center}

\end{titlepage}
\chapter*{Learning outcomes}

% The main requirements of a tutorial in the course is that it should teach the four points: How to use it, The theory of it, How it is implemented, and How to modify it. Therefore the list of learning outcomes is organized with those headers.\\[0.4cm]

\noindent The reader will learn:\\[0.4cm]

\noindent{\bf How to use it:}
\begin{itemize}
	\item how to use the reactingFoam solver, with complete understanding of necessary dictionaries required for the solver computations.
	\item how to make use of dynamic binding for loading OpenFOAM libraries in the application level.
\end{itemize}
{\bf The theory of it:}
\begin{itemize}
	\item the theory of finite-rate chemistry in reactive flow modeling.
    \item the importance of system Jacobian and derivatives to solve stiff system of ordinary differential equations (ODEs) relevant to chemical source term computations.
	\item the theory of Chemical Explosive Mode Analysis (CEMA) for local combustion mode characterization.
\end{itemize}
{\bf How it is implemented:}
\begin{itemize}
	\item how StandardChemistryModel is implemented in OpenFOAM.
    \item how ODE and chemistryModel libraries communicate in the lower code level for solving chemical source terms.
\end{itemize}
{\bf How to modify it:}
\begin{itemize}
	\item how to modify the templated StandardChemistryModel library to accommodate for analytical Jacobian and CEMA functionalities.
\end{itemize}
\chapter*{Prerequisites}

% This is an example for the prerequisites. 

The reader is expected to know the following in order to get maximum benefit out of this report:

\begin{itemize}

	% \item How to run standard document tutorials like damBreak tutorial.
	% \item Fundamentals of Computational Methods for Fluid Dynamics, Book by J. H. Ferziger and M.	Peric
	% \item How to customize a solver and do top-level application programming.
	% \item etc.

    \item Basic knowledge of OpenFOAM usage.
    % \item How to run standard OpenFOAM tutorials like counterFlowFlame2D tutorial.
    \item How to run standard OpenFOAM tutorials with proper knowledge of case files and dictionaries. %sub-directories
    \item Basics of ``Theoretical and Numerical Combustion'', Book by T. Poinsot and D. Veynante.
    \item Basics of ``Turbulent Combustion'', Book by N. Peters.
    \item Familiarization with dynamical system theory, along with the article by Lu et al. 2010, Journal of Fluid Mechanics, 652, 45-64, \url{https://doi.org/10.1017/S002211201000039X}.
    \item Basic understanding of object oriented programming, particularly the class inheritance and polymorphism.

\end{itemize}

\tableofcontents

\begin{thenomenclature} 
\nomgroup{A}
  \item [{CEM}]\begingroup Chemical Explosive Mode\nomeqref {0}\nompageref{4}
  \item [{CEMA}]\begingroup Chemical Explosive Mode Analysis\nomeqref {0}\nompageref{4}
  \item [{CFD}]\begingroup Computational Fluid Dynamics\nomeqref {0}\nompageref{4}
  \item [{CPU}]\begingroup Central Processing Unit\nomeqref {0}\nompageref{4}
  \item [{DI}]\begingroup Defectiveness Index\nomeqref {0}\nompageref{4}
  \item [{DRY}]\begingroup ``Don't Repeat Yourself'' concept\nomeqref {0}\nompageref{4}
  \item [{EI}]\begingroup Explosion Index\nomeqref {0}\nompageref{4}
  \item [{FVM}]\begingroup Finite Volume Method\nomeqref {0}\nompageref{4}
  \item [{GPU}]\begingroup Graphics Processing Unit\nomeqref {0}\nompageref{4}
  \item [{HRR}]\begingroup Heat Release Rate\nomeqref {0}\nompageref{4}
  \item [{NSE}]\begingroup Navier-Stokes Equations\nomeqref {0}\nompageref{4}
  \item [{PI}]\begingroup Participation Index\nomeqref {0}\nompageref{4}
  \item [{PISO}]\begingroup Pressure-Implicit with Splitting of Operators\nomeqref {0}\nompageref{4}
  \item [{SIMD}]\begingroup Single-Instruction Multiple Data\nomeqref {0}\nompageref{4}
  \item [{UML}]\begingroup Unified Modeling Language\nomeqref {0}\nompageref{4}
\nomgroup{E}
  \item [{$\bar{c}_k$}]\begingroup Concentration of species $k$\nomunit{[\si{\kmol\per\metre^3}]}\nomeqref {0}\nompageref{4}
  \item [{$\bar{R}$}]\begingroup Universal gas constant\nomunit{[\si{\joule\per\kmol\per\kelvin}]}\nomeqref {0}\nompageref{4}
  \item [{$\boldsymbol u$}]\begingroup Velocity vector\nomunit{[\si{\metre\per\second}]}\nomeqref {0}\nompageref{4}
  \item [{$\boldsymbol{\mathcal J}$}]\begingroup Jacobian matrix of stiff chemistry ODE system\nomunit{[\si{-}]}\nomeqref {0}\nompageref{4}
  \item [{$A_r$}]\begingroup Pre-exponential factor for elementary reaction $r$\nomunit{[varied]}\nomeqref {0}\nompageref{4}
  \item [{$c_{p_k}(T)$}]\begingroup Isobaric temperature-dependent specific heat capacity for species $k$\nomunit{[\si{\joule\per\kelvin\per\kilogram}]}\nomeqref {0}\nompageref{4}
  \item [{$D$}]\begingroup Mass diffusivity\nomunit{[\si{\metre^2\per\second}]}\nomeqref {0}\nompageref{4}
  \item [{$E_a$}]\begingroup Activation energy\nomunit{[\si{\joule\per\kmol}]}\nomeqref {0}\nompageref{4}
  \item [{$H$}]\begingroup Molar enthalpy\nomunit{[\si{\joule\per\kmol}]}\nomeqref {0}\nompageref{4}
  \item [{$h^0_f$}]\begingroup Enthalpy of formation (specific)\nomunit{[\si{\joule\per\kilogram}]}\nomeqref {0}\nompageref{4}
  \item [{$h_s$}]\begingroup Sensible enthalpy (specific)\nomunit{[\si{\joule\per\kilogram}]}\nomeqref {0}\nompageref{4}
  \item [{$h_t$}]\begingroup Total enthalpy (specific)\nomunit{[\si{\joule\per\kilogram}]}\nomeqref {0}\nompageref{4}
  \item [{$k_r$}]\begingroup Rate constant of reaction $r$\nomunit{[varied]}\nomeqref {0}\nompageref{4}
  \item [{$M$}]\begingroup Number of elements in species composition of chemical mechanism\nomunit{[\si{-}]}\nomeqref {0}\nompageref{4}
  \item [{$p$}]\begingroup Pressure\nomunit{[\si{\pascal}]}\nomeqref {0}\nompageref{4}
  \item [{$q_r$}]\begingroup Progress rate of reaction $r$\nomunit{[\si{\kmol\per\metre\cubed\per\second}]}\nomeqref {0}\nompageref{4}
  \item [{$S$}]\begingroup Molar entropy\nomunit{[\si{\joule\per\kmol\per\kelvin}]}\nomeqref {0}\nompageref{4}
  \item [{$T$}]\begingroup Temperature\nomunit{[\si{\kelvin}]}\nomeqref {0}\nompageref{4}
  \item [{$t$}]\begingroup Time\nomunit{\si{\second}}\nomeqref {0}\nompageref{4}
  \item [{$u_{\textrm{L}}$}]\begingroup Laminar burning velocity\nomunit{[\si{\metre\per\second}]}\nomeqref {0}\nompageref{4}
  \item [{$W_k$}]\begingroup molecular weight of species $k$\nomunit{[\si{\kilogram\per\kmol}]}\nomeqref {0}\nompageref{4}
  \item [{$Y_k$}]\begingroup Mass fraction of species $k$\nomunit{[\si{-}]}\nomeqref {0}\nompageref{4}
  \item [{\ce{CH4}}]\begingroup Methane species\nomunit{[\si{-}]}\nomeqref {0}\nompageref{4}
  \item [{\ce{CO2}}]\begingroup Carbon dioxide species\nomunit{[\si{-}]}\nomeqref {0}\nompageref{4}
  \item [{\ce{H2O}}]\begingroup Water (oxidane) species\nomunit{[\si{-}]}\nomeqref {0}\nompageref{4}
  \item [{\ce{N2}}]\begingroup Nitrogen species\nomunit{[\si{-}]}\nomeqref {0}\nompageref{4}
  \item [{\ce{O2}}]\begingroup Oxygen species\nomunit{[\si{-}]}\nomeqref {0}\nompageref{4}
\nomgroup{G}
  \item [{$\bar{\phi}$}]\begingroup Volumetric face flux in CFD\nomunit{[\si{\metre^3\per\second}]}\nomeqref {0}\nompageref{4}
  \item [{$\boldsymbol \Phi$}]\begingroup Thermochemical state vector comprising temperature and species mass fractions\nomunit{[varied]}\nomeqref {0}\nompageref{4}
  \item [{$\boldsymbol \Phi_c$}]\begingroup Thermochemical state vector comprising temperature and species concentrations\nomunit{[varied]}\nomeqref {0}\nompageref{4}
  \item [{$\boldsymbol \tau$}]\begingroup Viscous stress tensor\nomunit{[\si{\newton\per\meter^2}]}\nomeqref {0}\nompageref{4}
  \item [{$\delta t_\textrm{CFD}$}]\begingroup CFD timestep\nomunit{[\si{\second}]}\nomeqref {0}\nompageref{4}
  \item [{$\delta t_\textrm{ODE}$}]\begingroup ODE timestep\nomunit{[\si{\second}]}\nomeqref {0}\nompageref{4}
  \item [{$\delta_\textrm{L}$}]\begingroup Thermal thickness of laminar premixed flame\nomunit{[\si{\metre}]}\nomeqref {0}\nompageref{4}
  \item [{$\gamma$}]\begingroup Specific heat ratio\nomunit{[\si{-}]}\nomeqref {0}\nompageref{4}
  \item [{$\lambda_\textrm{exp}$}]\begingroup Explosive eigenvalue of Jacobian matrix\nomunit{[\si{-}]}\nomeqref {0}\nompageref{4}
  \item [{$\mu$}]\begingroup Fluid dynamic viscosity\nomunit{[\si{\pascal\second}]}\nomeqref {0}\nompageref{4}
  \item [{$\nabla$}]\begingroup Gradient operator\nomunit{[\si{1/\meter}]}\nomeqref {0}\nompageref{4}
  \item [{$\nu''_{k,r}$}]\begingroup Molar stoichiometric coefficients of species $k$ in reaction $r$ products\nomunit{[\si{-}]}\nomeqref {0}\nompageref{4}
  \item [{$\nu'_{k,r}$}]\begingroup Molar stoichiometric coefficients of species $k$ in reaction $r$ reactants\nomunit{[\si{-}]}\nomeqref {0}\nompageref{4}
  \item [{$\nu_{k,r}$}]\begingroup Net molar stoichiometric coefficients of species $k$ in reaction $r$\nomunit{[\si{-}]}\nomeqref {0}\nompageref{4}
  \item [{$\rho$}]\begingroup Fluid density\nomunit{[\si{\kilogram\per\meter^3}]}\nomeqref {0}\nompageref{4}
  \item [{${\dot{\omega}}_h$}]\begingroup Energy source term\nomunit{[\si{\joule\per\meter\cubed\per\second}]}\nomeqref {0}\nompageref{4}
  \item [{${\dot{\omega}}_k$}]\begingroup Chemical source term of species $k$\nomunit{[\si{\kilogram\per\meter\cubed\per\second}]}\nomeqref {0}\nompageref{4}
\nomgroup{S}
  \item [{$N_{r}$}]\begingroup Number of reactions\nomeqref {0}\nompageref{4}
  \item [{$N_{sp}$}]\begingroup Number of species\nomeqref {0}\nompageref{4}
\nomgroup{U}
  \item [{$0$}]\begingroup reference value\nomeqref {0}\nompageref{4}
  \item [{$\textrm{bw}$}]\begingroup backward\nomeqref {0}\nompageref{4}
  \item [{$\textrm{fw}$}]\begingroup forward\nomeqref {0}\nompageref{4}
  \item [{$\textrm{L}$}]\begingroup Laminar flame\nomeqref {0}\nompageref{4}
  \item [{$f$}]\begingroup formation\nomeqref {0}\nompageref{4}
  \item [{$h$}]\begingroup enthalpy\nomeqref {0}\nompageref{4}
  \item [{$k$}]\begingroup species index\nomeqref {0}\nompageref{4}
  \item [{$r$}]\begingroup reaction index\nomeqref {0}\nompageref{4}
  \item [{$s$}]\begingroup sensible\nomeqref {0}\nompageref{4}
  \item [{$t$}]\begingroup total\nomeqref {0}\nompageref{4}

\end{thenomenclature}

% ------------------------------------------------------------ %

% ------------------------- Chapters ------------------------- %
% \chapter{Introduction}
\chapter{Background and Introduction}
\label{ch:intro}

% \section{Background}

\begin{figure}[h!]
\centering
\includegraphics[width=0.8\textwidth]{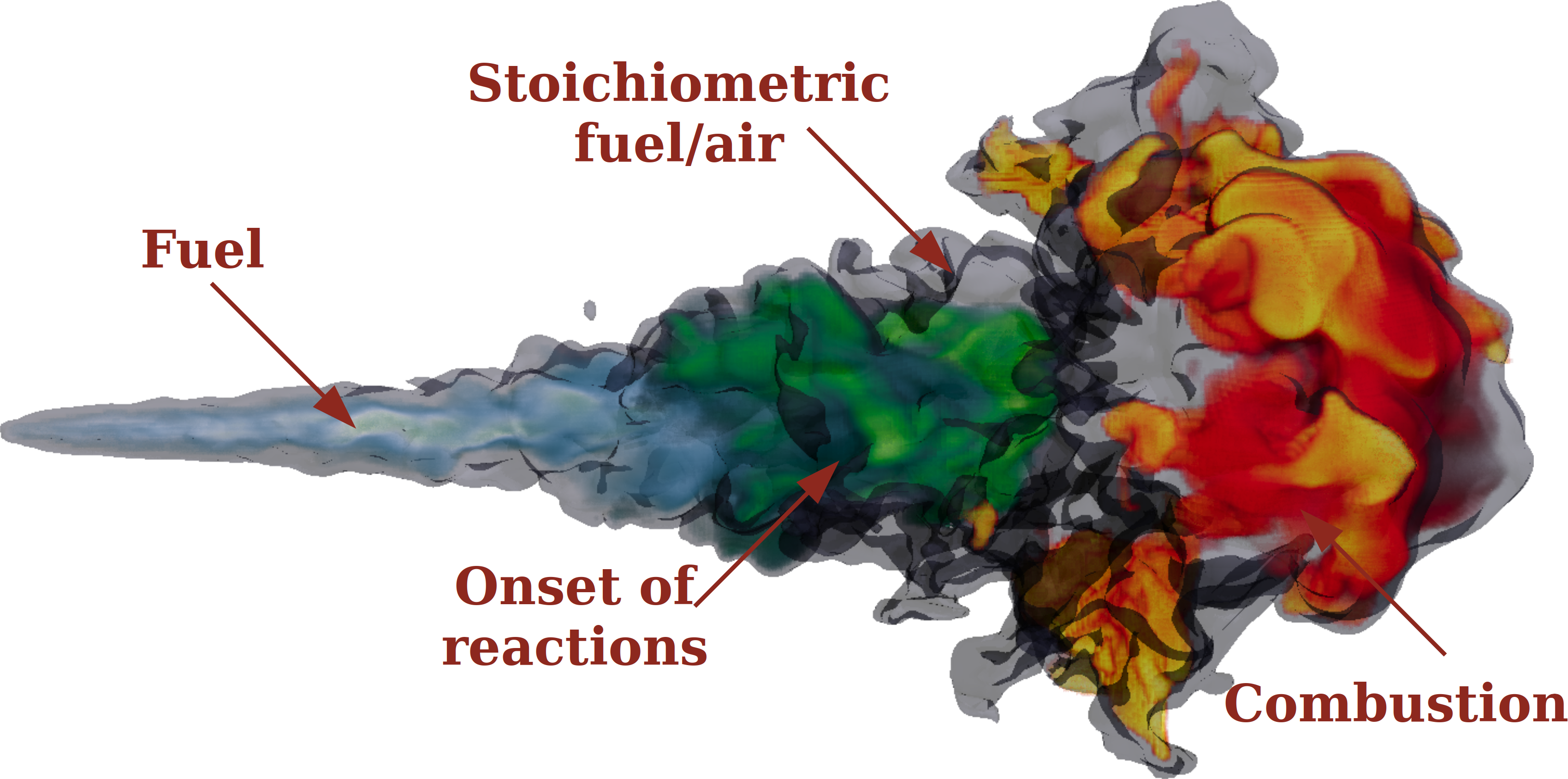}
\caption{Volume-rendered visualization of turbulent reactive flow.}
\label{fig:spray}
\end{figure}

Reactive flow modeling plays an essential role in the field of energy and combustion research~\cite{poinsot2011theoretical}.
Based on the constitutive laws of continuum mechanics and particularly in computational fluid dynamics (CFD), mathematical models are derived. Conservation laws for linear momentum, represented by the Navier-Stokes equations (NSE), are solved together with the conservation of total mass (i.e. continuity) and energy governing equations. In a chemically reactive system, convection and diffusion processes of the existent chemical species in a computational domain are governed by scalar transport equation for each species. More details on the modeling of turbulent reactive flows can be revised from the textbooks by Poinsot and Veynante~\cite{poinsot2011theoretical} and by N. Peters~\cite{peters2001turbulent}.
The species production and consumption due to chemical reactions are usually represented via source terms plugged into the species and energy equations. In \Cref{fig:spray}, we show a volume-rendered visualization of spray combustion simulation using OpenFOAM, where fuel is first injected into the domain, then it starts dissociation into lighter radicals and intermediate species, indicating the onset of chemical reactions. Further downstream, and after specific induction time, chain of exothermic reactions take place where heat release is generated with a thermal runaway, hence combustion.

It is of particular interest to identify and study the structure of reaction fronts, defined as the leading edge of the reaction zone, by experimental~\cite{Sick2013,Lee2003} and computational~\cite{Shan2012,Lu2010} means. Accordingly, the chemical explosive mode analysis (CEMA) has been developed by Lu et al.~\cite{Lu2010} with motivations from the earlier concept of computational singular perturbation~\cite{lam1985singular}. CEMA is considered a versatile computational diagnostics tool that enables the detection of various critical combustion features including reaction fronts, flame stabilization mechanisms, and auto-ignition and extinction zones~\cite{Shan2012}. Recently, it was further extended to account for diffusion~\cite{Xu2019} and evaporation~\cite{Mohaddes2021} processes to investigate their roles toward promoting or inhibiting chemistry and auto-ignition.

In this project, we aim at implementing the basic formulations of CEMA, which are responsible for the identification of pre- and post-ignition zones and subsequently the reaction fronts, into OpenFOAM. Such an implementation is considered the crucial part in the analysis tool development since it identifies the existence of chemical explosive mode (CEM). Once a CEM is detected, all subsequent developments can be implemented including the 
\begin{itemize}
    \item explosion index (EI): which quantifies the contribution of chemical species to a CEM,

    \item participation index (PI): which quantifies the contribution of chemical reactions to a CEM,

    \item defectiveness index (DI): which measures the defectiveness of thermo-chemical Jacobian matrix that might lead into solution singularity, and
     % that might lead into singularity of the eigenvector matrix,

    \item projection of local source terms including chemistry, diffusion, and possibly spray droplet evaporation, onto CEM to quantify and compare their roles toward promoting or inhibiting chemical reactions, hence the identification of local combustion modes.

\end{itemize}

With relevance to the latest point, in combustion systems, it is critically important to understand whether the mixture is burning in a deflagrative (i.e. slow combustion or a premixed flame propagation) or auto-ignitive (i.e. spontaneous combustion with an ignition front) type. In order to characterize this, CEMA becomes useful which offers an eigenvalue analysis tool to estimate the local mode of combustion~\cite{Xu2019}. In a turbulent spray assisted combustion, both of the spontaneous and deflagration modes of combustion are known to be present, which is the subject of ongoing studies.

Another important note is that, while the derivation of CEMA is based on an analytical formulation of the thermo-chemical Jacobian matrix comprising pressure, temperature and species concentrations~\cite{Lu2010}, in OpenFOAM v$2006$ the Jacobian matrix is only semi-analytical. Accordingly, the Python-based open-source library \verb|pyJac|~\cite{Niemeyer2017} is herein utilized. The \verb|pyJac| library is a source code generator which creates \verb|C| subroutines (or possibly in other programming languages) for analytical Jacobian matrix evaluation provided a particular chemical kinetic mechanism. Computationally, \verb|pyJac| is optimized to minimize operations and memory usage, in addition to being thread-safe and can be easily parallelized on single-instruction multiple data (SIMD) processors such as graphics processing units (GPUs). Moreover, the performance of \verb|pyJac| on central processing units (CPUs), as well as GPUs, is found to outperform other existing libraries for both finite-differencing and analytical evaluation of the Jacobian matrix. Despite the fact that other OpenFOAM releases feature implementations for analytical Jacobian evaluation, \verb|pyJac| was found to be superior particularly when it comes to computing the chemical source terms via stiff ordinary differential equation (ODE) integration with tight tolerances, which we encounter in this project. As a result, \verb|pyJac| will be herein employed for the generation of the analytical Jacobian matrix, and the corresponding CEMA results will be demonstrated and compared against those resulting from semi-analytical OpenFOAM routines for the Jacobian evaluation.

% Therefore, it becomes essential for the project scope to implement 

In the present report we first shed light on some standard OpenFOAM libraries that are responsible for the numerical modeling of combustion phenomena, with a particular focus on the \verb|chemistryModel| library responsible for chemical source terms and reaction rates. Then, we discuss the implementation details to replace particular OpenFOAM functionalities responsible for numerical Jacobian matrix approximation with the analytical formulation from \verb|pyJac|. After that, we utilize these implemented functionalities for analytical Jacobian formulation to develop CEMA for the identification of pre- and post-ignition zones as well as the detection of reaction fronts. Finally, a tutorial case is provided and discussed to validate and demonstrate the applicability of the developed model for analytical Jacobian and CEMA.

% with new functionalities for computational diagnostics of combustion using chemical explosive mode analysis (CEMA).

% Assuming subsonic propagation of flow and combustion, the fundamental combustion regimes are i) premixed combustion, ii) non-premixed combustion, and iii) volumetric auto-ignition. Furthermore, local zones can occur where transport processes can promote or inhibit the combustion, or possibly leading into flame extinctions. It is of particular interest to identify these local modes. In this report, we implement the famous Chemical Explosive Mode Analysis (CEMA) as an advanced flame diagnostics technique to be incorporated within OpenFOAM standard chemistry models.

% \section{reactingFoam solver}

% computational fluid dynamics (CFD) are employed to 

% \section{Background}

% \section{Combustion modeling}

% \section{Combustion mode characterization}

% \section{Present chemistry models in OpenFOAM}

% \section{Chemistry models in OpenFOAM}

% \subsection{Governing equations}

% % \subsection{Finite-rate chemistry}
% \subsection{Direct integration of finite-rate chemistry}

% \subsection{Local combustion mode characterization}

\chapter{Theory}
\label{ch:theory}

% This is a example of chapter creation using the \verb|\chapter| command.

\section{Governing equations}

The gaseous phase is herein described by the low Mach number compressible Navier-Stokes equations. Assuming laminar conditions, the corresponding formulations for the conservation of mass, momentum, species, and enthalpy are all presented as
\begin{align}
%%% Mass conservation %%%
\label{eq:mass}
\frac{\partial {\rho}}{\partial t} + \nabla \cdot ({\rho} \, {\boldsymbol u}) &= 0 ,\\
%%% Momentum conservation %%%
\label{eq:mom}
\frac{\partial {\rho} {\boldsymbol u}}{\partial t}
% + ({\rho} {\boldsymbol u} \cdot \nabla) {\boldsymbol u} &=
+ \nabla \cdot ({\rho} {\boldsymbol u} \otimes {\boldsymbol u}) &=
- \nabla {p}
+ \nabla \cdot {\boldsymbol \tau} ,\\
%%% Species conservation %%%
\label{eq:species}
\frac{\partial ({\rho} {Y}_k)}{\partial t}
+ \nabla \cdot ({\rho} {\boldsymbol u} {Y}_k)
&=
\nabla \cdot \big( 
{\rho} {D} \nabla {Y}_k
\big)
% + \circled{${\dot{\omega}}_k$} ,\\
+ \dot{\omega}_k ,\\
%%% Enthalpy conservation %%%
\label{eq:enthalpy}
\frac{\partial ({\rho} {h}_t)}{\partial t}
+ \nabla \cdot ({\rho} {\boldsymbol u} {h}_t)
&=
\frac{\partial {p}}{\partial t}
+ \nabla \cdot \big(
{\rho} {D} \nabla {h}_s
\big)
% + \circled{${\dot{\omega}_h}$} ,%\\
+ \dot{\omega}_h ,%\\
\end{align}

where ${\rho}$, ${\boldsymbol u}$, ${p}$, ${\boldsymbol \tau}$, ${Y}_k$, ${D}$, ${h}_s$, ${h}_t$ denote the density, velocity vector, pressure, viscous stress tensor, $k^{th}$ species mass fraction, mass diffusivity, sensible and total enthalpy, respectively. The total enthalpy is herein defined as the summation of sensible enthalpy and specific kinetic energy. The ($\otimes$) symbol refers to the outer product (dyadic), and the normal typeface 
symbols denote scalar quantities whereas bold ones represent higher order tensors. The viscous stress tensor for Newtonian fluids is defined as
\begin{equation}
\boldsymbol \tau = \mu \big( \nabla \boldsymbol u + (\nabla \boldsymbol u)^T - \frac{2}{3} (\nabla \cdot \boldsymbol u) \mathbf{I} \big) ,
\end{equation}

while the sensible enthalpy is defined according to the caloric state equation, i.e.

\begin{equation}
\label{eq:hs}
h_s = \sum_{k=1}^{N_{sp}} Y_k h_k = \sum_{k=1}^{N_{sp}} Y_k \big( h^0_{f,k} + \int_{T_0}^T c_{p_k}(T^*) dT^* \big) ,
\end{equation}

where $ h^0_{f,k}$ denotes $k^{th}$ species enthalpy of formation at reference temperature $T_0 = 298.15$~\si{K}, and $N_{sp}$ is the total number of species in the chemical mechanism. Moreover, $c_{p_k}(T^*)$ highlights temperature dependence of $c_{p_k}$ which is typically retrieved using tabulated polynomial fits, as it is explained in the following section.

\Cref{eq:mass,eq:mom,eq:species,eq:enthalpy} are closed by the ideal gas thermodynamic state equation, in addition to the pressure-momentum coupling using the reacting PISO (Pressure-Implicit with Splitting of Operators) algorithm~\cite{Issa1991}. The resulting set of equations is numerically discretized using the finite volume method (FVM) and it is solved using the OpenFOAM~\textsuperscript{\textregistered} framework. We note that additional terms have to be considered in the previous set of equations to account for subgrid scale modeling or the phase change. However, as explained earlier, our demonstration is based on the laminar and one-phase flow assumption, without loss of generality.

% The chemical source terms ${\dot{\omega}}_k$ and ${\dot{\omega}}_h$ denote the respective species net production/consumption rate and heat release rate (HRR).

The chemical source terms ${\dot{\omega}}_k$ and ${\dot{\omega}}_h$ denote the $k^\tx{th}$ species net production/consumption rate, and heat release rate (HRR), respectively. The definition of HRR is given by
\begin{equation}
\label{eq:hrr}
\dot{\omega}_h = \sum_{k=1}^{N_{sp}} h^0_{f,k}{\dot{\omega}}_k ,
\end{equation}
while the $k^\tx{th}$ species production rate represents the mass of the species produced per unit volume and unit time (i.e. production of $\rho Y_k$ per unit time), and it is the sum of production rates $\dot{\omega}_{k,r}$ produced by all $N_r$ reactions, that is
\begin{equation}
\label{eq:prod_rate}
\dot{\omega}_k = \sum_{r=1}^{N_r} \dot{\omega}_{k,r} = W_k \sum_{r=1}^{N_{r}} \nu_{k,r} q_r ,
\end{equation}
with $W_k$ denoting molecular weight of species $k$, and $\nu_{k,r} = \nu''_{k,r} - \nu'_{k,r}$ denoting net stoichiometric coefficient of species $k$ in reaction $r$, while $\nu'_{k,r}$ and $\nu''_{k,r}$ are the molar stoichiometric coefficients of species $k$ in reaction $r$ for the reactant ($\nu'$) and product ($\nu''$) sides, respectively. The term $q_r$ indicates the progress rate of reaction $r$ and it is discussed in the following section.
Further details on theoretical derivations of combustion governing equations under various conditions can be revised from Poinsot and Veynante~\cite{poinsot2005theoretical} or from the book by Peters~\cite{peters2001turbulent}.

\section{Finite-rate chemistry}

Over the past fifty years, knowledge about combustion chemistry has significantly grown, especially for the gaseous phase, and nowadays it has reached a sufficient level of maturity. Considering an oxidation process of hydrocarbon fuel, this typically occurs through a chain of intermediate reactions, which becomes longer depending on the complexity of the fuel molecular structure. A reaction rate determines the speed at which chemical reaction progresses, such that it is proportional to the increase of products concentrations and decrease of reactants concentrations. In the combustion literature, information regarding the important set reactions and their rate data are gathered in a so-called chemical kinetic mechanism, hence the term \emph{finite-rate chemistry} highlighting the finite set of intermediate \emph{elementary} reactions of a chain reaction process.

Considering an elementary reaction $r$ with equation \ce{a + b <=> c + d}, where the double arrow symbol denotes reversible reaction, the forward rate constant ($k_\textrm{fw}$) is defined according to the modified Arrhenius law as
\begin{equation}
\label{eq:arrhenius}
k_{\textrm{fw},r} = A_r T^b e^{-E_a / \bar{R} T} ,
\end{equation}
and the forward reaction rate is defined by
\begin{equation}
\label{eq:rr}
R_{\textrm{fw},r} = [a] [b] k_{\textrm{fw},r} ,
\end{equation}
with square brackets denoting molar concentrations of individual reactants $a$ and $b$.

The parameters $A_r$, $b$, $E_a$, $\bar{R}$ in \cref{eq:arrhenius} denote the pre-exponential factor, the temperature exponent, the activation energy, and the universal gas constant, respectively. The first three parameters are tunable for a given reaction and it is a research topic to fit the parameters for particular thermodynamic conditions when developing chemical kinetic mechanism for particular fuel (or fuels) oxidation. Additionally, we note that other types of elementary reactions may occur in the chain reaction and thereby be represented in the chemical mechanism, such as third-body, pressure-dependent, surface reactions, and others. However, a discussion of these reaction types is beyond the scope of this report.

The most common style, by far, to represent a chemical kinetic mechanism is through \mbox{CHEMKIN-II} format. Such a format provides information regarding chemical kinetics (\texttt{chem} file), species thermodynamic properties (\texttt{thermo} file), and species transport properties (\texttt{transport} file). The chemical kinetics file provides information for chemical elements, species, and reactions which are represented by their formulae and the tunable parameters $A$, $b$, and $E_a$ of the reaction, in addition to further parameters in case of pressure-dependent reaction types. Thermodynamics file provides a set of $14$ coefficients for every species ($7$ coefficients for each the lower and higher temperature interval) which are best fit polynomials (termed as NASA polynomials) to represent temperature dependent heat capacity $C_p(T)$, molar enthalpy $H(T)$, and molar entropy $S(T)$. The transport file provides information regarding species molecular transport properties, which are polynomial fits for the temperature dependent dynamic viscosity, thermal conductivity, and binary diffusion coefficients. Chemical mechanisms are usually publicly available. A widely common mechanism for natural gas combustion is the GRI-3.0 (which we will also use during our demonstration in the report) available at \url{http://combustion.berkeley.edu/gri-mech} and downloaded from \url{http://combustion.berkeley.edu/gri-mech/version30/text30.html}.

At this point, we may resume the derivation of the chemical source term (i.e. net production rate of $k^{th}$ species, $\dot{\omega}_k$) noted in the previous section. To compute \cref{eq:prod_rate}, 
% the production rate of reaction $r$ in a mechanism containing $N_r$ reactions is given by
the progress rate of reaction $r$ in a mechanism containing $N_r$ reactions is first evaluated using the following relation
% the reaction rates of progress at current state.

\begin{align}
\label{eq:qr}
\begin{split}
q_r &= R_{\textrm{fw},r} - R_{\textrm{bw},r} \\
 &= k_{\textrm{fw},r} \prod_{k=1}^{N_{sp}} \big( \frac{\rho Y_k}{W_k} \big)^{\nu'_{k,r}} - k_{\textrm{bw},r} \prod_{k=1}^{N_{sp}} \big( \frac{\rho Y_k}{W_k} \big)^{\nu''_{k,r}} ,
\end{split}
\end{align}
in which the species concentration $\bar{c}_k = \big( \frac{\rho Y_k}{W_k} \big)$ is represented in terms of its mass fraction for practical reasons, as $Y_k$ is usually an input in most numerical solvers. Again, we note that the previous equation is limited to elementary reactions, and some manipulation (third-body/pressure scaling) would be needed to account for other reaction types. Now, it is obvious the relation between this equation and \cref{eq:rr}. Moreover, this equation can be further expanded using \cref{eq:arrhenius}.

\subsection{Evaluating source terms via direct integration}

Due to the vast difference between chemistry and flow time scales, it is not feasible to limit the flow solver according to the smallest chemical time scale.
% just to compute source terms of \cref{eq:prod_rate,eq:hrr}. 
Instead, it is common to utilize operator splitting approach to separate the calculation of chemistry source terms from the flow in \cref{eq:species,eq:enthalpy}. In this way, chemistry source terms will represent the change of thermo-chemical composition within a CFD time step, as we further explain in the following.

The change in local thermo-chemical composition is obtained by solving (integrating) a stiff, nonlinear initial value problem described as

\begin{equation}
\label{eq:ode1}
\centering
\begin{cases}
\partial \boldsymbol \Phi / \partial t =  \left\{\dfrac{\partial T}{\partial t}, \dfrac{\partial Y_1}{\partial t}, \dfrac{\partial Y_2}{\partial t},..., \dfrac{\partial Y_{N_{sp}}}{\partial t}\right\}^\intercal = \boldsymbol f(\boldsymbol \Phi,t) , \\
\boldsymbol \Phi(t=t_0) = \boldsymbol \Phi_0 ,
\end{cases}
\end{equation}
wherein the rate of change of temperature and species mass fractions are combined into a state vector $\boldsymbol \Phi$ of size $N_{sp}+1$ in order to be solved while making use of \cref{eq:hrr,eq:prod_rate} for the nonlinear function in the right hand side. Moreover, the change of thermo-chemical state is related to chemical source terms by the relation
\begin{equation}
\label{eq:chem_source}
\frac{\partial Y_k}{\partial t } = \frac{1}{\rho} \dot{\omega}_k ,
\end{equation}
and for HRR by the relation
\begin{equation}
\label{eq:hrr_source}
\frac{\partial T}{\partial t } = \frac{-1}{\rho c_p} \sum_{k=1}^{N_{sp}} h_k \dot{\omega}_k ,
\end{equation}
which can be further expanded using \cref{eq:hs}. The relationship between a species mass fraction and its concentration is given by
\begin{equation}
\label{eq:conv_y_c}
\bar{c}_k = \frac{\rho_k Y_k}{W_k} .
\end{equation}

The aforementioned stiff ODE system in \cref{eq:ode1} is independent for every computational cell since the mixture of that cell is defined by the system state $\boldsymbol \Phi_0$ at time $t_0$. After that, time integration is performed to compute source terms for chemistry and enthalpy after marching a complete CFD timestep ($\delta t_\tx{CFD}$). In addition, due to the stiffness of such ODE systems, implicit and semi-implicit ODE solvers are preferred to ensure solution stability~\cite{Hairer1996}. Accordingly, such solvers attempt to solve the stiff ODE system over $\delta t_\tx{CFD}$ interval by marching over subintervals denoted by $\delta t_{ODE}$, which are smaller than $\delta t_\tx{CFD}$.

As in most implicit solvers for stiff ODE systems, a system Jacobian, $\boldsymbol{\mathcal J} = {\partial \boldsymbol{f}} / {\partial \boldsymbol \Phi}$, is utilized as it is demonstrated below for a single ODE solution subinterval, i.e.

% \begin{linenomath*}
\begin{equation}
\label{eq:ode2}
\begin{split}
    \boldsymbol \Phi_{n+1} &= \boldsymbol \Phi_n + \int_{t_n}^{t_n+\Delta t_{ODE}} \left( \boldsymbol f_n + \boldsymbol{\mathcal J}_n (\boldsymbol \Phi_{n+1} - \boldsymbol \Phi_n) + \mathcal{O} (\Delta t_{ODE}^2) \right) dt \\
    &= \boldsymbol \Phi_n + \boldsymbol f_n \Delta t_{ODE} + \boldsymbol{\mathcal J}_n (\boldsymbol \Phi_{n+1} - \boldsymbol \Phi_n) \Delta t_{ODE} + \mathcal{O} (\Delta t_{ODE}^2),
\end{split}
\end{equation}
% \end{linenomath*}
%
which is then linearized by neglecting the higher-order terms. As noted from the equation, the Jacobian is required for every subinterval, which can be an expensive procedure.
The direct solution of the previous equation for $\boldsymbol \Phi_{n+1}$ requires the inverse of $\boldsymbol{\mathcal J}$, which is commonly avoided through matrix factorization (i.e. LU decomposition) and back substitution techniques. 

In \Cref{ch:of} we demonstrate how the underlying derivations, herein noted, are realized in the OpenFOAM software for both the upper level and lower level implementations.

\section{Chemical Explosive Mode Analysis}
\label{sec:cema}

Chemical Explosive Mode Analysis (CEMA) is considered a powerful post-processing technique for the computational diagnostics of combustion events~\cite{Lu2010}. It is based on dynamical system theory~\cite{lawrence1991differential,la1976stability} and the wider concept of computational singular perturbation~\cite{Lam1989,lam1985singular}. 
Considering \cref{eq:species,eq:enthalpy} while using species molar concentrations instead of mass fractions, they can be combined and rewritten in the Lagrangian form as follows %\todo{!check if fine to use Eulerian form}

\begin{equation}
\label{eq:cema_1}
% \frac{\partial \boldsymbol \Phi_c} {\partial t} = \dot{\boldsymbol \omega} (\boldsymbol \Phi_c) + \boldsymbol c(\boldsymbol \Phi_c) + \boldsymbol s(\boldsymbol \Phi_c) \equiv \boldsymbol g(\boldsymbol \Phi_c),
\frac{D \boldsymbol \Phi_c} {D t} = \dot{\boldsymbol \omega} (\boldsymbol \Phi_c) + \boldsymbol s(\boldsymbol \Phi_c) \equiv \boldsymbol g(\boldsymbol \Phi_c),
\end{equation}
in which $\boldsymbol \Phi_c$ is the thermo-chemical state vector, similarly defined in \cref{eq:ode1} but using species concentrations, $\dot{\boldsymbol \omega}$ vector is the chemistry source terms in species and energy equations, and
% $\boldsymbol c$ and
$\boldsymbol s$ vector comprises the non-chemical 
% convective and 
diffusive transport terms
% s, respectively
. The Jacobian matrix of the above problem is $\boldsymbol{\mathcal J_g} = {\partial \boldsymbol{g}} / {\partial \boldsymbol \Phi_c}$, which could be used for integration using Newton's method. By linearity of operators, this Jacobian is decomposed into

\begin{equation}
\label{eq:cema_jac}
% \boldsymbol{\mathcal J_g} = \frac{\partial \boldsymbol{g}} {\partial \boldsymbol \Phi_c} = \frac{\partial \boldsymbol{\omega}} {\partial \boldsymbol \Phi_c} + \frac{\partial \boldsymbol{s}} {\partial \boldsymbol \Phi_c} \equiv  \boldsymbol{\mathcal J_\omega} + \boldsymbol{\mathcal J_c} + \boldsymbol{\mathcal J_s} ,
\boldsymbol{\mathcal J_g} = \frac{\partial \boldsymbol{g}} {\partial \boldsymbol \Phi_c} = \frac{\partial \dot{\boldsymbol \omega}} {\partial \boldsymbol \Phi_c} + \frac{\partial \boldsymbol{s}} {\partial \boldsymbol \Phi_c} \equiv  \boldsymbol{\mathcal J_\omega} + \boldsymbol{\mathcal J_s} ,
\end{equation}
where $\boldsymbol{\mathcal J_\omega}$ is comparable to the chemical Jacobian matrix in \cref{eq:ode2}, i.e. $\boldsymbol{\mathcal J} = {\partial \boldsymbol{f}} / {\partial \boldsymbol \Phi}$, depending on whether species concentrations or mass fractions are adopted for thermo-chemical state vector.~\footnote{In case of mass fraction based state vector, Jacobian $\boldsymbol{\mathcal J_\omega}$ from CEMA derivations in \cref{eq:cema_jac} is equivalent to $\boldsymbol{\mathcal J}$ from \cref{eq:ode1}.} By multiplying \cref{eq:cema_1} by $\mathcal J_\omega$ from the left side, we get the following chemical dynamical system

\begin{equation}
\label{eq:cema_2}
% \frac{\partial \dot{\boldsymbol \omega}} {\partial t} = \boldsymbol{\mathcal J_\omega} ( \dot{\boldsymbol \omega} + \boldsymbol c + \boldsymbol s) .
% \frac{\partial \dot{\boldsymbol \omega}} {\partial t} = \boldsymbol{\mathcal J_\omega} ( \dot{\boldsymbol \omega} + \boldsymbol s) .
\frac{D \dot{\boldsymbol \omega}} {D t} = \boldsymbol{\mathcal J_\omega} ( \dot{\boldsymbol \omega} + \boldsymbol s) .
\end{equation}

CEMA is conceptually based on the stability of the aforementioned dynamical system. This is indicated by eigenvalue analysis of the Jacobian matrix $\boldsymbol{\mathcal J_\omega}$. Despite time dependence of the non-linear dynamical system, it has been widely accepted in the literature to use simple eigendecomposition such that a fully decoupled set of modes (or basis vectors) is achieved~\cite{Maas1992,Lu2001,Lu2010}. Therefore, the diagonal matrix $\boldsymbol \Lambda$ denoting the Jacobian eigenvalues is obtained by

\begin{equation}
\label{eq:cema_2}
\boldsymbol \Lambda = \boldsymbol B \cdot \boldsymbol{\mathcal J_\omega} \cdot \boldsymbol A ,
\end{equation}
wherein the matrices $\boldsymbol A$ and $\boldsymbol B$ are composed of column and row basis vectors, respectively, with \mbox{$\boldsymbol A = \boldsymbol B^{-1}$}. Moreover, the time dependence of matrix $\boldsymbol B$ is herein neglected, and $\boldsymbol \Lambda$ is diagonal provided that $\boldsymbol{\mathcal J_\omega}$ is not defective such that the modes are fully decoupled with leading order accuracy.

The chemical explosive subspace (i.e. unstable subspace) of the Jacobian matrix is spanned by the basis vectors representing Chemical Explosive Modes (CEMs). CEMs are associated with eigenvalues of real part that is positive, i.e.
\begin{equation}
\mathrm{Re}(\lambda_\mathrm{exp}) > 0 ,
\end{equation}
and $b_\mathrm{exp}$ and $a_\mathrm{exp}$ are left and right eigenvectors, respectively, of the Jacobian $\boldsymbol{\mathcal J_\omega}$ corresponding to $\lambda_\mathrm{exp}$. Typically, when multiple CEMs are present, $\lambda_\mathrm{exp}$ is chosen as the eigenvalue of the largest real part ---while excluding conservation modes--- hence denoting the fastest CEM along which chemical explosion occurs. On the other hand, non explosive modes are defined by eigenvalues of negative real part, i.e. stable modes of the dynamical system.

From a physical perspective, the existence of CEM indicates the propensity of a local mixture to auto-ignite if isolated, i.e. considering adiabatic and constant volume environment. Simply, a CEM is present in pre-ignition mixtures and absent in post-ignition mixtures. The transition of a CEM from explosive to non-explosive, i.e. zero-crossing eigenvalues, is strongly correlated to critical flame features such as ignition, extinction, and premixed flame front locations~\cite{Lu2010}.

One important note is that, due to conservation of energy and species elements, the chemical Jacobian $\boldsymbol{\mathcal J_\omega}$ is typically constituted of $M+1$ zero eigenvalues, where $M$ is the number of elements in species composition. The element conservation modes are typically identified by examining the magnitude of the eigenvalues.
The energy conservation mode is, however, more challenging to identify and distinguish from other
slow chemical modes. The main reason is the non linear temperature-dependent thermodynamics including heat capacity, which could induce nonlinear effects from eigenvector rotation. 
% Accordingly, for applications where dynamics of the energy conservation mode is not of particular interest, including the time dependence of heat capacity, then such effects are dynamic neglected and the energy conservation mode is trivial and corresponds to zero eigenvalue.
Accordingly, for applications where dynamics of the energy conservation mode is not of particular interest, for instance when time dependence of heat capacity is not relevant, then the energy conservation mode becomes trivial and corresponds to zero eigenvalue. This is further clarified during the present implementation of CEMA in \Cref{ch:cema}, where $M+1$ modes are to be skipped.

\chapter{Existing OpenFOAM features}
\label{ch:of}

\section{High-level overview of \texttt{reactingFoam} solver}

In this section, we highlight the high-level implementations of the \verb|reactingFoam| solver as well as the mathematical equations of flow and thermo-chemistry discussed in \Cref{ch:theory}. 

\subsection{Included header files}

We start with the \verb|reactingFoam.C| file and walk it through, first, by listing the included header files with a brief description as shown in \Cref{ls_st1,ls_st2}.

% Copy the \verb|pitzDaily| case in your \verb|run| directory with this command:
% \begin{verbatim}
% cp -r  $FOAM_TUTORIALS/incompressible/simpleFoam/pitzDaily
% \end{verbatim}

\begin{lstlisting}[label={ls_st1}, style=cpp, caption=Initial header files in the \texttt{reactingFoam} solver,captionpos=t, linerange={37-46},firstnumber=37]
#include "fvCFD.H"    
#include "turbulentFluidThermoModel.H" // Typedef for turbulence models (compressibleTurbulenceModels)
#include "psiReactionThermo.H" // Declare class psiReactionThermo (reactionThermophysicalModels)
#include "CombustionModel.H" // Combustion models for templated thermodynamics (combustionModels)
#include "multivariateScheme.H" // Generic multi-variate discretization scheme (finiteVolume)
#include "pimpleControl.H" // Convergence information/checks for PIMPLE loop (finiteVolume)
#include "pressureControl.H" // Provide controls for the pressure reference (finiteVolume)
#include "fvOptions.H" // Finite-volume options (finiteVolume)
#include "localEulerDdtScheme.H" // First-order Euler implicit/explicit ddt scheme (finiteVolume)
#include "fvcSmooth.H" // Smooth and redistribute fields during face interpolation
\end{lstlisting}

% \setlength{\unitlength}{1cm}
% \thicklines
% \begin{picture}(1,1)
% \newcommand*{\ytop}{4.25}
% \newcommand*{\dy}{0.33}
% % \line(x-slope,y-slope){length}
% \put(5.5,\ytop){\line(1,0){0.5} \scriptsize Typedefs for turbulence models (\texttt{compressibleTurbulenceModels})}
% % \newcommand{\yone}{\pgfmathparse{\ytop-1*\dy}\pgfmathresult}
% % \yone
% \put(4.25,3.92){\line(1,0){1.75} \scriptsize Declare class psiReactionThermo (\texttt{reactionThermophysicalModels})}
% % \newcommand{\ytwo}{\pgfmathparse{\ytop-2*\dy}\pgfmathresult}
% \put(4.0,3.59){\line(1,0){2.0} \scriptsize Combustion models for templated thermodynamics (\texttt{combustionModels})}
% % \newcommand{\ythree}{\pgfmathparse{\ytop-3*\dy}\pgfmathresult}
% \put(4.5,3.26){\line(1,0){1.5} \scriptsize Generic multi-variate discretisation scheme (\texttt{finiteVolume})}
% % \newcommand{\yfour}{\pgfmathparse{\ytop-4*\dy}\pgfmathresult}
% \put(3.5,2.93){\line(1,0){2.5} \scriptsize Convergence information/checks for PIMPLE loop (\texttt{finiteVolume})}
% \put(3.75,2.56){\line(1,0){2.25} \scriptsize Provide controls for the pressure reference (\texttt{finiteVolume})}
% \put(3.0,2.27){\line(1,0){3.0} \scriptsize Finite-volume options (\texttt{finiteVolume})}
% \put(4.5,1.94){\line(1,0){1.5} \scriptsize First-order Euler implicit/explicit ddt scheme (\texttt{finiteVolume})}
% \put(3.0,1.61){\line(1,0){3.0} \scriptsize Smooth and redistribute field during face interpolation (\texttt{finiteVolume})}
% \put(2,2.2){\line(1,0){6}}
% \put(10,10){TEST}
% \end{picture}
% \vspace{-15mm}

\begin{lstlisting}[label={ls_st2}, style=cpp,caption=Headers inclusion inside \texttt{fvCFD.H},captionpos=t, firstnumber=4]
#include "parRun.H" // routines for initializing the parallel run and for finalizing it

#include "Time.H"   // controls the information of Time during the simulations
#include "fvMesh.H" // topological and geometric information related to mesh for FV discretization
#include "fvc.H" // ''explicit'' calculus operations and geometric field discretization, return geometricField (contribute to source term b in Ax=b)
#include "fvMatrices.H" 
#include "fvm.H" // ''implicit'' calculus operations and geometric field discretization, return fvMatrix (contribute to coefficient matrix A in Ax=b)
#include "linear.H" // central-differencing interpolation scheme
#include "uniformDimensionedFields.H"
#include "calculatedFvPatchFields.H" // macro, add BC to run-time selection table, set debug switch
#include "extrapolatedCalculatedFvPatchFields.H" // similar to above
#include "fixedValueFvPatchFields.H"   // similar to above
#include "zeroGradientFvPatchFields.H" // similar to above
#include "fixedFluxPressureFvPatchScalarField.H" // class for BC
#include "constrainHbyA.H"     // relevant to PISO algorithm
#include "constrainPressure.H"
#include "adjustPhi.H"
#include "findRefCell.H"       // find reference cell nearest to given index
#include "IOMRFZoneList.H"     // list of MRF zones
#include "constants.H"         // various constants defined within FOAM namespace
#include "gravityMeshObject.H" // gravitational acceleration vector

#include "columnFvMesh.H"      // generates a 1D column representation of a mesh

#include "OSspecific.H" // functions specific to POSIX compliant operating systems
#include "argList.H" // extract command arguments and options from the supplied from argc and argv
#include "timeSelector.H" // allows selecting list of time ranges in simulation
\end{lstlisting}

After that, in the main function of \verb|reactingFoam.C|, we have additional header files that are included together with brief descriptions, as shown in \Cref{ls_solver_2}.

\begin{lstlisting}[label=ls_solver_2, style=cpp, caption=Included header files after main() in the \texttt{reactingFoam} solver,captionpos=t, linerange={57-67},firstnumber=57]
#include "postProcess.H" // Execute application functionObjects to post-process existing results

#include "addCheckCaseOptions.H" // allow dry-run in supplied command-line flags
#include "setRootCaseLists.H" // setRootCase to check validity of case dir, (with additional solver-related listings)
#include "createTime.H"       // constructs the runTime object of the class Time
#include "createMesh.H"       // constructs mesh object (reference) of class fvMesh
#include "createControl.H"    // allow control options for PISO, SIMPLE, PIMPLE
#include "createTimeControls.H" // read control parameters used by setDeltaT (adjustTimeStep, maxCo, maxDeltaT)
#include "initContinuityErrs.H" // declare and initialise cumulative continuity error
#include "createFields.H"
#include "createFieldRefs.H"
\end{lstlisting}

\subsection{Created fields with glimpse into libraries}

Here, we discuss the created fields noted in \verb|createFields.H| file of the \verb|reactingFoam| solver directory. 
% It is straight forward the geometricFields that are created and stored to object registry, namely the density, velocity, pressure, face flux $\bar{\phi}$,
% % $ = \verb|linearInterpolate(rho*U) & mesh.Sf()|$ defined in \verb|$(FOAM_SRC)/finiteVolume/cfdTools/compressible|, 
% transient pressure \verb|dpdt|, kinetic energy \verb|K|, and HRR (i.e. ${\dot{\omega}_h}$, c.f. \cref{eq:enthalpy}) defined by \verb|Qdot| in OpenFOAM.
The creation of the \verb|GeometricField| objects is fairly straight forward and we only show one example for the velocity field $U$ in \Cref{ls_create_field}~\footnote{Note that \texttt{volVectorField} is just typedef of the \texttt{GeometricField<vector, fvPatchField, volMesh>} templated class, and it is defined in \texttt{volFieldsFwd.H}.}. The remaining \verb|GeometricField| objects that are created and stored to object registry are the density $\rho$, pressure $p$, face flux $\bar{\phi}$,
% $ = \verb|linearInterpolate(rho*U) & mesh.Sf()|$ defined in \verb|$(FOAM_SRC)/finiteVolume/cfdTools/compressible|, 
transient pressure \verb|dpdt|, kinetic energy $K$, and HRR (i.e. ${\dot{\omega}_h}$, c.f. \cref{eq:enthalpy}) defined by \verb|Qdot| in OpenFOAM.

\begin{lstlisting}[label={ls_create_field}, style=cpp,caption=Example of reading geometric fields in \texttt{createFields.H},captionpos=t, firstnumber=43]
volVectorField U // variable type and name
(
    IOobject
    (
        "U",
        runTime.timeName(),
        mesh,
        IOobject::MUST_READ, // return error if ``U'' field not found in case time dir
        IOobject::AUTO_WRITE
    ),
    mesh
);
\end{lstlisting}

% volFieldsFwd.H
% typedef GeometricField<vector, fvPatchField, volMesh> volVectorField;

One important note is the selector function \verb|New()| in the runtime selection mechanism in OpenFOAM.
It is implemented in the base class with the purpose of creating an object of derived classes
according to an input that we provide during runtime. In \Cref{ls_selector}, we have three selectors
for thermophysical model, turbulence model, and reaction model. 

\begin{lstlisting}[label={ls_selector}, style=cpp,caption=Selector functions in \texttt{createFields.H} to create object of derived classes during runTime,captionpos=t, firstnumber=3, ,mathescape=true]
Info<< "Reading thermophysical properties\n" << endl;
autoPtr<psiReactionThermo> pThermo(psiReactionThermo::New(mesh));
psiReactionThermo& thermo = pThermo();

basicSpecieMixture& composition = thermo.composition();
PtrList<volScalarField>& Y = composition.Y(); $\lstsetnumber{\ldots}$
$\lstresetnumber\setcounter{lstnumber}{51}$ 
Info << "Creating turbulence model.\n" << nl;
autoPtr<compressible::turbulenceModel> turbulence
(
    compressible::turbulenceModel::New
    (
        rho,
        U,
        phi,
        thermo
    )
);

Info<< "Creating reaction model\n" << endl;
autoPtr<CombustionModel<psiReactionThermo>> reaction
(
    CombustionModel<psiReactionThermo>::New(thermo, turbulence())
);
\end{lstlisting}

In the thermophysical model case, the function \verb|New()| is called from the derived class \verb|psiReactionThermo| which overrides implementation in the base class \verb|basicThermo| and it is implemented in the library \verb|libreactionThermophysicalModels|.
% The operator \verb|()| returns a constant reference to the object for modification. 

% In the second case, the pointer \verb|turbulence| is constructed and calls the function \verb|New()| belonging to the namespace \verb|Foam::compressible| and the class \verb|turbulenceModel| which is a typeDef of \verb|ThermalDiffusivity<CompressibleTurbulenceModel<fluidThermo>>|, and together with the \verb|New()| function they are declared in \verb|turbulentFluidThermoModel.H| with implementation in the shared library \verb|libcompressibleTurbulenceModels|. 
In the turbulence model case, the pointer \verb|turbulence| is constructed and it calls \verb|New()| function as argument. Such a function \verb|New()| belongs to the namespace \verb|Foam::compressible| and the class \verb|turbulenceModel| which is typedef \verb|ThermalDiffusivity<CompressibleTurbulenceModel<fluidThermo>>|. Both the class \verb|turbulenceModel| and the \verb|New()| function are declared in \verb|turbulentFluidThermoModel.H| with implementation in the corresponding source file that is compiled to create the shared library \verb|libcompressibleTurbulenceModels|.

% In the third case, the variable \verb|reaction| is a templated pointer with the parameter being CombustionModel class that is also templated by the parameter psiReactionThermo class. 
In the reaction model case, the variable \verb|reaction| is a pointer to an object of a templeted class, with the parameter being CombustionModel class that is also templated by the parameter psiReactionThermo class. 
The selector function \verb|New()| belongs to the templated class \verb|CombustionModel| declared in \verb|CombustionModel.H| and implemented in the corresponding source code of same name, while the library \verb|libcombustionModels| contains binaries of the compiled code. The selector function takes two non-default arguments of type \verb|ReactionThermo| and \verb|compressibleTurbulenceModel|. This implies that the combustion model depends on both reactive thermodynamics and turbulence properties
%and it is standard selection based on fvMesh

Finally, we note that the implementation of \verb|New()| functions is quite complicated and the present report does not provide detailed discussions. Therefore, we advise interested readers to further dig through the code themselves or to find other material that better covers such details.

% \clearpage

% To sum up, the fv::convectionScheme<Type>::New followed by () operator returns an
% object of one of convectionScheme’s sub-classes based on input at the runtime.

% autoPtr<psiReactionThermo> pThermo(psiReactionThermo::New(mesh));
% psiReactionThermo& thermo = pThermo();
% basicSpecieMixture& composition = thermo.composition();
% PtrList<volScalarField>& Y = composition.Y();
%% Fields for rho, U, p, 
% autoPtr<CombustionModel<psiReactionThermo>> reaction
% (
%     CombustionModel<psiReactionThermo>::New(thermo, turbulence())
% );
% Qdot

\subsection{Linear systems in the time loop}

Now, we proceed with \verb|reactingFoam.C| and take a look into the main part of time loop (\Cref{ls_solver}). There, we find inclusion of \verb|rhoEqn.H| (located in \verb|$(FOAM_SRC)/finiteVolume/cfdTools/compressible|) and the files \verb|UEqn.H|, \verb|YEqn.H|, \verb|EEqn.H|, and \verb|pEqn.H| (located in the solver directory) which are responsible for solving the governing equations for mass, momentum, species, energy, and pressure-momentum coupling, respectively. We will not go through all details, but we show only the relevant parts that reflect discussions from the previous chapter.

% By examining the previously mentioned files, we find that the governing equations are represented as in the following listing. We can find that such equations for mass, momentum, species, and energy are comparable with those in \cref{eq:mass,eq:mom,eq:species,eq:enthalpy} if we discard additional terms from \verb|MRF| or \verb|fvOptions|. We note that \verb|reaction->R(Yi)| denotes species overall production rate ${\dot{\omega}_k}$ which will be further discussed in \Cref{sec:walkstandard}.
The previously mentioned files are represented in \Cref{ls_solver_rhoEqn,ls_solver_UEqn,ls_solver_YEqn,ls_solver_EEqn}, while the pressure-momentum coupling implementation is detailed in \Cref{ls_solver_pEqn}. The finite-volume system equations for mass, momentum, species, and energy are comparable with those conservation equations in \cref{eq:mass,eq:mom,eq:species,eq:enthalpy}, assuming that we discard additional terms from e.g. \verb|MRF| and \verb|fvOptions|. Moreover, we note that \verb|reaction->R(Yi)| (in \verb|YEqn.H|) denotes species overall production rate ${\dot{\omega}_k}$ which is further discussed in \Cref{sec:walkstandard}.

\begin{lstlisting}[label=ls_solver, style=cpp, caption=Solving conservation equations inside time loop of \texttt{reactingFoam.C},captionpos=t, linerange={95-124},firstnumber=95]
        ++runTime;

        Info<< "Time = " << runTime.timeName() << nl << endl;

        #include "rhoEqn.H"

        while (pimple.loop())
        {
            #include "UEqn.H"
            #include "YEqn.H"
            #include "EEqn.H"

            // --- Pressure corrector loop
            while (pimple.correct())
            {
                if (pimple.consistent())
                {
                    #include "pcEqn.H"
                }
                else
                {
                    #include "pEqn.H"
                }
            }

            if (pimple.turbCorr())
            {
                turbulence->correct();
            }
        }
\end{lstlisting}

\begin{lstlisting}[label=ls_solver_rhoEqn, style=cpp, caption=Finite volume system for mass conservation (continuity) included from \texttt{rhoEqn.H},captionpos=t, linerange={35-47},firstnumber=35]
    fvScalarMatrix rhoEqn
    (
        fvm::ddt(rho)
      + fvc::div(phi)
      ==
        fvOptions(rho)
    );

    fvOptions.constrain(rhoEqn);

    rhoEqn.solve();

    fvOptions.correct(rho);
\end{lstlisting}

\begin{lstlisting}[label=ls_solver_UEqn, style=cpp, caption=Finite volume system for momentum conservation included from \texttt{UEqn.H},captionpos=t, linerange={5-25},firstnumber=5]
tmp<fvVectorMatrix> tUEqn
(
    fvm::ddt(rho, U) + fvm::div(phi, U)
  + MRF.DDt(rho, U)
  + turbulence->divDevRhoReff(U)
 ==
    fvOptions(rho, U)
);
fvVectorMatrix& UEqn = tUEqn.ref();

UEqn.relax();

fvOptions.constrain(UEqn);

if (pimple.momentumPredictor())
{
    solve(UEqn == -fvc::grad(p));

    fvOptions.correct(U);
    K = 0.5*magSqr(U);
}
\end{lstlisting}

\begin{lstlisting}[label=ls_solver_YEqn, style=cpp, caption=Finite volume system for species conservation included from \texttt{YEqn.H},captionpos=t, linerange={13-47},firstnumber=13]
    reaction->correct(); // Here chemical source term is computed
    Qdot = reaction->Qdot(); // Here HRR is computed
    volScalarField Yt(0.0*Y[0]);

    forAll(Y, i)
    {
        if (i != inertIndex && composition.active(i))
        {
            volScalarField& Yi = Y[i];

            fvScalarMatrix YiEqn
            (
                fvm::ddt(rho, Yi)
              + mvConvection->fvmDiv(phi, Yi)
              - fvm::laplacian(turbulence->muEff(), Yi)
             ==
                reaction->R(Yi)
              + fvOptions(rho, Yi)
            );

            YiEqn.relax();

            fvOptions.constrain(YiEqn);

            YiEqn.solve(mesh.solver("Yi"));

            fvOptions.correct(Yi);

            Yi.max(0.0);
            Yt += Yi;
        }
    }

    Y[inertIndex] = scalar(1) - Yt;
    Y[inertIndex].max(0.0);
\end{lstlisting}

\begin{lstlisting}[label=ls_solver_EEqn, style=cpp, caption=Finite volume system for energy conservation included from \texttt{EEqn.H},captionpos=t, linerange={2-32},firstnumber=2]
    volScalarField& he = thermo.he();

    fvScalarMatrix EEqn
    (
        fvm::ddt(rho, he) + mvConvection->fvmDiv(phi, he)
      + fvc::ddt(rho, K) + fvc::div(phi, K)
      + (
            he.name() == "e"
          ? fvc::div
            (
                fvc::absolute(phi/fvc::interpolate(rho), U),
                p,
                "div(phiv,p)"
            )
          : -dpdt
        )
      - fvm::laplacian(turbulence->alphaEff(), he)
     ==
        Qdot
      + fvOptions(rho, he)
    );

    EEqn.relax();

    fvOptions.constrain(EEqn);

    EEqn.solve();

    fvOptions.correct(he);

    thermo.correct();
\end{lstlisting}

\begin{lstlisting}[label=ls_solver_pEqn, style=cpp, caption=Equations for pressure-momentum coupling included from \texttt{pEqn.H},captionpos=t]
rho = thermo.rho();

volScalarField rAU(1.0/UEqn.A());
surfaceScalarField rhorAUf("rhorAUf", fvc::interpolate(rho*rAU));
volVectorField HbyA(constrainHbyA(rAU*UEqn.H(), U, p));

if (pimple.nCorrPISO() <= 1)
{
    tUEqn.clear();
}

if (pimple.transonic())
{
    surfaceScalarField phid
    (
        "phid",
        fvc::interpolate(psi)
       *(
            fvc::flux(HbyA)
          + MRF.zeroFilter
            (
                rhorAUf*fvc::ddtCorr(rho, U, phi)/fvc::interpolate(rho)
            )
        )
    );

    MRF.makeRelative(fvc::interpolate(psi), phid);

    while (pimple.correctNonOrthogonal())
    {
        fvScalarMatrix pEqn
        (
            fvm::ddt(psi, p)
          + fvm::div(phid, p)
          - fvm::laplacian(rhorAUf, p)
         ==
            fvOptions(psi, p, rho.name())
        );

        pEqn.solve(mesh.solver(p.select(pimple.finalInnerIter())));

        if (pimple.finalNonOrthogonalIter())
        {
            phi == pEqn.flux();
        }
    }
}
else
{
    surfaceScalarField phiHbyA
    (
        "phiHbyA",
        (
            fvc::flux(rho*HbyA)
          + MRF.zeroFilter(rhorAUf*fvc::ddtCorr(rho, U, phi))
        )
    );

    MRF.makeRelative(fvc::interpolate(rho), phiHbyA);

    // Update the pressure BCs to ensure flux consistency
    constrainPressure(p, rho, U, phiHbyA, rhorAUf, MRF);

    while (pimple.correctNonOrthogonal())
    {
        fvScalarMatrix pEqn
        (
            fvm::ddt(psi, p)
          + fvc::div(phiHbyA)
          - fvm::laplacian(rhorAUf, p)
         ==
            fvOptions(psi, p, rho.name())
        );

        pEqn.solve(mesh.solver(p.select(pimple.finalInnerIter())));

        if (pimple.finalNonOrthogonalIter())
        {
            phi = phiHbyA + pEqn.flux();
        }
    }
}

#include "rhoEqn.H"
#include "compressibleContinuityErrs.H"

// Explicitly relax pressure for momentum corrector
p.relax();

U = HbyA - rAU*fvc::grad(p);
U.correctBoundaryConditions();
fvOptions.correct(U);
K = 0.5*magSqr(U);

if (pressureControl.limit(p))
{
    p.correctBoundaryConditions();
}

rho = thermo.rho();

if (thermo.dpdt())
{
    dpdt = fvc::ddt(p);
}
\end{lstlisting}

% \begin{lstlisting}[style=cpp,caption=System matrix equations to be solved in \texttt{reactingFoam.C} expanded from various headers,captionpos=t]
% fvScalarMatrix rhoEqn
% (
%     fvm::ddt(rho)
%   + fvc::div(phi)
%   ==
%     fvOptions(rho)
% );

% tmp<fvVectorMatrix> tUEqn
% (
%     fvm::ddt(rho, U) + fvm::div(phi, U)
%   + MRF.DDt(rho, U)
%   + turbulence->divDevRhoReff(U)
%  ==
%     fvOptions(rho, U)
% );
% //... some other code
% if (pimple.momentumPredictor())
% {
%     solve(UEqn == -fvc::grad(p));

%     fvOptions.correct(U);
%     K = 0.5*magSqr(U);
% }

% forAll(Y, i)
% {
%     //... some other code
%         fvScalarMatrix YiEqn
%         (
%             fvm::ddt(rho, Yi)
%           + mvConvection->fvmDiv(phi, Yi)
%           - fvm::laplacian(turbulence->muEff(), Yi)
%          ==
%             reaction->R(Yi)
%           + fvOptions(rho, Yi)
%         );
%     //... some other code
% }

% fvScalarMatrix EEqn
% (
%     fvm::ddt(rho, he) + mvConvection->fvmDiv(phi, he)
%   + fvc::ddt(rho, K) + fvc::div(phi, K)
%   + (
%         he.name() == "e"
%       ? fvc::div
%         (
%             fvc::absolute(phi/fvc::interpolate(rho), U),
%             p,
%             "div(phiv,p)"
%         )
%       : -dpdt
%     )
%   - fvm::laplacian(turbulence->alphaEff(), he)
%  ==
%     Qdot
%   + fvOptions(rho, he)
% );

% fvScalarMatrix pEqn
% (
%     fvm::ddt(psi, p)
%   + fvm::div(phid, p)
%   - fvm::laplacian(rhorAUf, p)
%  ==
%     fvOptions(psi, p, rho.name())
% );

% \end{lstlisting}

A particular focus in this report is to understand how chemical source terms are evaluated, as we are interested in manipulating the solution algorithm with our developed library. In order to achieve this, we have to dive deeper in the libraries, particularly the interaction between \verb|combustionModel|, \verb|chemistryModel|, and \verb|reactionThermo|, before which we manipulate the standard chemistryModel implementations.

% \section{Walk through standard chemistry models}
\section[Walk-through to compute chemistry source terms]{A walk-through to compute chemistry source terms}
\label{sec:walkstandard}

Here, we discuss the standard implementation of finite-rate chemistry for computing the chemical source terms, ${\dot{\omega}}_k$, and subsequently HRR, ${\dot{\omega}}_h$. First, let us start from the solver level, again, and try to track back the function call \verb|reaction->correct()| from \verb|YEqn.H| (\Cref{ls_solver_YEqn}). As previously noted, \verb|reaction| is just a pointer to the employed combustion model. In the present development, we consider the implementation of finite-rate chemistry under \emph{laminar} conditions, i.e. no subgrid modeling for turbulence-chemistry interactions. Therefore, we check the implementation of \verb|correct()| function in such a combustion model from \verb|$(FOAM_SRC)/combustionModels/laminar/laminar.C|. We see that the implementation is conditioned when \verb|active| flag is enabled in the \verb|combustionProperties| dictionary (c.f. constructor in \verb|combustionModel.C|). Furthermore, the following condition is for integrating the reaction rate over CFD time step ($\delta t_\tx{CFD}$). This is activated by default if the laminar model is chosen in the \verb|combustionProperties| and the keyword \verb|integrateReactionRate| is not added in the dictionary (c.f. constructor in \verb|laminar.C|). The third condition is for the \verb|ddt| scheme, whether or not it is first order Euler implicit/explicit ddt, since the reciprocal of local time step will have to be looked up from database. In combustion simulations it is common to use a higher order \verb|ddt| schemes so this will leave us with the only function call
\begin{verbatim}
this->chemistryPtr_->solve(this->mesh().time().deltaTValue());
\end{verbatim}

Now, we need to track the pointer \verb|chemistryPtr_|. In fact, it is a pointer to chemistry model, and it is declared in \verb|ChemistryCombustion.H| with type \verb|autoPtr<BasicChemistryModel<ReactionThermo>>|. Moreover, it is initialized
 % in \verb|ChemistryCombustion.C|
using selector function \verb|BasicChemistryModel<ReactionThermo>::New(thermo)|.
This gives us hint that the function \verb|solve(this->mesh().time().deltaTValue())| should be defined in some derived class of the \verb|BasicChemistryModel| that is chosen during runTime.

Let us now go and check possible derived classes of \verb|BasicChemistryModel|.
% , located in \verb|thermophysicalModels/chemistryModel/chemistryModel| in \verb|src|
We only see that possible models are the templated \verb|StandardChemistryModel| denoting standard and direct integration of the chemistry ODE system, and the \verb|TDACChemistryModel| denoting the tabulated dynamic adaptive model, which is a tabulation-based strategy for chemistry calculations. The present report is based on developments on top of the standard chemistry model, so let us proceed with this approach.

Now, after checking for the function \verb|solve()| we find that it is defined in various forms inside \verb|StandardChemistryModel.C|. Through a cross comparison, we realize that the function call \verb|solve(this->mesh().time().deltaTValue())| from laminar combustion model calls the function \verb|solve(const scalar deltaT)| in standard chemistry model. Accordingly, we further proceed with the implementation of this function while showing the interesting part in \Cref{ls_sttmp}.

\begin{lstlisting}[label={ls_sttmp}, style=cpp, caption=Part of the \texttt{solve()} function definition in \texttt{StandardChemistryModel.C},captionpos=t, linerange={706-750},firstnumber=706]
    tmp<volScalarField> trho(this->thermo().rho());
    const scalarField& rho = trho();

    const scalarField& T = this->thermo().T();
    const scalarField& p = this->thermo().p();

    scalarField c0(nSpecie_);

    forAll(rho, celli)
    {
        scalar Ti = T[celli];

        if (Ti > Treact_)
        {
            const scalar rhoi = rho[celli];
            scalar pi = p[celli];

            for (label i=0; i<nSpecie_; i++)
            {
                c_[i] = rhoi*Y_[i][celli]/specieThermo_[i].W();
                c0[i] = c_[i];
            }

            // Initialise time progress
            scalar timeLeft = deltaT[celli];

            // Calculate the chemical source terms
            while (timeLeft > SMALL)
            {
                scalar dt = timeLeft;
                this->solve(c_, Ti, pi, dt, this->deltaTChem_[celli]);
                timeLeft -= dt;
            }

            deltaTMin = min(this->deltaTChem_[celli], deltaTMin);

            this->deltaTChem_[celli] =
                min(this->deltaTChem_[celli], this->deltaTChemMax_);

            for (label i=0; i<nSpecie_; i++)
            {
                RR_[i][celli] =
                    (c_[i] - c0[i])*specieThermo_[i].W()/deltaT[celli];
            }
        }
\end{lstlisting}

Here, we may realize few notes. First, inside the cells loop, a scalar \verb|timeLeft| is initialized with, probably, $\delta t_\tx{CFD}$ of a computational cell. Second, as long as the value of \verb|timeLeft| did not vanish, another function named \verb|solve()| is also called, which is inherited from another class, and it takes the pressure, temperature, and species concentrations, hence the thermo-chemical composition of that computational cell, in addition to two more variables related to time steps which are not CFD time steps. Moreover, the comment above the \verb|while| loop states that such a scope is responsible for computing chemical source terms. Based on our educated guess, this part of the code implements is relevant to the ODE solution routines which are previously noted in \cref{eq:ode1,eq:chem_source,eq:hrr_source,eq:ode2}. More importantly, the last loop in the function definition updates the chemistry source terms for each computational cell, which is exactly \cref{eq:chem_source} while converting mass fractions into concentrations using \cref{eq:conv_y_c}.

Now, let us continue our investigation to track the function call 
\begin{verbatim}
this->solve(c_, Ti, pi, dt, this->deltaTChem_[celli]).
\end{verbatim}
In fact, this function form has been declared as pure abstract in \verb|StandardChemistryModel.H|. After quick check, we find that a definition of this function exists in the derived class \verb|chemistrySolver|. Therefore, we understand now that the function will be executed according to the chemistry solver. In this report, we base our development on ODE approach (i.e. not implicit Euler), hence we check the \verb|solve()| function definition in \verb|ode.C| which is presented in the following \Cref{ls_ode}.

\begin{lstlisting}[style=cpp, label=ls_ode, caption=Definition of the \texttt{solve()} function in \texttt{ode.C} of chemistrySolver,captionpos=t, linerange={51-86},firstnumber=51]
template<class ChemistryModel>
void Foam::ode<ChemistryModel>::solve
(
    scalarField& c,
    scalar& T,
    scalar& p,
    scalar& deltaT,
    scalar& subDeltaT
) const
{
    // Reset the size of the ODE system to the simplified size when mechanism
    // reduction is active
    if (odeSolver_->resize())
    {
        odeSolver_->resizeField(cTp_);
    }

    const label nSpecie = this->nSpecie();

    // Copy the concentration, T and P to the total solve-vector
    for (int i=0; i<nSpecie; i++)
    {
        cTp_[i] = c[i];
    }
    cTp_[nSpecie] = T;
    cTp_[nSpecie+1] = p;

    odeSolver_->solve(0, deltaT, cTp_, subDeltaT);

    for (int i=0; i<nSpecie; i++)
    {
        c[i] = max(0.0, cTp_[i]);
    }
    T = cTp_[nSpecie];
    p = cTp_[nSpecie+1];
}
\end{lstlisting}

The interesting part of this function is that it first constructs a solve vector \verb|cTp_| comprising thermo-chemical composition (concentrations, temperature, pressure). After that, it calls another \verb|solve()| function using the pointer \verb|odeSolver_| which is of mutable type \verb|autoPtr<ODESolver>| and it is initialized using \verb|New()| selector function from \verb|ODESolver| class in \verb|$(FOAM_SRC)/ODE|. At this level, the class \verb|ODE| does not know (or need) any information about the chemistry or thermophysics of the problem. It is purely mathematical procedure at this point.

In this report, we are not interested in the implementation details of the stiff ODE solvers, nor the various algorithms and their comparative performance. Instead, we note an important point which is that \verb|StandardChemistryModel| inherits from both \verb|BasicChemistryModel| and the abstract class \verb|ODESystem|. The reason of the latter inheritance is that chemistry model implements three important functions which are declared pure virtual in \verb|ODESystem| as provided in the following \Cref{ls_ode_system}.

\begin{lstlisting}[style=cpp, label=ls_ode_system, caption=Abstract class \texttt{ODESystem} with 3 functions implemented in \texttt{chemistryModel},captionpos=t, linerange={50-88},firstnumber=50]
class ODESystem
{

public:

    // Constructors

        //- Construct null
        ODESystem()
        {}


    //- Destructor
    virtual ~ODESystem() = default;


    // Member Functions

        //- Return the number of equations in the system
        virtual label nEqns() const = 0;

        //- Calculate the derivatives in dydx
        virtual void derivatives
        (
            const scalar x,
            const scalarField& y,
            scalarField& dydx
        ) const = 0;

        //- Calculate the Jacobian of the system
        //  Need by the stiff-system solvers
        virtual void jacobian
        (
            const scalar x,
            const scalarField& y,
            scalarField& dfdx,
            scalarSquareMatrix& dfdy
        ) const = 0;
};
\end{lstlisting}

As it will be explained in the following chapter, we are interested in replacing the difference-based implementation of the Jacobian function with another analytical formulation using the external source code generator \verb|pyJac|. Accordingly, replacing the Jacobian function will result into various functions that would need to be updated to accommodate for the changed Jacobian matrix. We provide detailed discussion in the next chapter.

% \section{Summary of classes interactions}
% \section[UML sequence diagram of chemistry source term function calls]{Summary of function calls for chemistry source terms}
\section[UML sequence diagram]{Summary of chemistry source terms function calls}
\label{sec:uml}
% \todo{UML diagrams for interactions between CombustionModel, chemistryModel, ODE, and solver to compute chemistry source terms}

\begin{figure}[h!]
\centering
\includegraphics[width=\textwidth]{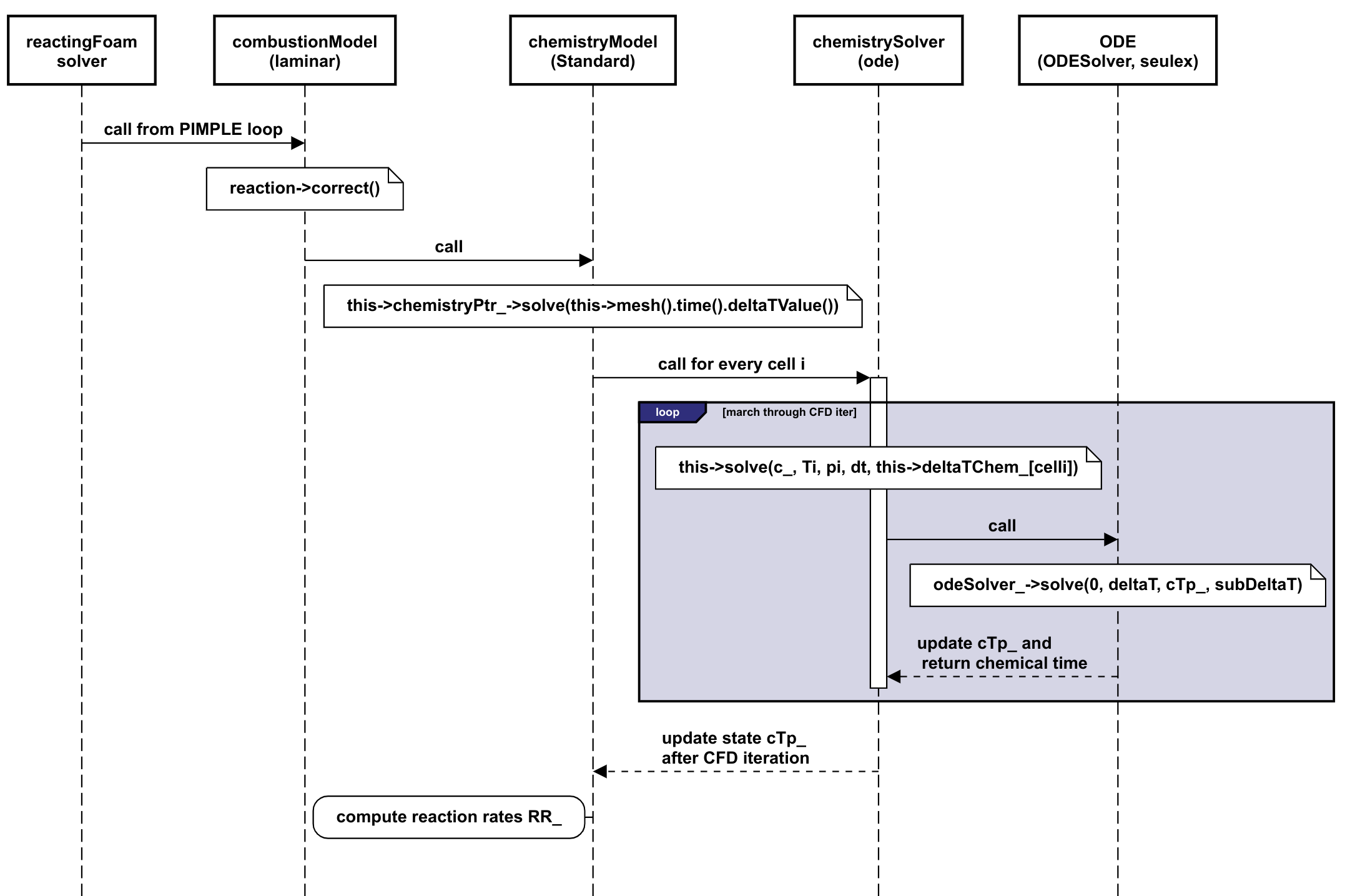}
\caption{Sequence diagram of function calls to compute chemistry source terms.}
\label{fig:uml}
\end{figure}

Here, we provide a unified modeling language (UML) sequence diagram, depicted in \Cref{fig:uml}, to back track the implementation and function calls relevant to computing the chemical source terms, starting from \verb|reaction->correct()| in \verb|YEqn.H| of solver level. Additionally, we also note that \verb|ODESystem| abstract class includes methods for \verb|nEqns()|, \verb|derivatives()|, and \verb|jacobian()|, which are all implemented in the \verb|chemistryModel|.
% of \verb|chemistryModel::solve()| calls \verb|chemistrySolver::solve()| calls \verb|ODE::solve()|

        % IMPLEMENTS 
        % //- Return the number of equations in the system
        % virtual label nEqns() const = 0;

        % //- Calculate the derivatives in dydx
        % virtual void derivatives
        % (
        %     const scalar x,
        %     const scalarField& y,
        %     scalarField& dydx
        % ) const = 0;

        % //- Calculate the Jacobian of the system
        % //  Need by the stiff-system solvers
        % virtual void jacobian
        % (
        %     const scalar x,
        %     const scalarField& y,
        %     scalarField& dfdx,
        %     scalarSquareMatrix& dfdy
        % ) const = 0;

% FROM ILYA - NUMBERING MULTIPLE BLOCKS FROM SAME CODE FILE
% \begin{lstlisting}[style=cpp,title={File ``createMesh.H"},captionpos=t,mathescape=true]
% Foam::autoPtr<Foam::fvMesh> meshPtr(nullptr);$\lstsetnumber{\ldots}$
% $\lstresetnumber\setcounter{lstnumber}{55}$
%     meshPtr.reset
%     (
%         new Foam::fvMesh
%         (
%             Foam::IOobject
%             (
%                 regionName,
%                 runTime.timeName(),
%                 runTime,
%                 Foam::IOobject::MUST_READ
%             )
%         )
%     );
% \end{lstlisting}
% \chapter{Incorporating CEMA and analytical Jacobian evaluation in OpenFOAM}
\chapter{Analytical Jacobian for OpenFOAM}
\label{ch:jac}

\section{Motivation}

% Here the actual main motivation would need to be a REMINDER what is the new thing we do here. In CEMA we want to understand the relative importance of X and Y and now we do it in a way which is more general...... etc

The main motive behind utilizing an analytical Jacobian for the chemistry problem is that it provides i) better accuracy to the solution of stiff ODE chemistry problem compared with numerical Jacobian, ii) accurate predictions of CEMA results, since CEMA computations are sensitive to significant digits of the Jacobian matrix and thereby information might be lost due to truncation errors in numerical Jacobian~\cite{Lu2010}. Relevant details are provided in \Cref{sec:cema_results2}.

As previously discussed in \Cref{ch:theory}, the Jacobian matrix $\boldsymbol{\mathcal J_\omega}$ is required in the Newton solver when computing the chemical source terms. The Jacobian can be computed either analytically via algebraic summation of contributions from all reactions, or numerically via perturbation of thermo-chemical state vector. In numerical simulations involving implicit or semi-implicit solvers (such as the present case for \verb|Seulex| algorithm), the Jacobian operations involving evaluation and factorization are computationally expensive. 
According to Lu and Law~\cite{Lu2009}, evaluation of the numerical Jacobian scales quadratically as ($N_{sp} \times N_r$) and consequently as ${N_{sp}}^2$. On the other hand, the analytical Jacobian evaluation scales linearly as $N_r$. Therefore, usage of the analytical Jacobian becomes computationally recommended whenever possible.

Besides computational performance, various computational diagnostics techniques including CEMA~\cite{Lu2010}, CSP~\cite{Lam1989,Lam1993}, and Jacobian analysis~\cite{turanyi1990reduction} would require accurate evaluation of the chemical Jacobian matrix, hence the advantage of analytical formulation. However, since Jacobian factorization operation scales cubically as ${N_{sp}}^3$, therefore it should be performed {only when needed}.
% posteriori 

The presently employed OpenFOAM version (v$2006$) utilizes a semi-analytical formulation of the concentration-based chemical Jacobian. We note that there has been implementations of fully analytical Jacobian formulations introduced in recent OpenFOAM versions (e.g. OpenFOAM-$8$). However, thus far the Open-source library \verb|pyJac| seems to comprise robust implementation techniques that minimize computational and memory operations. Moreover, \verb|pyJac| was reported to outperform other existing routines for analytical Jacobian evaluation either analytically or through finite-differencing~\cite{Niemeyer2017}.
% pyJac is src code generator given chemical kinetic mechanism.

As a brief note, \verb|pyJac| is a software package which is intended to generate source code files of the analytical Jacobian formulation and other helper functionalities tailored for a specific chemical kinetic mechanism provided by the user. In order to generate the source files, we need to have the chemical mechanism in either \verb|Chemkin-II| format (the most widely used format) or in \verb|Cantera| formats.~\footnote{It is possible to convert between \texttt{Chemkin-II} and \texttt{Cantera} formats through \texttt{ck2cti} and \texttt{ck2yaml} utilities, as described in \url{https://cantera.org/tutorials/ck2cti-tutorial.html}.}

% \todo{We can also convert the format between Chemkin and .cti formats, which is further elaborated in the appendix}.

In this chapter, we demonstrate how to link \verb|pyJac| with \verb|OpenFOAM| to provide fully algebraic analytical formulation of the chemical Jacobian, which will be also used for CEMA computations.

% Analytic Jacobian only
% needs to be generated
% once for each
% mechanism
%  Analytic Jacobian is
% more accurate
%  Analytic Jacobian should
% always be used

\section{Creating \texttt{cemaPyjacChemistryModel}}
% \section[shorter version of the title]{very very very very very very very very very very very very very very long section title}

% \verb|cemaPyjacChemistryModel|
%% INHERIT BASIC %%%%
Aiming for simplicity over a DRY (Don't Repeat Yourself) but possibly complicated code, we create our custom chemistry model by copying the \verb|StandardChemistryModel| and adding or modifying the functionalities of interest. This implies that our new
chemistry model inherits from \verb|BasicChemistryModel| and \verb|ODESystem| and non-modified methods and class attributes will be duplicated. Indeed, a better way is to declare our model inherited from \verb|StandardChemistryModel| while overriding and/or extending the desired functionalities. However, as mentioned we will proceed in the most straight forward way.

After sourcing \verb|OpenFOAM|-v$2006$, we execute the following \verb|bash| commands through a \verb|Linux| terminal for creating a user-defined library \verb|cemaPyjacChemistryModel| copied from \verb|StandardChemistryModel| while cleaning unnecessary files.

\begin{verbatim}
> foam
> cp -r --parents src/thermophysicalModels/chemistryModel $WM_PROJECT_USER_DIR
> cd $WM_PROJECT_USER_DIR/src/thermophysicalModels/chemistryModel/chemistryModel
> rm -r TDACChemistryModel BasicChemistryModel basicChemistryModel
> mv StandardChemistryModel cemaPyjacChemistryModel; cd cemaPyjacChemistryModel
> mv StandardChemistryModel.C cemaPyjacChemistryModel.C
> mv StandardChemistryModel.H cemaPyjacChemistryModel.H
> mv StandardChemistryModelI.H cemaPyjacChemistryModelI.H
> sed -i 's/Standard/cemaPyjac/g' *
\end{verbatim}

and then we also change the TypeName for the new model in the runTime selection table
\begin{verbatim}
> sed -i 's/standard/cemaPyjac/g' cemaPyjacChemistryModel.H
\end{verbatim}

Now, we need to update description in the header files (optional), and then to modify the \verb|Make/options| file (\Cref{ls_options_nopyj}) to pass information to the compiler for header inclusion and library linkings. 
Since our chemistry model is copied from standard model which is part of the \verb|chemistryModel| library (\verb|libchemistryModel.so|), there will be possibly dependencies on other components of that library when we want to compile our model separately. Therefore, we need to include the path of headers (or their symbolic links) that are included by \verb|StandardChemistryModel|, as well as linking to the corresponding library to enable all implementations of standard model to be also realized within our custom model.

\begin{lstlisting}[style=cpp, label=ls_options_nopyj, caption=Make/options \textbf{without} provided path for \texttt{pyJac} header files,captionpos=t]
EXE_INC = \
    -I$(LIB_SRC)/finiteVolume/lnInclude \
    -I$(LIB_SRC)/meshTools/lnInclude \
    -I$(LIB_SRC)/ODE/lnInclude \
    -I$(LIB_SRC)/transportModels/compressible/lnInclude \
    -I$(LIB_SRC)/thermophysicalModels/reactionThermo/lnInclude \
    -I$(LIB_SRC)/thermophysicalModels/basic/lnInclude \
    -I$(LIB_SRC)/thermophysicalModels/specie/lnInclude \
    -I$(LIB_SRC)/thermophysicalModels/functions/Polynomial \
    -I$(LIB_SRC)/thermophysicalModels/thermophysicalFunctions/lnInclude \
    -I$(LIB_SRC)/thermophysicalModels/chemistryModel/lnInclude

LIB_LIBS = \
    -lfiniteVolume \
    -lmeshTools \
    -lODE \
    -lcompressibleTransportModels \
    -lfluidThermophysicalModels \
    -lreactionThermophysicalModels \
    -lspecie \
    -lchemistryModel
\end{lstlisting}

As we see, the only modifications we made in \verb|Make/options| is adding the lines \verb|-lchemistryModel| and \verb|-I$(LIB_SRC)/thermophysicalModels/chemistryModel/lnInclude| for each \verb|LIB_LIBS| and \verb|EXE_INC| variables, respectively. This means that we allow the compiler to find all included header files needed from standard model as well as implementations of the library into our custom model. We note that we also need to add one more path related to header files from \verb|pyJac| for declarations of function that we will use, as it is shown in \Cref{ap:tree} for \verb|pyjacInclude| subdirectory. Basically, such header files are provided for the sake of model compilation, and they can be acquired by generating source code through \verb|pyJac| for an arbitrary mechanism.
% but this will be shown later. %in \Cref{sec:pyjac_inc}. % WE NEED PROPER SECTION HERE!!

Since chemistry model is a templated library by default, we also need to allow our custom chemistry model instances to be created for all possible thermodynamic and compressibility models. We achieve this thanks to the \emph{macro} \verb|makeChemistryModelType| similarly used in \verb|BasicChemistryModels.C| from standard model. Accordingly, we copy the macros definitions and modify their usage accordingly into our own model directory as in the following.

\begin{verbatim}
> cd $WM_PROJECT_USER_DIR/src/thermophysicalModels/chemistryModel
> cp $FOAM_SRC/thermophysicalModels/chemistryModel/chemistryModel/makeChemistryModel.H .
> thermophysicalModels=$FOAM_SRC/thermophysicalModels
> BasicChemistryModel=chemistryModel/chemistryModel/BasicChemistryModel
> thermophysicalBasicChemistryModel=$(thermophysicalModels)/$(BasicChemistryModel)
> cp $(thermophysicalBasicChemistryModel)/BasicChemistryModels.C makeChemistryModels.C
> sed -i 's/Standard/cemaPyjac/g' makeChemistryModels.C
\end{verbatim}
% cp $FOAM_SRC/thermophysicalModels/chemistryModel/chemistryModel/BasicChemistryModel/BasicChemistryModels.C makeChemistryModels.C

After that, we \emph{clean} all lines of code relevant to TDAC model from \verb|makeChemistryModels.C| as we did not include it to our development, otherwise our model will not compile. Next, we repeat the same procedure but with chemistry solver which is also templated on thermodynamics. Moreover, we modify the name of the ODE chemistry solver to \verb|odePyjac| to reflect our intention to modify such class with functionalities from \verb|pyJac|. Further discussions and demonstrations are depicted in \Cref{sec:cema_make} and \Cref{ap:pyjac_cmake}.

\begin{verbatim}
> cd $WM_PROJECT_USER_DIR/src/thermophysicalModels/chemistryModel
> chemistryModel=$FOAM_SRC/thermophysicalModels/chemistryModel
> cp $chemistryModel/chemistrySolver/chemistrySolver/makeChemistrySolver* .
> sed -i 's/ode,/odePyjac,/g' makeChemistrySolverTypes.H
> sed -i 's/ode.H/odePyjac.H/g' makeChemistrySolverTypes.H
\end{verbatim}

Similar to the chemistry model, we herein remove all lines of code relevant to TDAC, including the header inclusion, from \verb|makeChemistrySolverTypes.H|.
Finally, we modify the \verb|Make/files| (\Cref{ls_make_files_0}) to compile the proper source files using defined macros, and we specify the executable name and location.

\begin{lstlisting}[label=ls_make_files_0,style=cpp,caption=Make/files,captionpos=t]
chemistryModel/makeChemistryModels.C
chemistryModel/makeChemistrySolvers.C

LIB = $(FOAM_USER_LIBBIN)/libcemaPyjacChemistryModel
\end{lstlisting}

Next, we dive deeper into the new chemistry model in order to incorporate functionalities from \verb|pyJac|.

\section{Using \texttt{pyJac} functionalities}
\label{sec:pyj_func}

In this section, we focus on implementing an analytical formulation of the chemical Jacobian matrix along with other exact formulations such as temporal derivatives of the thermo-chemical state vector by making use of \verb|pyJac|. We also discuss the particular features of \verb|pyJac| which must be considered upon implementation since they are rather different from stock \verb|OpenFOAM| implementation. In brief, the special considerations are that i) \verb|pyJac| generates Jacobian matrix based on the mass fraction-based state vector ($\boldsymbol \Phi$) while standard \verb|OpenFOAM| operates on concentration-based state vector ($\boldsymbol \Phi_c$), and ii) \verb|pyJac| considers ($N_{sp}-1$) species in the state vector while dumping all numerical residuals into the last species, which is chosen as inert (or most abundant) species of the chemical mechanism such as Nitrogen. Hence, $Y_\mathrm{last} = 1.0 - \sum_{i=1}^{N_{sp}-1}Y_i$.
In the following, we discuss how to download and compile the \verb|pyJac| package.

% The consequences of using analytical Jacobian from PyJac is the following:
% - We need to update functions jacobian() and direvatives() required in the odePyjac::solve()

% - We need to work using Y instead of C, since PyJac jacobian is dfdy not dfdc, therefore we need to replace everywhere

% - We need to update solution vector into PTY instead of cTp. We put Y at the end so that it becomes easier to operate on N-1

% \subsection{Generating source code from PyJac and available features}

% Download mechanism (GRI-30) files from:
% http://combustion.berkeley.edu/gri-mech/version30/text30.html
% DISCLAIMER: THERE IS ALREADY A BUG IN OPENFOAM TUTORIAL CASES WHEREIN 
% $FOAM_TUTORIALS/combustion/reactingFoam/RAS/DLR_A_LTS/chemkin HAS TRANSPORT PROPERTIES CONSTANT FOR MOST SPECIES FROM WHICH, ALTHOUGH MIGHT HAVE INSIGNIFICANT EFFECTS ON Cp EVALUATION, IT DIFFERS FROM TRANSPORT DATA OF GRI-30 IN ORIGINAL CHEMKIN FORMAT

% cd $WM_PROJECT_USER_DIR/src/thermophysicalModels/chemistryModel/chemistryModel/cemaPyjacChemistryModel
% cp -r $FOAM_RUN/premixedFlame1D_standard/mechanism/chemkin_orig/out pyjacInclude

% python3 -m pyjac --lang c --last_species N2 --input grimech30.cti

\subsection{Download and compile \texttt{pyJac}}
\label{sec:pyj_install}

The \verb|pyJac| package is publicly available on Github through the following link \url{https://github.com/SLACKHA/pyJac}. As it is indicated from the documentation, \verb|pyJac| package can be installed via different means. Besides building the package from the source code (while considering all dependencies, and most importantly \verb|Cantera|)
\begin{verbatim}
> python setup.py install
\end{verbatim}
it can also be installed via the \verb|Conda| package manager
\begin{verbatim}
> conda install -c slackha pyjac
\end{verbatim}
or from the Python package index \verb|PyPI| using the \verb|pip| tool
\begin{verbatim}
> pip install pyjac
\end{verbatim}

% through \verb|Conda| package manager as follows

% \begin{verbatim}
% > conda install -c slackha pyjac
% \end{verbatim}

% or it can be installed from the Python package index \verb|PyPI| as follows

% \begin{verbatim}
% > pip install pyjac
% \end{verbatim}

% Finally, it can be also downloaded and installed from the source code as follows

% \begin{verbatim}
% > python setup.py install
% \end{verbatim}

The necessary header files from \verb|pyJac| that are required for library compilation are attached in \Cref{ap:pyjac_include}.
In the following, we discuss details for updating methods of Jacobian, derivatives, and heat release rate. %, and also on providing \verb|pyJac| header files necessary for the model compilation.

\subsection{Updating Jacobian method}
\label{sec:jac_update}

The Jacobian function of the model is declared as in the following \Cref{ls_jac_00}. 
% \begin{verbatim}
% template<class ReactionThermo, class ThermoType>
% void Foam::cemaPyjacChemistryModel<ReactionThermo, ThermoType>::jacobian
% (
%     const scalar t,
%     const scalarField& c,
%     scalarField& dcdt,
%     scalarSquareMatrix& dfdc
% ) const
% \end{verbatim}
\begin{lstlisting}[label=ls_jac_00, style=cpp,caption=\texttt{jacobian()} function arguments in \texttt{cemaPyjacChemistryModel.C},captionpos=t, firstnumber=357]
template<class ReactionThermo, class ThermoType>
void Foam::cemaPyjacChemistryModel<ReactionThermo, ThermoType>::jacobian
(
    const scalar t,
    const scalarField& c,
    scalarField& dcdt,
    scalarSquareMatrix& dfdc
) const
\end{lstlisting}

The parameters \verb|c| and \verb|dfdc| of the previous listing correspond to the state vector and Jacobian, respectively. It is important to note that \verb|c| vector originally denotes species concentrations, temperature, and pressure (i.e. \verb|cTp| variable which we will also see in the chemistry solver). However, in our implementation we use pressure, temperature, and species mass fraction.

As we previously noted, the original chemistry model operates on $\boldsymbol \Phi_c$ which is concentration based. Since \verb|pyJac| uses species mass fractions without last inert species, the parameter \verb|c| (acting as $\boldsymbol \Phi$ of size of $N_{sp}+2$) would be used to prepare the input variable for evaluating Jacobian from \verb|pyJac|. Moreover, the Jacobian matrix \verb|dfdc| is mass fraction based and holding size of ($N_{sp} \times N_{sp}$) for temperature and species partial derivatives while excluding last species. The Jacobian function call from \verb|pyJac| looks like the following
\begin{verbatim}
void eval_jacob (const double t, const double pres, const double * __restrict__ y, \
                 double * __restrict__ jac)
\end{verbatim}
in which the parameters \verb|t|, \verb|pres|, and \verb|y| denote current system time, pressure, and state vector of temperature and species mass fractions of size $N_{sp}$. Therefore, we need to prepare the proper input data to be passed to the \verb|pyJac| function call so that the parameter \verb|jac| is populated with analytical Jacobian data. Additionally, we need to include proper header file (\emph{mechanism-independent}) for function declaration while using dynamic binding for linking the implementation (\emph{mechanism-dependent}) during runtime. The corresponding header file for derivatives declaration is \verb|dydt.h| which is included as in the following \Cref{ls_header_jac}.

% \lstinputlisting[style=cpp, caption=Declarations for the \texttt{jacobian()}, \texttt{derivatives()}, and helper functions in \texttt{cemaPyjacChemistryModel.H},captionpos=t, linerange={50-54},firstnumber=50]{Codes/lib_cema/chemistryModel/chemistryModel/cemaPyjacChemistryModel/cemaPyjacChemistryModel_noCema.H}
\begin{lstlisting}[style=cpp, label=ls_header_jac, caption=Declarations of the \texttt{jacobian()} and \texttt{derivatives()} methods in addition to helper functions in \texttt{cemaPyjacChemistryModel.C},captionpos=t, linerange={50-54},firstnumber=54]
extern "C" {
    #include "chem_utils.h"
    #include "dydt.h"
    #include "jacob.h"
};
\end{lstlisting}

In our implementation, we replace the whole definition of the semi-analytical Jacobian method with the following

% \lstinputlisting[style=cpp, caption=Implementation of \texttt{jacobian()} without CEMA, defined in \texttt{cemaPyjacChemistryModel.C},captionpos=t, linerange={178-180},firstnumber=178]{Codes/lib_cema/chemistryModel/chemistryModel/cemaPyjacChemistryModel/cemaPyjacChemistryModel_noCema.C}

% \lstinputlisting[style=cpp, caption=Definition of the \texttt{solve()} function in \texttt{ode.C} of chemistrySolver,captionpos=t, linerange={51-86},firstnumber=51]{Codes/lib_standard/chemistryModel/chemistrySolver/ode/ode.C}

\begin{lstlisting}[style=cpp, caption=Definition of the \texttt{jacobian()} method in \texttt{cemaPyjacChemistryModel.C} -- \textbf{without CEMA},captionpos=t, linerange={164-221},firstnumber=174]
template<class ReactionThermo, class ThermoType>
void Foam::cemaPyjacChemistryModel<ReactionThermo, ThermoType>::jacobian
(
    const scalar t,
    const scalarField& c,
    scalarField& dcdt,
    scalarSquareMatrix& dfdc
) const
{
    std::vector<double> TY(nSpecie_+1, 0.0);
    std::vector<double> dfdy(nSpecie_*nSpecie_, 0.0);

    const scalar p = c[0];
    const scalar T = c[1];

    scalar csum = 0.0;
    forAll(c_, i)
    {
        c_[i] = max(c[i+2], 0.0);
        csum += c_[i];
    }
    // Then we exclude last species from csum and instead dump all
    // residuals into last species to ensure mass conservation
    csum -= c_[nSpecie_-1];
    c_[nSpecie_-1] = 1.0 - csum;

    dfdc = Zero;

    TY[0] = T;
    forAll(c_, i)
    {
        TY[i+1] = c_[i];
    }

    eval_jacob(0, p, TY.data(), dfdy.data());

    // Back substitution to update dfdc
    // Assign first row and column to zero since they correspond to const pressure
    for (label j = 0; j < nSpecie_ + 1; ++j)
    {
        dfdc(0,j) = 0.0;
        dfdc(j,0) = 0.0;
    }

    label k = 0;
    // Loop cols
    for (label j = 1; j < nSpecie_+1; ++j)
    {
        // Loop rows
        for (label i = 1; i < nSpecie_+1; ++i)
        {
            dfdc(i,j) = dfdy[k + i - 1];
        }
        k += nSpecie_;
    }
    // Note that dcdt is not needed in most ODE solvers so here we just return 0
    dcdt = Zero;
}
\end{lstlisting}

\subsection{Updating derivatives method}

Here, we replace the standard implementation for temporal derivatives of state vector with that provided from \verb|pyJac|. The \verb|derivatives()| method has the following form

% \begin{verbatim}
\begin{lstlisting}[style=cpp,caption=\texttt{derivatives()} function arguments in \texttt{cemaPyjacChemistryModel.C},captionpos=t, firstnumber=312]
template<class ReactionThermo, class ThermoType>
void Foam::cemaPyjacChemistryModel<ReactionThermo, ThermoType>::derivatives
(
    const scalar time,
    const scalarField& c,
    scalarField& dcdt
) const
% \end{verbatim}
\end{lstlisting}
in which the parameters \verb|time|, \verb|c| and \verb|dcdt| correspond respectively to system time, state vector, and derivatives. As in the \verb|Jacobian()| method, the state vector \verb|c| herein denotes pressure, temperature, and species mass fraction. The derivatives function from \verb|pyJac| is declared as follows

\begin{verbatim}
void dydt (const double t, const double pres, const double * __restrict__ y, \
           double * __restrict__ dy)
\end{verbatim}
in which the parameters \verb|t|, \verb|pres|, and \verb|y| are similar as discussed in \verb|eval_jacob| from previous section, while the array \verb|dy| of size ($N_{sp}$) is to be populated by derivatives. The function \verb|dydt| is defined in \verb|dydt.c| which is mechanism dependent so we only include the header \verb|dydt.h| (c.f. \Cref{ls_header_jac}) for function declaration while using dynamic binding for function definition at runtime.

% \lstinputlisting[label=ls_, style=cpp, caption=Declarations of the \texttt{jacobian()} and \texttt{derivatives()} methods in addition to helper functions in \texttt{cemaPyjacChemistryModel.C},captionpos=t, linerange={50-54},firstnumber=54]{Codes/lib_cema/chemistryModel/chemistryModel/cemaPyjacChemistryModel/cemaPyjacChemistryModel_noCema.H}

In our implementation, we replace the whole definition of the derivatives method with the following

\begin{lstlisting}[style=cpp, caption=Definition of the \texttt{derivatives()} method in \texttt{cemaPyjacChemistryModel.C},captionpos=t, linerange={126-162},firstnumber=128]
{
    // Arrays to be passed into PyJac function call for derivatives
    std::vector<double> TY(nSpecie_+1, 0.0);
    // if TY has N+1 elements, diff(TY) has N elements
    std::vector<double> dTYdt(nSpecie_, 0.0);
    // state vector has now PTY instead of cTp
    const scalar p = c[0];
    const scalar T = c[1];

    scalar csum = 0.0;
    forAll(c_, i)
    {
        c_[i] = max(c[i+2], 0.0);
        csum += c_[i];
    }
    // Then we exclude last species from csum and dump all residuals into last
    // species to ensure mass conservation
    csum -= c_[nSpecie_-1];
    c_[nSpecie_-1] = 1.0 - csum;

    TY[0] = T;
    forAll(c_, i)
    { 
        TY[i+1] = c_[i];
    }

    // call pyJac RHS function
    dydt(0, p, TY.data(), dTYdt.data());

    // dp/dt = 0
    dcdt[0] = 0.0;
    // Back substitute into dcdt (dcdt has nSpecie+1 elements for diff(PTY))
    for (label i = 0; i < nSpecie_; ++i)
    {
        dcdt[i+1] = dTYdt[i];
    }
}
\end{lstlisting}

\subsection{Updating heat release method}

We also need to update the \verb|Qdot()| method which computes the heat release rate, $\dot{\omega}_h$, defined in \cref{eq:hrr}. We herein just replace the chemical enthalpy (i.e. enthalpy of formation) with the corresponding data from \verb|pyJac|. The function responsible for chemical enthalpy from \verb|pyJac|, \verb|eval_h()|, is declared in \verb|chem_utils.h|. Such a header declares many features that can be used (e.g. specific heat capacity) for a larger reliance on \verb|pyJac|, however, we only herein demonstrate the usage of chemical enthalpy. Linking to the mechanism dependent function definition is left for runtime dynamic binding. While declarations have been included through \Cref{ls_header_jac}, the corresponding \verb|pyJac| function for enthalpy evaluation has the following form
\begin{verbatim}
void eval_h (const double T, double * __restrict__ h)
\end{verbatim}
% and the declaration is included in the following block
% \lstinputlisting[style=cpp, caption=Declarations of the \texttt{jacobian()} and \texttt{derivatives()} methods in addition to helper functions in \texttt{cemaPyjacChemistryModel.C},captionpos=t, linerange={50-54},firstnumber=54]{Codes/lib_cema/chemistryModel/chemistryModel/cemaPyjacChemistryModel/cemaPyjacChemistryModel_noCema.H}

Since the enthalpy of formation for a given mechanism at reference temperature is not going to change throughout the simulation, perhaps it could be better to just keep the data as member of the class. Accordingly, we declare the scalarField \verb|sp_enth_form| as shown in \Cref{ls_sp_enth}.

\begin{lstlisting}[label=ls_sp_enth, style=cpp, caption=Declaration of \texttt{sp\_enth\_form} for species enthalpy of formation, captionpos=t, linerange={126-127},firstnumber=126]
        // Enthalpy of formation for every species, from PyJac
        scalarList sp_enth_form;
\end{lstlisting}

Then, we initialize it with size of $N_{sp}$ as shown in \Cref{ls_sp_enth_init}.
\begin{lstlisting}[label=ls_sp_enth_init, style=cpp, caption=Initialization of \texttt{sp\_enth\_form} scalarField with size of number of species,captionpos=t, linerange={68-68},firstnumber=68]
    sp_enth_form(nSpecie_)
\end{lstlisting}

After that, in the class constructor, we update it with data from \verb|pyJac| as shown in \Cref{ls_sp_enth_update}.
\begin{lstlisting}[label=ls_sp_enth_update, style=cpp, caption=Update \texttt{sp\_enth\_form} with data from \texttt{pyJac} in the class constructor,captionpos=t, linerange={96-106},firstnumber=96]
    if (this->chemistry_) {
        Info << "\n Evaluating species enthalpy of formation using PyJac\n" << endl;
        //- Enthalpy of formation for all species
        std::vector<scalar> sp_enth_form_(nSpecie_, 0.0);
        //- Enthalpy of formation is taken from pyJac at T-standard (chem_utils.h)
        eval_h(298.15, sp_enth_form_.data());
        for (label i = 0; i < nSpecie_; ++i)
        { 
            sp_enth_form[i] = sp_enth_form_[i];
        }
    }
\end{lstlisting}

Finally, the implementation for $\dot{\omega}_h$ is updated to incorporate the \verb|pyJac| data as shown in \Cref{ls_hrr_mod}.

\begin{lstlisting}[label=ls_hrr_mod, style=cpp, caption=Modify heat release rate implementation to incorporate formation enthalpy from \texttt{pyJac},captionpos=t, linerange={249-256},firstnumber=249]
        forAll(Y_, i)
        {
            forAll(Qdot, celli)
            {
                scalar hi = sp_enth_form[i];
                Qdot[celli] -= hi*RR_[i][celli];
            }
        }
\end{lstlisting}

% \subsection{Number of species and chemical reactions}

\subsection{Updating solve method from chemistry model}

The most important implications due to \verb|pyJac| usage is the mass fraction based Jacobian matrix. This implies that, according to \cref{eq:ode2}, the state vector needs to be also based on mass fractions instead of concentrations.
This is realized in the \verb|solve()| method of the chemistryModel, wherein the data passed to the function call of chemistrySolver are pressure, temperature, and mass fractions, c.f. \Cref{sec:uml}.
In the code, replacement of concentrations into mass fractions is adopted for the data to be passed to \verb|solve()| function call of chemistrySolver as in the following \Cref{ls_solve_c_y}.

\begin{lstlisting}[label=ls_solve_c_y, style=cpp, caption=Use species mass fractions instead of concentrations in \texttt{cemaPyjacChemistryModel.C},captionpos=t, linerange={295-300},firstnumber=295]
            for (label i=0; i<nSpecie_; i++)
            {
                // c_[i] = rhoi*Y_[i][celli]/specieThermo_[i].W();                
                c_[i] = Y_[i][celli];
                c0[i] = c_[i];
            }
\end{lstlisting}

Then, as shown in \Cref{ls_solve_c_y_2}, the concentration-mass conversion also needs to be achieved when computing the source terms, noting that ${\dot{\omega}}_k$ units must be consistent in \cref{eq:species}, i.e. ($c_i W_i = \rho Y_i$).

\begin{lstlisting}[label=ls_solve_c_y_2, style=cpp, caption=Use species mass fractions instead of concentrations in chemistry source terms,captionpos=t, linerange={319-324},firstnumber=319]
            for (label i=0; i<nSpecie_; i++)
            {
                // CHEMICAL SOURCE TERM PER SPECIES
                // RR_[i][celli] = (c_[i] - c0[i])*specieThermo_[i].W()/deltaT[celli];
                this->RR_[i][celli] = rhoi*(this->c_[i] - c0[i])/deltaT[celli];
            }
\end{lstlisting}

At this point, we discussed all modifications necessary to incorporate chemical Jacobian matrix from \verb|pyJac| in the chemistry model. In the following section we discuss modifications required in the ODE approach of the chemistry solver.

% \subsection[Updating odePyjac::solve method]{Updating odePyjac::solve method from chemistry solver}
\subsection{Updating solve method from chemistry solver}

Prior to discussion, we note that the following modifications concern the ODE approach of the \verb|chemistrySolver| class, i.e. within the \verb|chemistryModel| library and not the \verb|ODE| library of \verb|OpenFOAM|.

It is demonstrated in the previous section that species mass fractions are used instead of concentrations. The mass fraction based state vector, defined by \verb|c_|, \verb|Ti|, and \verb|pi| variables is used for calling the \verb|odePyjac::solve()| method in chemistrySolver class as shown in \Cref{ls_solve_c_y_solver}.

\begin{lstlisting}[label=ls_solve_c_y_solver, style=cpp, caption=Using species mass fractions instead of concentrations while calling odePyjac::solve() in chemistrySolver,captionpos=t, linerange={306-312},firstnumber=306]
            while (timeLeft > SMALL)
            {
                scalar dt = timeLeft;
                // Calls ode::solve() from chemistrySolver
                this->solve(c_, Ti, pi, dt, this->deltaTChem_[celli]);
                timeLeft -= dt;
            }
\end{lstlisting}

Here, we provide insight toward the required modifications in the \verb|odePyjac::solve()| method as a response to the adopted changes in the state vector. First, as the Jacobian matrix is evaluated from \verb|pyJac| for species mass fractions excluding last species, we have a total of $N_{sp}+1$ equations to be solved when also considering pressure and temperature. Accordingly, we modify the total solve vector \verb|cTp_| to have a size of $N_{sp}+1$ before which the ODE solver is called, c.f. \Cref{sec:uml}. This is achieved by modifying the inline method to exclude last species from computations of the solve vector as shown in \Cref{ls_modify_nspecies}.

% Since the standard implementation solves for $N_{sp}+2$, we have to modify the number of equations to exclude last species from computations. This is achieved by modifying the inline method as follows

\begin{lstlisting}[label=ls_modify_nspecies, style=cpp, caption=Modify number of equations to exclude solving for last species,captionpos=t, linerange={33-39},firstnumber=319]
template<class ReactionThermo, class ThermoType>
inline Foam::label
Foam::cemaPyjacChemistryModel<ReactionThermo, ThermoType>::nEqns() const
{
    // nEqns = number of species (N-1) + temperature + pressure
    return nSpecie_ + 1;
}
\end{lstlisting}

The updated size ($N_{sp}+1$) is used to initialize the total solve vector in the class constructor as follows
\begin{lstlisting}[style=cpp, caption=Initialize total solve vector with size of $N_{sp}+1$ during construction of class \texttt{odePyjac},captionpos=t, linerange={38-38},firstnumber=38]
    cTp_(this->nEqns())
\end{lstlisting}

Finally, the full implementation of the \verb|odePyjac::solve()| method is presented in \Cref{ls_solve_def}.

\begin{lstlisting}[label=ls_solve_def, style=cpp, caption=Definition of \texttt{odePyjac::solve()} method in chemistrySolver,captionpos=t, linerange={51-94},firstnumber=61]
template<class ChemistryModel>
void Foam::odePyjac<ChemistryModel>::solve
(
    scalarField& c,
    scalar& T,
    scalar& p,
    scalar& deltaT,
    scalar& subDeltaT
) const
{
    // Reset the size of the ODE system to the simplified size when mechanism
    // reduction is active
    if (odeSolver_->resize())
    {
        odeSolver_->resizeField(cTp_);
    }

    const label nSpecie = this->nSpecie();

    // Copy the concentration, T and P to the total solve-vector (N+1)
    cTp_[0] = p;
    cTp_[1] = T;
    // Update for N-1 species
    for (label i=0; i<nSpecie-1; i++)
    {
        cTp_[i+2] = c[i];
    }

    // Here, we call ODE solver...
    odeSolver_->solve(0, deltaT, cTp_, subDeltaT);

    // Back substitute results
    p = cTp_[0];
    T = cTp_[1];
    scalar csum = 0;

    for (label i=0; i<nSpecie-1; i++)
    {
        c[i] = max(0.0, cTp_[i+2]);
        csum += c[i];
    }
    // Last species
    c[nSpecie-1] = 1.0 - csum;
}
\end{lstlisting}

\chapter{CEMA for OpenFOAM}
\label{ch:cema}

\section{Implementing CEMA}

In order to implement CEMA, as presented in \Cref{sec:cema}, first we need to store the chemical Jacobian matrix as member of the class so that we can operate on the data using separate methods. After that, we need to create a new field to be populated using CEMA. Accordingly, we declare the square matrix \verb|chemJacobian_| for the Jacobian matrix, and the field \verb|cem_| as it follows in \Cref{ls_cem1}.

\begin{lstlisting}[label={ls_cem1}, style=cpp, caption=Declarations of the Jacobian matrix and CEMA field as members of the class \texttt{cemaPyjacChemistryModel},captionpos=t, linerange={135-139},firstnumber=135]
    	label nElements_;

        // Jacobian from chemistry problem, from pyJac
        mutable scalarSquareMatrix chemJacobian_;
\end{lstlisting}

After that, we initialize both \verb|chemJacobian_| and \verb|cem_| as shown in \Cref{ls_cem2}.

\begin{lstlisting}[label={ls_cem2}, style=cpp, caption=Initialization of \texttt{chemJacobian\_} and \texttt{cem\_} variables in \texttt{cemaPyjacChemistryModel.C},captionpos=t, linerange={70-84},firstnumber=69]
    sp_enthalpy_(nSpecie_),
    nElements_(BasicChemistryModel<ReactionThermo>::template get<label>("nElements")),
    chemJacobian_(nSpecie_),
    cem_
    (
        IOobject
        (
            "cem",
            this->mesh_.time().timeName(),
            this->mesh_,
            IOobject::NO_READ,
            IOobject::AUTO_WRITE
        ),
        this->mesh_,
        dimensionedScalar("cem", dimless, 0),
\end{lstlisting}

Here, we note that the chemical Jacobian has size of $N_{sp}$ in order to hold information for temperature and species mass fractions excluding last species. After that, we read the number of elements of the chemical mechanism as specified by the user in \verb|chemistryProperties| dictionary using the keyword \verb|nElements|. This information is necessary in order to exclude the $M+1$ insignificant eigenvalues due to conservation modes, as discussed in \Cref{sec:cema}. This is achieved through the following declaration and definition of the variable \verb|nElements_| in \Cref{ls_cem2_2} and \Cref{ls_cem2_2}.
% Here, we note that the developed chemistry model inherits from \verb|get()| function from \verb|IOdictionary| class is used, since chemistry model inherits from 

\begin{lstlisting}[label={ls_cem2_1}, style=cpp, caption=Declaration of \texttt{nElements\_} variable in \texttt{cemaPyjacChemistryModel.H},captionpos=t, linerange={133-133},firstnumber=133]

\end{lstlisting}

\begin{lstlisting}[label={ls_cem2_2}, style=cpp, caption=Definition of \texttt{nElements\_} variable in \texttt{cemaPyjacChemistryModel.C},captionpos=t, linerange={69-69},firstnumber=69]
    dcdt_(nSpecie_),
\end{lstlisting}

Now, in order to perform eigendecomposition, we can use implementations from \verb|EigenMatrix| class. All we need is to include declaration headers, as shown in \Cref{ls_cem3}, and the implementations are defined in standard \verb|OpenFOAM| library which is already linked.
\begin{lstlisting}[label={ls_cem3}, style=cpp, caption=Include declaration header file for \texttt{EigenMatrix} class,captionpos=t, linerange={50-50},firstnumber=50]
#include "simpleMatrix.H"
\end{lstlisting}

Finally, the full implementation is presented in the following \Cref{ls_cem4}.

\begin{lstlisting}[label={ls_cem4}, style=cpp, caption=Definition of the \texttt{cema()} method to compute $\lambda_\mathrm{exp}$,captionpos=t, linerange={752-781},firstnumber=752]
}

template<class ReactionThermo, class ThermoType>
void Foam::cemaPyjacChemistryModel<ReactionThermo, ThermoType>::cema
(
    scalar& cem
) const
{

    const Foam::EigenMatrix<scalar> EM(chemJacobian_);
    DiagonalMatrix<scalar> EValsRe(EM.EValsRe());
    DiagonalMatrix<scalar> EValsIm(EM.EValsIm());

    DiagonalMatrix<scalar> EValsMag(EValsRe.size(), 0.0);
    forAll(EValsRe, i)
    {
        EValsMag[i] = (EValsRe[i]*EValsRe[i] + EValsIm[i]*EValsIm[i]);
    }

    // Sort eigenvalues in ascending order, and track indices
    const auto ascend = [&](scalar a, scalar b){ return a < b; };
    const List<label> permut(EValsMag.sortPermutation(ascend));

    // Skip conservation modes for elements and temperature
    for (label i=0; i<nElements_+1; ++i) 
    {
        label idx = permut[i];
        EValsRe[idx] = -1E30;
    }
\end{lstlisting}

\section{Building \texttt{cemaPyjacChemistryModel} with Make}
\label{sec:cema_make}

In order to build the library \verb|cemaPyjacChemistryModel|, we use the OpenFOAM \verb|wmake| compilation script. The corresponding \verb|Make/files| and \verb|Make/options| are depicted in \Cref{ls_make_files} and \Cref{ls_make_options}, respectively.

\begin{lstlisting}[label={ls_make_files}, style=cpp, caption=Make (\texttt{files}) for the \texttt{cemaPyjacChemistryModel} library compilation,captionpos=t]
makeChemistryModels.C
makeChemistrySolvers.C

LIB = $(FOAM_USER_LIBBIN)/libcemaPyjacChemistryModel
\end{lstlisting}

\begin{lstlisting}[label={ls_make_options}, style=cpp, caption=Make (\texttt{options}) for the \texttt{cemaPyjacChemistryModel} library compilation,captionpos=t]
EXE_INC = \
    -I$(LIB_SRC)/finiteVolume/lnInclude \
    -I$(LIB_SRC)/meshTools/lnInclude \
    -I$(LIB_SRC)/ODE/lnInclude \
    -I$(LIB_SRC)/transportModels/compressible/lnInclude \
    -I$(LIB_SRC)/thermophysicalModels/reactionThermo/lnInclude \
    -I$(LIB_SRC)/thermophysicalModels/basic/lnInclude \
    -I$(LIB_SRC)/thermophysicalModels/specie/lnInclude \
    -I$(LIB_SRC)/thermophysicalModels/functions/Polynomial \
    -I$(LIB_SRC)/thermophysicalModels/thermophysicalFunctions/lnInclude \
    -I$(LIB_SRC)/thermophysicalModels/chemistryModel/lnInclude \
    -IpyjacInclude

LIB_LIBS = \
    -lfiniteVolume \
    -lmeshTools \
    -lODE \
    -lcompressibleTransportModels \
    -lfluidThermophysicalModels \
    -lreactionThermophysicalModels \
    -lspecie \
    -lchemistryModel
\end{lstlisting}

As it is observed, the Make files are actually templates to create instances of the chemistry models and chemistry solvers templated on the type of thermodynamics. The macros required to instantiate chemistry models and chemistry solvers according to compressibility and transport types, as well as adding them to runTime selection table, are depicted respectively in \Cref{ls_make_1,ls_make_2}.

\begin{lstlisting}[label={ls_make_1}, style=cpp, caption=Macros for chemistry models based on compressibility and transport types,captionpos=t, linerange={45-72},firstnumber=45]
#define makeChemistryModel(Comp)                                               \
                                                                               \
    typedef BasicChemistryModel<Comp> BasicChemistryModel##Comp;               \
                                                                               \
    defineTemplateTypeNameAndDebugWithName                                     \
    (                                                                          \
        BasicChemistryModel##Comp,                                             \
        "BasicChemistryModel<"#Comp">",                                        \
        0                                                                      \
    );                                                                         \
                                                                               \
    defineTemplateRunTimeSelectionTable                                        \
    (                                                                          \
        BasicChemistryModel##Comp,                                             \
        thermo                                                                 \
    );


#define makeChemistryModelType(SS, Comp, Thermo)                               \
                                                                               \
    typedef SS<Comp, Thermo> SS##Comp##Thermo;                                 \
                                                                               \
    defineTemplateTypeNameAndDebugWithName                                     \
    (                                                                          \
        SS##Comp##Thermo,                                                      \
        (#SS"<"#Comp"," + Thermo::typeName() + ">").c_str(),                   \
        0                                                                      \
    );
\end{lstlisting}

\begin{lstlisting}[label={ls_make_2}, style=cpp, caption=Macros for chemistry solvers based on compressibility and transport types,captionpos=t, linerange={41-80},firstnumber=41]
#define makeChemistrySolverType(SS, Comp, Thermo)                              \
                                                                               \
    typedef SS<cemaPyjacChemistryModel<Comp, Thermo>> SS##Comp##Thermo;         \
                                                                               \
    defineTemplateTypeNameAndDebugWithName                                     \
    (                                                                          \
        SS##Comp##Thermo,                                                      \
        (#SS"<" + word(cemaPyjacChemistryModel<Comp, Thermo>::typeName_()) + "<"\
        + word(Comp::typeName_()) + "," + Thermo::typeName() + ">>").c_str(),  \
        0                                                                      \
    );                                                                         \
                                                                               \
    BasicChemistryModel<Comp>::                                                \
        add##thermo##ConstructorToTable<SS##Comp##Thermo>                      \
        add##SS##Comp##Thermo##thermo##ConstructorTo##BasicChemistryModel##Comp\
##Table_;


#define makeChemistrySolverTypes(Comp, Thermo)                                 \
                                                                               \
    makeChemistrySolverType                                                    \
    (                                                                          \
        noChemistrySolver,                                                     \
        Comp,                                                                  \
        Thermo                                                                 \
    );                                                                         \
                                                                               \
    makeChemistrySolverType                                                    \
    (                                                                          \
        EulerImplicit,                                                         \
        Comp,                                                                  \
        Thermo                                                                 \
    );                                                                         \
                                                                               \
    makeChemistrySolverType                                                    \
    (                                                                          \
        odePyjac,                                                             \
        Comp,                                                                  \
        Thermo                                                                 \
    );                                                                         \
\end{lstlisting}

Finally, the templated instances of the chemistry models and chemistry solvers, which are specified in \verb|Make/files|, c.f.~\Cref{ls_make_files}, create the instances for all combinations of OpenFOAM thermodynamic and transport models. Due to their lengthy code, definitions of the corresponding files are shown in \Cref{ap:make}.

\chapter{Tutorial on 1D planar premixed flame}

% This is a example of chapter creation using the \verb|\chapter| command.
In this tutorial, we demonstrate the applicability of the developed model on a laminar one-dimensional unstrained planar premixed flame comprising methane and air at equivalence ratio of $\phi=0.5$ and thermodynamic conditions of $T=900$~\si{K} and $p=1$~\si{atm}. The core file structure for this demonstration is based on the default OpenFOAM tutorial \verb|counterFlowFlame2D_GRI| which employs the GRI-3.0 mechanism~\cite{gri} that we are going to use as well.
% , so that we do not build the case from scratch.
Accordingly, The tutorial case is copied to a convenient workspace together with some initial cleaning as follows.

\begin{verbatim}
> cp -r $FOAM_TUTORIALS/combustion/reactingFoam/laminar/counterFlowFlame2D_GRI \
$FOAM_RUN/cemaPyjac_tutorial
> rm $FOAM_RUN/cemaPyjac_tutorial/0/alphat
> rm $FOAM_RUN/cemaPyjac_tutorial/constant/reactions
> rm $FOAM_RUN/cemaPyjac_tutorial/constant/thermo.compressibleGas
\end{verbatim}

In the following, we first discuss the generation and compilation of the relevant \verb|pyJac| subroutines for the GRI-3.0 mechanism. After that, we proceed by discussing the case setup in terms of initial and boundary conditions, mesh generation, finite-volume settings, and IO and control options. After executing all necessary commands to run the simulation using the newly developed model, the results are presented with validations and demonstrations of the \verb|pyJac| analytical Jacobian and subsequently CEMA fields.

% Here, we provide the crucial parts of the \verb|controlDict| and \verb|thermophysicalProperties| corresponding to dynamic linking of \verb|PyJac| and the developed model, and providing the path for reading of chemistry data, respectively. The complete set of input data and dictionaries are provided separately.

% \section{Generating GRI-30 mechanism using \texttt{pyJac}}
\section{Preparation of the chemical mechanism}
\label{sec:tut_mech}

First, it is expected from the reader to have successfully installed \verb|pyJac| on their computer. The package is hosted by \verb|GitHub| and can be accessed through \url{https://github.com/SLACKHA/pyJac}. Moreover, brief instructions on the package installation are found in \Cref{sec:pyj_install}, while detailed instructions can be found through the online documentation \url{http://slackha.github.io/pyJac}.

After navigating to the tutorial workspace, we find that the mechanism reaction file (\verb|reactionsGRI|) and thermodynamic file(\verb|thermo.compressibleGasGRI|), which are both located in \verb|constant| directory, are in OpenFOAM format. In order to generate corresponding \verb|pyJac| subroutines, the mechanism files in either \verb|Chemkin| or \verb|Cantera| format should be provided. Accordingly, we download the original files of the GRI-3.0 mechanism from the website \url{http://combustion.berkeley.edu/gri-mech/version30/text30.html} and we store the $3$ files, namely \verb|grimech30.dat|, \verb|thermo30.dat|, and \verb|transport.dat| in directory named \verb|mechanism|~\footnote{Remember to save the files in \texttt{*.dat} format and not \texttt{*.dat.txt}.} in the case folder.

Now, we are ready to use \verb|pyJac| to generate the corresponding subroutines. This can be executed as follows.
\begin{verbatim}
> cd $FOAM_RUN/cemaPyjac_tutorial/mechanism
> python3 -m pyjac --lang c --input grimech30.dat --thermo thermo30.dat --last_species N2
\end{verbatim}

The previous \verb|Python| command with \verb|-m| switch executes the \verb|pyJac| module through $\underline{\hspace{3mm}}$\verb|main|$\underline{\hspace{3mm}}$\verb|.py|. The remaining arguments of the command are \verb|pyJac| related, and they are responsible for the programming language of the output source files (i.e. \verb|--lang|), the input mechanism filename (i.e. \verb|--input|), the thermodynamic database (i.e. \verb|--thermo|), and the species name to be set as the last species for dumping all numerical residuals (i.e. \verb|--last\_species|), c.f.~\autoref{sec:pyj_func}.

The result of the previous command is a new subdirectory named \verb|out|, located inside \verb|mechanism| directory. We can see that definition of analytical Jacobian evaluation of the GRI-3.0 mechanism is realized in \verb|eval_jacob()| function located in \verb|out/jacob.c| and \verb|out/jacobs/jacob_*.c|. Also, functionalities for derivatives and enthalpy of formation are realized in \verb|dydt()| and \verb|eval_h()|, located in \verb|out/dydt.c| and \verb|out/chem_utils.c|, respectively.

Therefore, we need compile all these source code files into a shared object to be dynamically linked to OpenFOAM, providing all function definitions previously declared when compiling the model~\footnote{Recall that all function declarations of \texttt{pyJac} are mechanism independent.}. The compilation process can be achieved through \verb|CMake|, and the corresponding set of directives and instructions for the source code files and targets are described in the following \verb|CMakeLists.txt|.

\begin{lstlisting}[label={pyj_cmakelists}, style=cpp, caption=\texttt{CMakeLists.txt} to compile mechanism source code files generated from \texttt{pyJac},captionpos=t]
cmake_minimum_required(VERSION 2.6)

project(pyJac)

set(CMAKE_BUILD_TYPE Release)

enable_language(C)

set(CMAKE_C_FLAGS "-std=c99 -O3 -mtune=native -fPIC")

include_directories(out)
include_directories(out/jacobs)

file(GLOB_RECURSE SOURCES "out/*.c")

add_library(c_pyjac SHARED ${SOURCES})

install(TARGETS c_pyjac DESTINATION .)

\end{lstlisting}

After creating \verb|CMakeLists.txt| inside the \verb|mechanism| directory, the build process can be achieved by executing the following commands.

\begin{verbatim}
> cd $FOAM_RUN/cemaPyjac_tutorial/mechanism
> mkdir build
> cd build && cmake .. -DCMAKE_C_COMPILER=cc
> make
> cp libc_pyjac.so ../../constant/
\end{verbatim}

Here, the target name of the resulting shared object is \verb|libc_pyjac.so| and it is located inside \verb|build| directory, so we conveniently copy it to \verb|constant| so that all mechanism files are there. The shared library object can be dynamically linked, together with the developed model, through the \verb|controlDict| by specifying them as follows.

\begin{verbatim}
libs (
    "libcemaPyjacChemistryModel.so"
    "$FOAM_CASE/constant/libc_pyjac.so"
    );
\end{verbatim}

It is worth noting that the decision of locating the shared object \verb|libc_pyjac.so| inside the case directory (i.e. to be case specific) is to avoid interference of mechanism function definitions when running multiple OpenFOAM cases using different chemical mechanisms. In simple words, locating the \verb|pyJac| mechanism object into the conventional OpenFOAM user libraries path (i.e. \verb|$FOAM_USER_LIBBIN|) might result into erroneous results when other simulations, using the developed model, run while using different mechanisms than the one compiled in \verb|libc_pyjac.so|.

At this point, the \verb|constant| directory contains the original OpenFOAM reaction and thermodynamic mechanism files for GRI-3.0, along with the \verb|pyJac| analytical based functionalities. There are two important notes to herein consider. 
First, reaction type and Arrhenius coefficients are no longer required when \verb|pyJac| is used. The reason is that \verb|pyJac| already contains all the information regarding reaction rates, progress rate of reactions, 
% species overall production rates, 
thermo-chemical Jacobian, among others, which are necessary for computing the chemistry source terms.
Second, the ordering of chemical species in the mechanism reaction file must follow the same ordering to that in \verb|pyJac|.
This is particularly important since we specified \ce{N2} to be the last species (i.e. using \verb|--last\_species| switch) to handle all residuals, c.f.~\autoref{sec:pyj_func}. The correct species ordering that must be followed can be retrieved from \verb|mechanism/out/mechanism.h|.
Therefore, we create a mechanism reaction file \verb|reactionsGRIPyjac|, modified from \verb|constant/reactionsGRI|, and it is depicted in \Cref{reac_mech}.

\begin{lstlisting}[label={reac_mech}, style=OpenFOAMDict, caption=Reaction mechanism file \texttt{reactionsGRIPyjac} located in \texttt{constant},captionpos=t]
elements
5
(
O
H
C
N
Ar
)
;

species
53
(
H2
H
O
O2
OH
H2O
HO2
H2O2
C
CH
CH2
CH2(S)
CH3
CH4
CO
CO2
HCO
CH2O
CH2OH
CH3O
CH3OH
C2H
C2H2
C2H3
C2H4
C2H5
C2H6
HCCO
CH2CO
HCCOH
N
NH
NH2
NH3
NNH
NO
NO2
N2O
HNO
CN
HCN
H2CN
HCNN
HCNO
HOCN
HNCO
NCO
AR
C3H7
C3H8
CH2CHO
CH3CHO
N2
)
;

reactions
{
}
\end{lstlisting}

A final note of this section is to remind the user to manually specify ---again--- the number of elements of the chemical mechanism (defined in \verb|reactionsGRIPyjac|) into the \verb|chemistryProperties| dictionary, so as to be used by CEMA subroutine in the chemistry model. The keyword \verb|nElements| needs to be used as follows for this particular mechanism.

\begin{lstlisting}[label={n_elements}, style=OpenFOAMDict, caption=Number of elements of the reaction mechanism file \texttt{reactionsGRIPyjac} to be specified in \texttt{chemistryProperties} for CEMA,captionpos=t, linerange={35-35},firstnumber=35]
nElements        5;
\end{lstlisting}

After reaching this point, we are ready to proceed with the case setup.

\section{Case setup}

In this section, we discuss the choices for initial and boundary conditions, domain discretization, finite-volume numerical schemes and solver settings, and the IO control options.

% On the other hand, if we wish to have stabilized flame, then reaction front needs to be initialized and the velocity is set equivalent to the laminar burning velocity such that the front is spatially fixed. In this report, we choose the latter option and therefore profiles of temperature and key species are initialized with the reaction front located about the middle of the domain.

\subsection{Internal and boundary conditions}

The present tutorial aims at simulating a one-dimensional ($1$D) unstrained planer premixed flame of methane and air mixture at equivalence ratio of $\phi=0.5$ and at temperature and pressure of $T=900$~[\si{K}] and $p=1$~[\si{atm}], respectively. In order to ensure a quasi-stabilized flame, we initialize the domain with non-uniform internal fields for the temperature and key species, namely \ce{CH4}, \ce{O2}, \ce{N2}, \ce{CO2}, and \ce{H2O}, such that an ignition event is initialized.
The spatial distribution of the aforementioned profiles is set to attain temperature rise after nearly halfway of the domain length from inlet boundary. Stabilization of the reaction front is achieved by setting the inlet velocity equivalent to the laminar burning velocity $u_\textrm{L} \approx 1.39$~[\si{m/s}] . A data interpolation is performed on the initialized profiles of temperature and the aforementioned key species to match the desired grid points, as it is further elaborated in \autoref{sec:tut_mesh}. The boundary conditions are presented in \autoref{tab:tut_bc}. 

% It is important to carefully set up the case so that boundary conditions and the internal field values are physically meaningful. In the author's opinion, there seems to be two approaches to achieve that. The first approach is to set the inlet boundary such that it introduces fresh fuel and air mixture at the designated thermodynamic conditions. Additionally, the internal field should be initialized as hot and inert, while the outlet boundary is defined by zero gradient conditions. The downside of this approach is that it requires inlet velocity to be chosen with a proper value that is larger than laminar burning velocity so that reaction front will occur somewhere in the middle of the domain. The downside of this approach is that flame is not stabilized and eventually will escape the domain (in case of zero gradient outlet) after some time. The second approach is have a stabilized flame by setting velocity at the inlet boundary equivalent to laminar burning velocity. However, this requires initializing the reaction front somewhere in the internal field by providing proper profiles for the spatial distribution of temperature and key species. In this report, we choose the latter approach.

% \begin{table}[!htb]
\begin{table}[h!]
\caption{Internal and boundary conditions of the tutorial case. Symbol $\gamma$ denotes specific heat ratio.}
\label{tab:tut_bc}
\small
\begin{center}
\begin{tabular}{l l l l}
\hline
\bf Field variable & {\bf Inlet} & {\bf Internal} & {\bf Outlet} \\
\hline
% Pressure~[\si{Pa}] & \texttt{waveTransmissive} ($\gamma^\text{a} = 1.318654$) & $101325$ & \texttt{waveTransmissive} ()$\gamma = 1.272352$) \\
Pressure~[\si{Pa}] & \texttt{waveTransmissive} ($\gamma = 1.32$) & $101325$ & \texttt{waveTransmissive} ($\gamma = 1.27$) \\
Temperature~[\si{K}] & $900$ & non-uniform~$^\text{a}$ & zero gradient \\
\ce{CH4} molar $\%$ (mass $\%$) & $4.98812$ ($2.83654$) & non-uniform & zero gradient \\
\ce{O2} molar $\%$ (mass $\%$) & $19.9525$ ($22.631$) & non-uniform & zero gradient \\
\ce{N2} molar $\%$ (mass $\%$) & $75.05938$ ($74.53246$) & non-uniform & zero gradient \\
\ce{CO2} molar $\%$ (mass $\%$) & $0.0$ ($0.0$) & non-uniform & zero gradient \\
\ce{H2O} molar $\%$ (mass $\%$) & $0.0$ ($0.0$) & non-uniform & zero gradient \\
Equivalence ratio~$^\text{b}$ ($\phi$) & $0.5$ & $0.5$ & $0.5$ \\
Velocity~[\si{m/s}] & ($1.456$, $0$, $0$) & $0.0$ & zero gradient \\
\end{tabular}
\end{center}
\footnotesize
% $^\text{a}$ Similar to the ECN Spray A, \ce{n-C12H26} is injected continuously at T=363 K and p=150 MPa, and the nominal nozzle diameter is 90 $\mu$m. \\
% $^\text{a}$ Specific heat ratio.\\
$^\text{a}$ Spatial distribution profiles are interpolated. Corresponding scripts are provided in \Cref{ap:utilities}.\\
$^\text{b}$ For unity Lewis number of all species, which is the case in the present simulation, local equivalence ratio is expected to be rather constant and equals to that in the reactants. For simulations with non-unity Lewis number, variation is expected to lower down across the reaction front, before approaching again the local value of the reactants after crossing the front toward product side~\cite{Lee2021}. This is beyond the scope of this project.
\end{table}

Regarding the corresponding files for initial and boundary conditions, they are presented in \Cref{ls_init_T,ls_init_ch4} for temperature and methane (\ce{CH4}), respectively. The remaining key species are specified in an analogous manner to that for \ce{CH4} while modifying the uniform value of the inlet boundary by using \Cref{tab:tut_bc}, and updating the species names in the two occurrences of the dictionary. The interpolated data that is included using the command \verb|#include"CH4.dat"| are provided in the tutorial case folder under \verb|0| directory. Moreover, it is possible to re-generate the data using the attached helper scripts in \verb|utilities| directory. The utilities scripts are written in \verb|Python| and they require \verb|Cantera| package to be installed. After executing \verb|compute_inital_fields.py| and then \verb|interpolate_inital_fields.py|, the generated \verb|*.dat| files will be located inside \verb|out_states| subdirectory which are then required to be copied into \verb|0| directory of the tutorial case.

% Nevertheless, helper scripts for the interpolator and the input files generator is also attached in \Cref{ap:utilities}. \todo{ENSURE ATTACHED!}

\begin{lstlisting}[label={ls_init_T}, style=OpenFOAMDict, caption=Initial and boundary conditions for temperature field,captionpos=t, linerange={8-41},firstnumber=8]
FoamFile
{
    version     2.0;
    format      ascii;
    class       volScalarField;
    location    "0";
    object      T;
}
// * * * * * * * * * * * * * * * * * * * * * * * * * * * * * * * * * * * * * //

dimensions      [0 0 0 1 0 0 0];

internalField   nonuniform
#include"T.dat"
;

boundaryField
{
    inlet
    {
        type            fixedValue;
        value           uniform 900;
    }

    outlet
    {
        type              zeroGradient;
    }

    frontAndBack
    {
        type            empty;
    }
}
\end{lstlisting}

\begin{lstlisting}[label={ls_init_ch4}, style=OpenFOAMDict, caption=Initial and boundary conditions for methane field,captionpos=t, linerange={8-41},firstnumber=8]
FoamFile
{
    version     2.0;
    format      ascii;
    class       volScalarField;
    location    "0";
    object      CH4;
}
// * * * * * * * * * * * * * * * * * * * * * * * * * * * * * * * * * * * * * //

dimensions      [0 0 0 0 0 0 0];

internalField   nonuniform
#include"CH4.dat"
;

boundaryField
{
    inlet
    {
        type            fixedValue;
        value           uniform 0.0283654;
    }

    outlet
    {
        type            zeroGradient;
    }

    frontAndBack
    {
        type            empty;
    }
}
\end{lstlisting}

% U is set to correspond ---approximately--- to laminar flame speed, so that reaction front should be spatially fixed and thereby flame is stabilized.

% Accordingly, we need at least temperature field to be properly dis
It is worth mentioning that the chosen approach of initializing the ignition event by interpolating spatial profiles of temperature and key species is made for simulation feasibility. In particular, for the initial conditions of pressure and temperature presented in \Cref{tab:tut_bc}, ignition delay time (IDT) of methane/air mixture at $\phi=0.5$ is about $5.39$~[\si{s}] which is considered too long, considering the relatively small temporal and spatial length scales of the simulation as it will be further elaborated in \Cref{sec:tut_mesh,sec:io}.

% \subsection{Computational mesh}
\subsection{Domain discretization}
\label{sec:tut_mesh}

Aiming for resolving the flame thermal thickness ($\delta_\textrm{L} \approx 3.72 \times 10^{-4}$~[\si{m}] for the conditions specified in \Cref{tab:tut_bc}), the grid spacing is chosen to be $1.5 \times 10^{-5}$~[\si{m}]. Therefore, thermal thickness is resolved with more than $20$ grid points. The domain length is set to $2 \times 10^{-2}$~[\si{m}] hence a total amount of $1350$ grid points, in which the number is inserted to the interpolator to generate arrays \verb|0/*.dat| of $1350$ internal grid points in the discretized domain. The \verb|blockMesh| dictionary is depicted in \Cref{ls_blockmesh}.

\begin{lstlisting}[label={ls_blockmesh}, style=OpenFOAMDict, caption=\texttt{blockMesh} dictionary,captionpos=t, linerange={8-74},firstnumber=17]
FoamFile
{
    version     2.0;
    format      ascii;
    class       dictionary;
    object      blockMeshDict;
}
// * * * * * * * * * * * * * * * * * * * * * * * * * * * * * * * * * * * * * //

scale   0.01;

vertices
(
    (0          0       0)
    (2.0        0       0)
    (2.0        0.25    0)
    (0          0.25    0)
    (0          0       0.25)
    (2.0        0       0.25)
    (2.0        0.25    0.25)
    (0          0.25    0.25)
);

blocks
(
    hex (0 1 2 3 4 5 6 7) (1350 1 1) simpleGrading (1 1 1)
);

edges
(
);

boundary
(
    inlet
    {
        type patch;
        faces
        (
           (4 7 3 0)
        );
    }
    outlet
    {
        type patch;
        faces
        (
            (1 2 6 5)
        );
    }

    frontAndBack
    {
        type empty;
        faces
        (
            (0 3 2 1)
            (4 5 6 7)
            (7 6 2 3)
            (1 0 4 5)
        );
    }
);

mergePatchPairs
(
);
\end{lstlisting}

% // For p=1 atm
% // Flamespeed = 1.7668476341920771
% // Flame thickness = 0.0003964743031231495 => Need dx=1.5e-05 to resolve with 25 points
% // CORRECT VALUES BELOW
% // Flamespeed = 1.6494969950659772
% // Flame thickness = 0.0003727488703227711

% 0.02 m with 1350 grid points, so that we ensure resolving flame thermal thickness with at least 20 grid points

% block mesh dict

% 0.000373 m

\subsection{Finite volume settings}

Modifications in the finite-volume numerical schemes and solver settings, with respect to default ones from \verb|counterFlowFlame2D_GRI| tutorial, should not have major impact on the results. Nevertheless, our preferences for the numerical schemes and solver options are herein presented. Modifications on numerical schemes are adopted through the following commands

\begin{verbatim}
> cd $FOAM_RUN/cemaPyjac_tutorial
> sed -i 's/Euler/backward/g' system/fvSchemes
> sed -i 's/limitedLinearV/limitedLinear/g' system/fvSchemes #not necessary
> sed -i 's/Gauss linear orthogonal/Gauss linear corrected/g' system/fvSchemes
\end{verbatim}

For solver settings, the updated \verb|fvSolution| dictionary is depicted in \Cref{ls_fvsol}.

\begin{lstlisting}[label={ls_fvsol}, style=OpenFOAMDict, caption=\texttt{fvSolution} dictionary,captionpos=t, linerange={8-81},firstnumber=8]
FoamFile
{
    version     2.0;
    format      ascii;
    class       dictionary;
    location    "system";
    object      fvSolution;
}
// * * * * * * * * * * * * * * * * * * * * * * * * * * * * * * * * * * * * * //

solvers
{
    rho
    {
        solver           PCG;
        preconditioner   DIC;
        tolerance        1e-8;
        relTol           0.005;
    }

    rhoFinal
    {
        $rho;
        tolerance 1e-8;
        relTol           0.0;
    }
    p
    {
        solver           PCG;
        preconditioner   DIC;
        tolerance        1e-8;
        relTol           0.005;
    }

    pFinal
    {
        $p;
        tolerance 1e-8;
        relTol           0.0;
    }

    "(U|h|k|epsilon)"
    {
        solver          PBiCGStab;
        preconditioner  DILU;
        tolerance       1e-7;
        relTol          0.005;
    }

    "(U|h|k|epsilon)Final"
    {
        $U;
        relTol          0;
    }

    Yi
    {
        $hFinal;
    }

    YiFinal
    {
        $Yi;
    }

}

PIMPLE
{
    momentumPredictor no;
    nOuterCorrectors  3;
    nCorrectors     4;
    nNonOrthogonalCorrectors 0;
}
\end{lstlisting}

% Changes in the fvSolution should not have significant impact on the results. Nevertheless, we present the chosen settings in the following listing

\subsection{Chemistry and thermophysical properties}

At this point, the tutorial case files are updated for the mechanism files (inside \verb|constant| directory, copied from \verb|mechanism| directory), the initial and boundary conditions (inside \verb|0| directory), the mesh data (inside \verb|constant| directory), and the finite-volume settings (inside \verb|system| directory). Now, we update the chemistry and thermophysical properties in the \verb|constant| directory. The \verb|chemistryProperties| dictionary needs to be properly updated to allow using \verb|cemaPyjac| chemistry model, while modifying stiff ODE solver tolerances for faster computations. The modifications can be adopted through the following commands.

\begin{verbatim}
> sed -i 's/ode;/odePyjac;\nmethod\t\tcemaPyjac;/g' constant/chemistryProperties
> sed -i 's/initialChemicalTimeStep 1e-7/initialChemicalTimeStep 1e-8/g' \
  constant/chemistryProperties
> sed -i 's/1e-12/1e-08/g' constant/chemistryProperties #absTol
> sed -i 's/1e-1/1e-05/g' constant/chemistryProperties #relTol
\end{verbatim}

Another note is to remember specifying the number of elements (using \verb|nElements| keyword) of the chemical mechanism, which can be looked up from \verb|reactionsGRIPyjac|. This is particularly important for CEMA computations to skip the conservative modes and thereby accurately identify the CEM. 
The final form of the \verb|chemistryProperties| dictionary is depicted in \Cref{ls_chemprop}.

\begin{lstlisting}[label={ls_chemprop}, style=OpenFOAMDict, caption=\texttt{chemistryProperties} dictionary,captionpos=t, linerange={8-35},firstnumber=8]
FoamFile
{
    version     2.0;
    format      ascii;
    class       dictionary;
    location    "constant";
    object      chemistryProperties;
}
// * * * * * * * * * * * * * * * * * * * * * * * * * * * * * * * * * * * * * //

chemistryType
{
    solver            odePyjac; //ode; odePyjac;
    method	      cemaPyjac; //standard; cemaPyjac;
}

chemistry       on;

initialChemicalTimeStep 1e-8;

odeCoeffs
{
    solver          seulex;
    absTol          1e-08;
    relTol          1e-05;
}

nElements        5;
\end{lstlisting}

The \verb|thermophysicalProperties| dictionary needs to be updated only to specify the path of the modified reaction mechanism file (i.e. with species correct ordering and without reaction rate data, c.f.~\Cref{sec:tut_mech}). This is achieved through the following command.

\begin{verbatim}
> sed -i 's/reactionsGRI/reactionsGRIPyjac/g' constant/thermophysicalProperties
\end{verbatim}

The updated \verb|thermophysicalProperties| dictionary should look like the following \Cref{ls_therphysprop}. The remaining dictionaries in \verb|constant| directory for turbulence and combustion properties are kept as \verb|laminar|, similar to the original \verb|counterFlowFlame2D_GRI| tutorial.

\begin{lstlisting}[label={ls_therphysprop}, style=OpenFOAMDict, caption=\texttt{thermophysicalProperties} dictionary,captionpos=t, linerange={8-33},firstnumber=8]
FoamFile
{
    version     2.0;
    format      ascii;
    class       dictionary;
    location    "constant";
    object      thermophysicalProperties;
}
// * * * * * * * * * * * * * * * * * * * * * * * * * * * * * * * * * * * * * //

thermoType
{
    type            hePsiThermo;
    mixture         reactingMixture;
    transport       sutherland;
    thermo          janaf;
    energy          sensibleEnthalpy;
    equationOfState perfectGas;
    specie          specie;
}

inertSpecie N2;

chemistryReader foamChemistryReader;
foamChemistryFile "<constant>/reactionsGRIPyjac";
foamChemistryThermoFile "<constant>/thermo.compressibleGasGRI";
\end{lstlisting}

\subsection{IO control and dynamic linking}
\label{sec:io}

When it comes to IO control through \verb|controlDict| file, it is always up to user's personal preferences to decide on the total simulation time as well as the data writing format and frequency. However, the essential parts to be noted are that the numerical time step and total simulation time both need to be adequate with regard to domain characteristic length and time scales. More over, the simulation time should be specified sufficiently long to allow the reactive flow to fully develop in terms of species concentrations and reaction front establishment. 
Compared with the default \verb|controlDict| from \verb|counterFlowFlame2D_GRI|, the modifications can be adopted through the following commands

\begin{verbatim}
> sed -i 's/0.5/0.002/g' system/controlDict #endTime
> sed -i 's/1e-6/1e-8/g' system/controlDict #deltaT
> sed -i 's/adjustable/adjustableRunTime/g' system/controlDict #writeControl
> sed -i 's/0.05/1e-04/g' system/controlDict #writeInterval
> sed -i 's/0.4/0.3/g' system/controlDict #maxCo
\end{verbatim}

The shared object for the chemistry model and for \verb|pyJac| are both dynamically linked as previously discussed in \Cref{sec:tut_mech}. The \verb|controlDict| file after all modifications is depicted in \Cref{ls_controldict}

\begin{lstlisting}[label={ls_controldict}, style=OpenFOAMDict, caption=\texttt{controlDict},captionpos=t, linerange={8-55},firstnumber=8]
FoamFile
{
    version     2.0;
    format      ascii;
    class       dictionary;
    location    "system";
    object      controlDict;
}
// * * * * * * * * * * * * * * * * * * * * * * * * * * * * * * * * * * * * * //

application     reactingFoam;

startFrom       startTime;

startTime       0;

stopAt          endTime;

endTime         0.002;

deltaT          1e-8;

writeControl    adjustableRunTime;

writeInterval   1e-04;

purgeWrite      0;

writeFormat     ascii;

writePrecision  6;

writeCompression off;

timeFormat      general;

timePrecision   6;

runTimeModifiable true;

adjustTimeStep  yes;

maxCo           0.3;

libs (
      "libcemaPyjacChemistryModel.so"
      "$FOAM_CASE/constant/libc_pyjac.so"
     );
\end{lstlisting}

% \lstinputlisting[style=OpenFOAMDict, caption=Dynamic linking of libraries during run time through \texttt{controlDict},captionpos=t, linerange={52-55},firstnumber=52]{Codes/tutorial/system/controlDict}

% \lstinputlisting[style=OpenFOAMDict, caption=specify path of chemistry data files,captionpos=t, linerange={18-33},firstnumber=18]{Codes/tutorial/constant/thermophysicalProperties}

\clearpage

At this point, the tutorial case is ready to be run. To summarize what we have done, we list the procedure in brief as in the following.
\begin{enumerate}
    \item Copy the original tutorial \verb|counterFlowFlame2D_GRI| of \verb|reactingFoam| solver to a desired workspace.

    \item Download chemical mechanism (GRI-3.0) in Chemkin format from the original website. While having \verb|pyJac| installed, generate source code for the GRI-3.0 mechanism involving analytical Jacobian using \verb|pyJac|. After that, build the files using \verb|CMake| and copy the shared object \verb|libc_pyjac.so| to somewhere inside the case directory such as \verb|constant|.

    \item Modify original chemical reaction file such that reaction rate information are omitted, and species index ordering is consistent with \verb|pyJac|, which can be retrieved from \verb|out/mechanism.h| in the generated mechanism source code.

    \item Adjust case setup in terms of initial and boundary conditions, domain discretization, and finite-volume settings as desired, or according to the prescribed settings in the report.

    \item Modify \verb|chemistryProperties| dictionary to use the developed chemistry model \verb|cemaPyjac| while specifying \verb|odePyjac| for the analytical Jacobian based stiff ODE solver. Moreover, specify number of elements of the chemical mechanism, retrieved from mechanism reaction file, using the keyword \verb|nElements|.

    \item Modify \verb|thermophysicalProperties| dictionary to specify location of the updated mechanism reaction file, using keyword \verb|foamChemistryFile|, with the corrected species index ordering and omitted reaction rate data.

    \item Perform dynamic binding of the developed model shared object \verb|libcemaPyjacChemistryModel.so| and the mechanism's analytical Jacobian functionality definitions in \verb|libcemaPyjacChemistryModel.so|. Both objects are dynamically linked at runtime through \verb|controlDict|.

\end{enumerate}

Finally, the simulation can be executed through the following commands, while the results are shown in the following section.

\begin{verbatim}
# Remember to source OpenFOAM-v2006 and to compile CEMAFoam
> blockMesh
> reactingFoam
\end{verbatim}

% \section{Allrun}

% COPY 
% mkdir mechanism; cd mechanism
% cp -r $FOAM_TUTORIALS/combustion/reactingFoam/laminar/counterFlowFlame2D_GRI/constant/reactionsGRI .
% cp -r $FOAM_TUTORIALS/combustion/reactingFoam/laminar/counterFlowFlame2D_GRI/constant/thermo.compressibleGasGRI .
% NOTE: TRANSPORT PROPS IN THERMO FILES ARE BUGGED: MENTION FIXES BY HKAHILA
% cp -r $FOAM_TUTORIALS/combustion/reactingFoam/RAS/DLR_A_LTS/chemkin .

% CONVERT CHEMKIN TO CTI
% https://cantera.org/tutorials/ck2cti-tutorial.html

%%% IMPORTANT: CHEMKIN TO FOAM: USING MECHANISM IN OPENFOAM TUTORIALS (ALREADY BUGGED IN TRANSPORT)
%%% IMPORTANT: CHEMKIN TO CTI: MUST USE ORIGINAL MECHANISM AS THERMO/TRANSPORT GENERATES SOME ERROR AND RESULTS ARE INCOMPLETE
% python3 -m cantera.ck2cti --input=grimech30.dat --thermo=thermo30.dat --transport=transport.dat

\clearpage
% JACOBIAN FILE GENERATION

\section{Results-I: Analytical Jacobian validation}

\begin{figure}[htb!]
\centering
\includegraphics[width=0.95\textwidth]{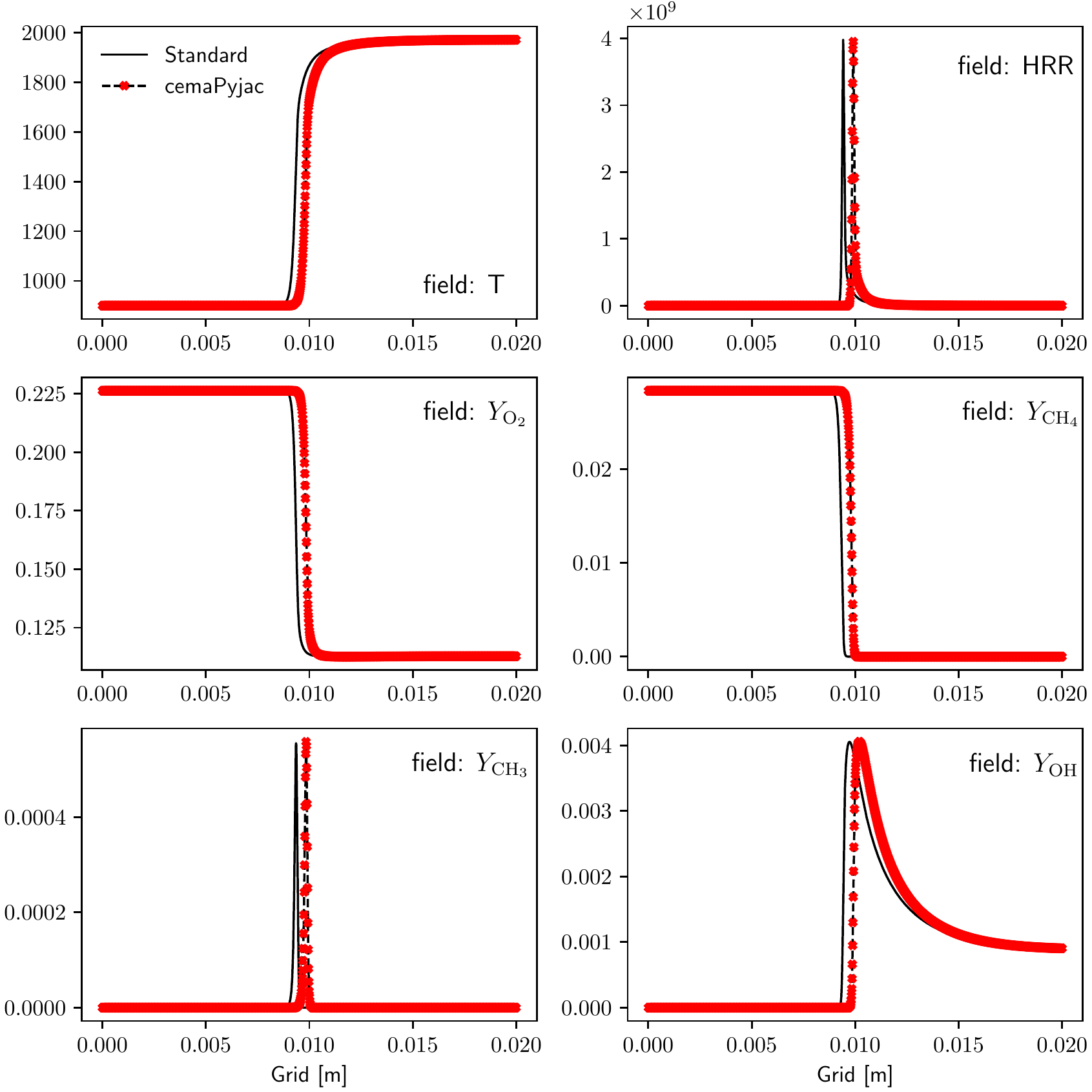}
\caption{Validation of the developed \texttt{cemaPyjacChemistryModel} to computing chemical source terms using analytical Jacobian formulations. Data are sampled after running the simulation for $0.002$~[\si{s}]. Small shift is noted for the reaction zone indicating smaller burning rate for the case adopting \texttt{cemaPyjac} model with analytical Jacobian.}
\label{fig:valid_pyj}
\end{figure}

The validation results of the simulation after $0.002$~[\si{s}] are depicted in \Cref{fig:valid_pyj}. Here, profiles of temperature, heat release rate (HRR), in addition to mass fractions of \ce{O2}, \ce{CH4}, \ce{CH3}, and \ce{OH} are plotted and compared with \verb|standard| chemistry model (without \verb|pyJac|). The presented results verify the implementation accuracy of the developed \texttt{cemaPyjac} chemistry model with the analytical Jacobian formulations from \verb|pyJac|. We note a small shift in the reaction zone which indicates a slower burning rate in the case of using an analytical Jacobian. A possible reasons behind such a shift could be the mass-based versus concentration-based Jacobian matrix which, in turn, might possibly lead into variations in the iterative procedure and solution convergence rate for both cases. This justification could seem reasonable especially that the relative velocity between inlet flow and laminar burning rate is not exactly zero (i.e. there is slight slip in the front location), therefore the front location is not fixed with respect to its initial location.
Next, we present validations of the developed chemistry model for CEMA.

% Simulation is run for some time so that all the $54$ species are properly updated. Moreover, another case with same simulation settings but using the \verb|Standard| chemistry model is performed for validation purposes.

% \subsection{Results using StandardChemistryModel}

% \subsection{Results using cemaPyjacChemistryModel}

% Plot HRR and compare results

\clearpage
\section{Results-II: CEMA demonstration}
\label{sec:cema_results2}

\begin{figure}[htb!]
\centering
\includegraphics[width=0.6\textwidth]{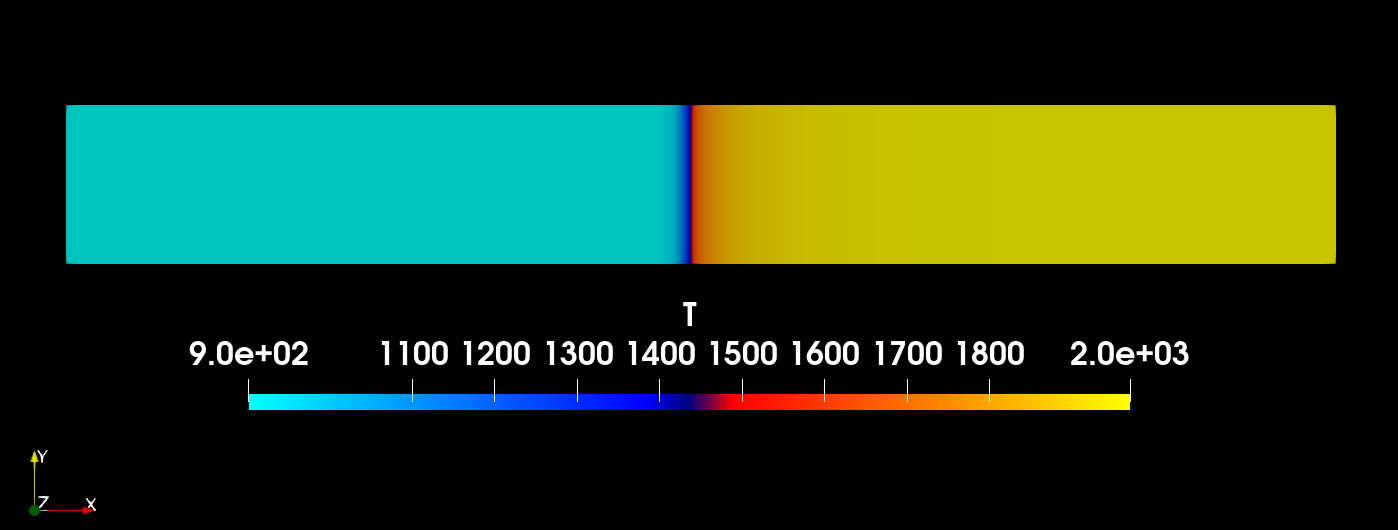}

\vspace{2mm}
\includegraphics[width=0.6\textwidth]{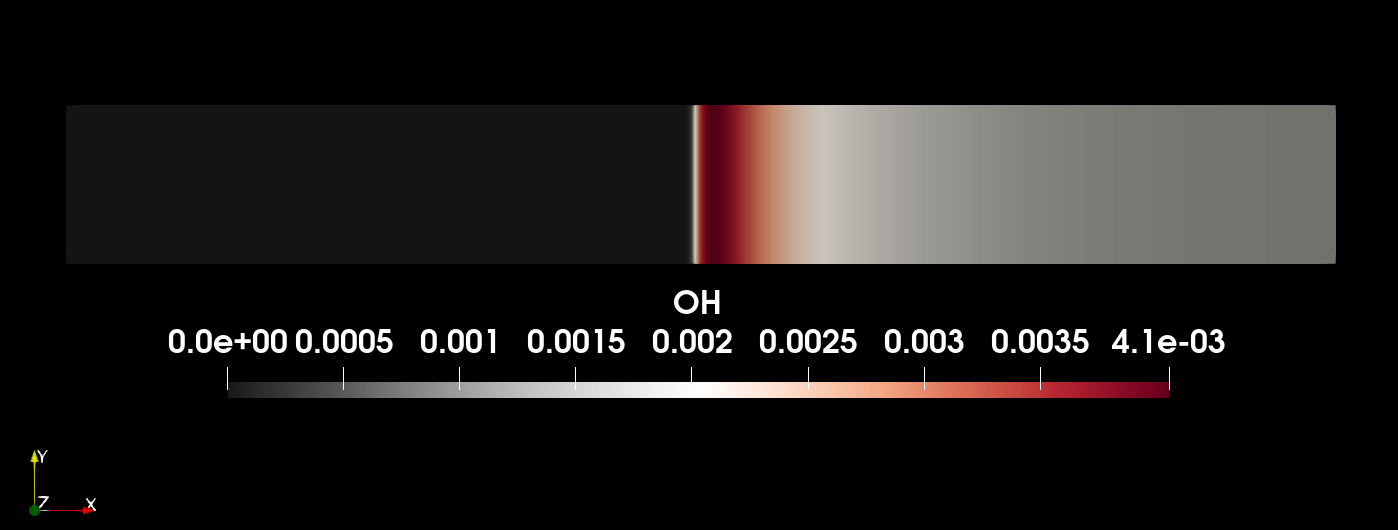}

% \vspace{2mm}
% \includegraphics[width=0.6\textwidth]{Figures/hrr_dfdy.png}
% \includegraphics[width=0.6\textwidth]{Figures/gri30_ch3_dfdy_final.png}

\vspace{2mm}
\includegraphics[width=0.6\textwidth]{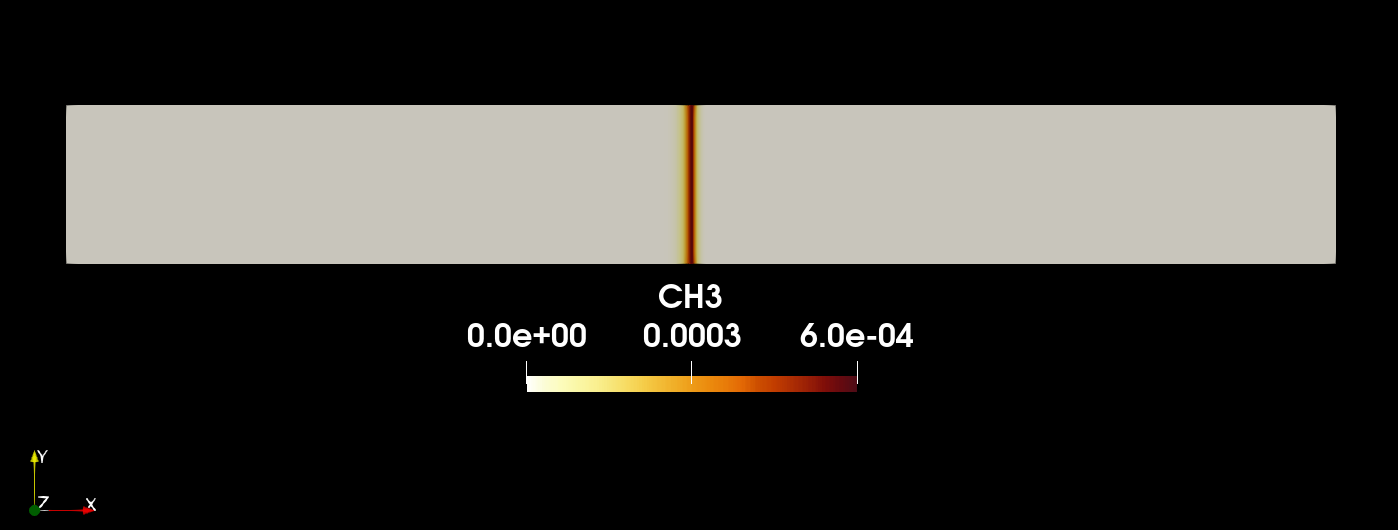}

\vspace{2mm}
\includegraphics[width=0.6\textwidth]{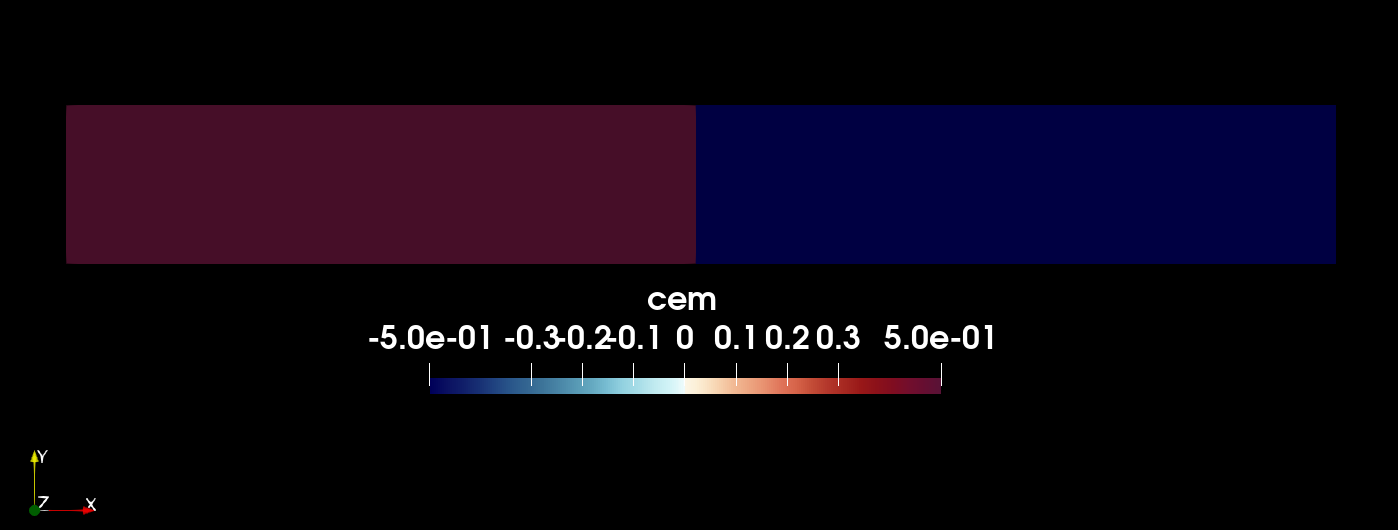}

\caption{Validation of the developed \texttt{cemaPyjacChemistryModel} for CEMA using either analytical Jacobian. Zero-crossing CEMA field clearly indicates the reaction front.}
\label{fig:valid_cem1}
\end{figure}

In this section, we demonstrate results from CEMA functionality of the developed model. The variable \verb|cem|~\footnote{Note that the variable \texttt{cem} denotes leading eigenvalue (i.e. $\lambda_\textrm{exp}$) and it is not the chemical explosive mode (CEM) denoting the corresponding eigenvector.} is the leading non-conservative eigenvalue of the thermo-chemical analytic Jacobian matrix. In \Cref{fig:valid_cem1}, we present field plots of temperature, \ce{OH} and \ce{CH3} mass fractions, in addition to the \verb|cem| field. The positive values of \verb|cem| indicate pre-ignition zones (unstable mode of the dynamical system) and the negative values indicate post-ignition zones (stable mode of the dynamical system), whereas zero-crossing interface can be regarded as the reaction front. As previously mentioned in \Cref{ch:intro}, the identification of pre- and post-ignition zones as well as zero-crossing is a crucial analysis for various subsequent developments as it is briefly demonstrated in \Cref{sec:cema_results3}.

% Here, we need to post process for one snapshot. Therefore, we 

\clearpage

\begin{figure}[htb!]
\centering
\includegraphics[width=0.485\textwidth]{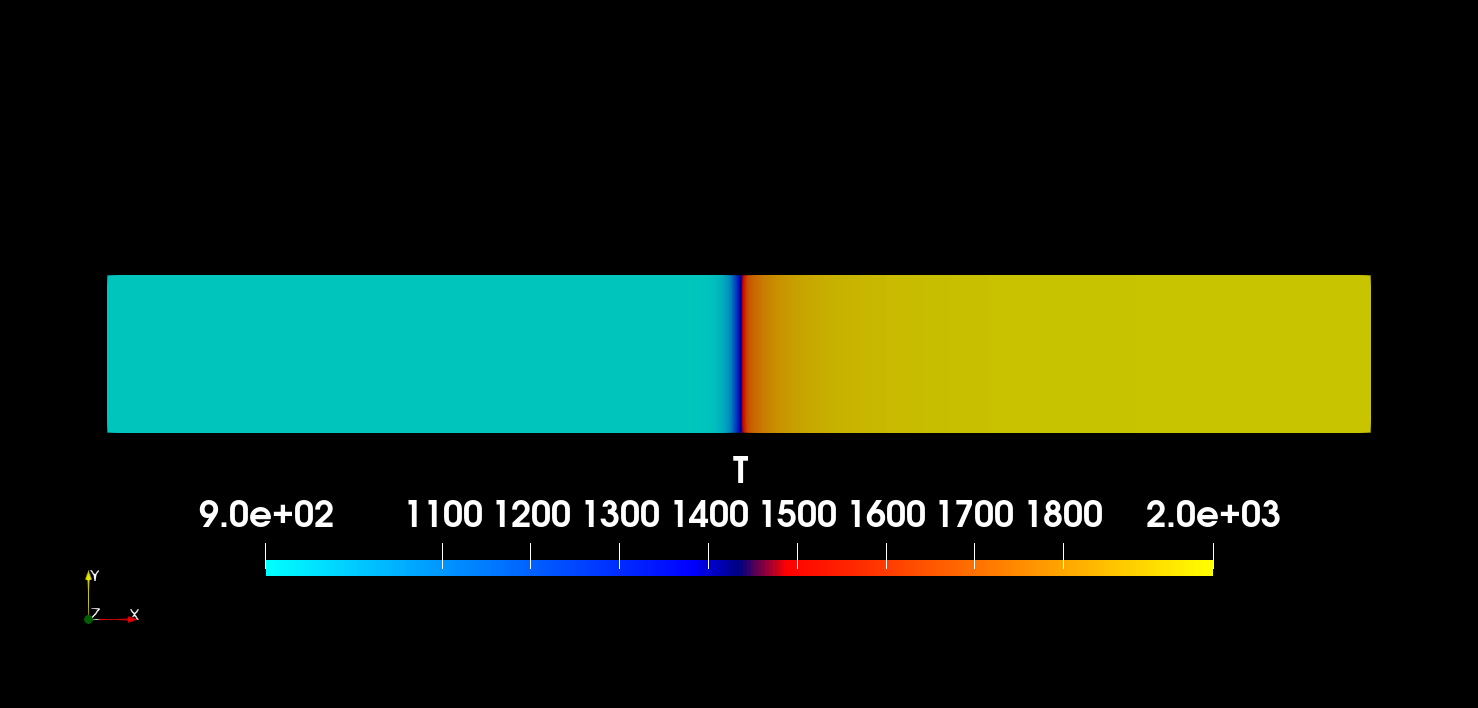} \hfill
\includegraphics[width=0.485\textwidth]{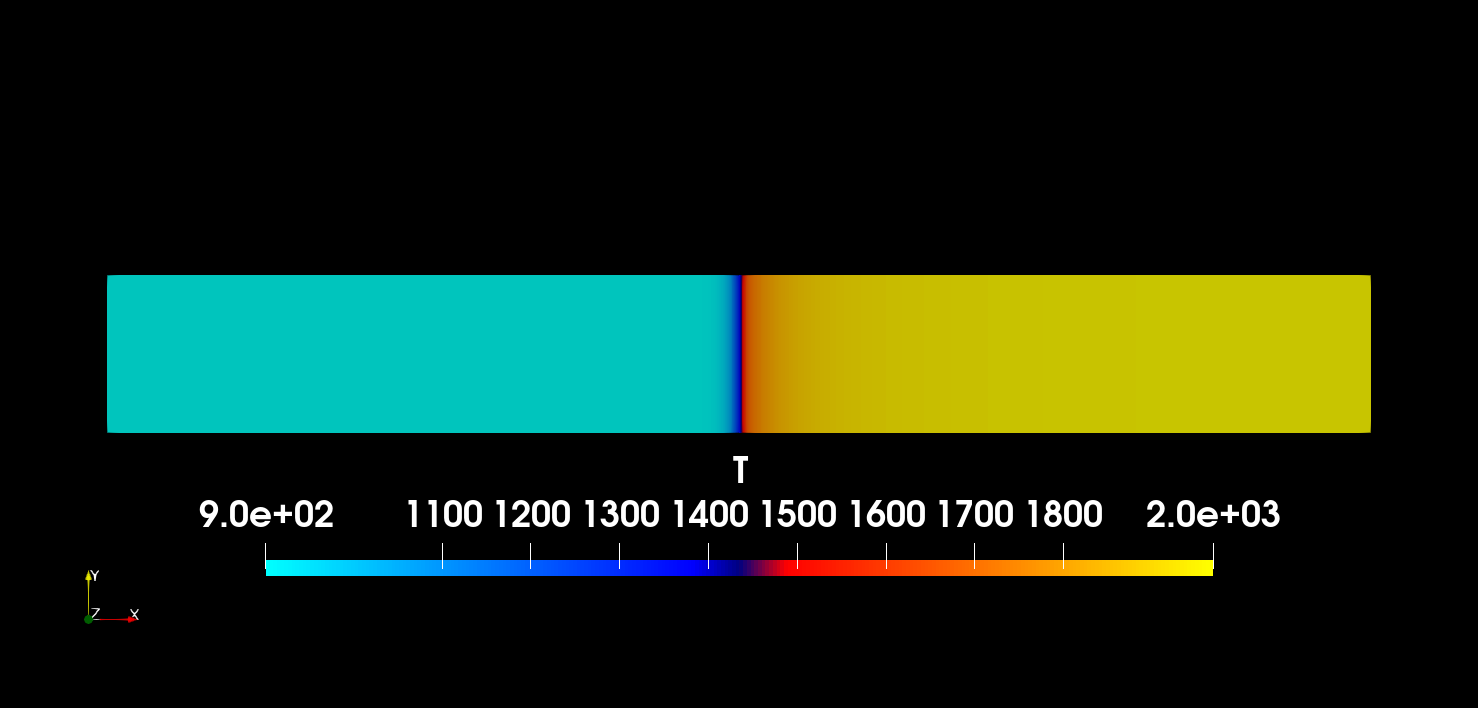}

\vspace{2mm}
\includegraphics[width=0.485\textwidth]{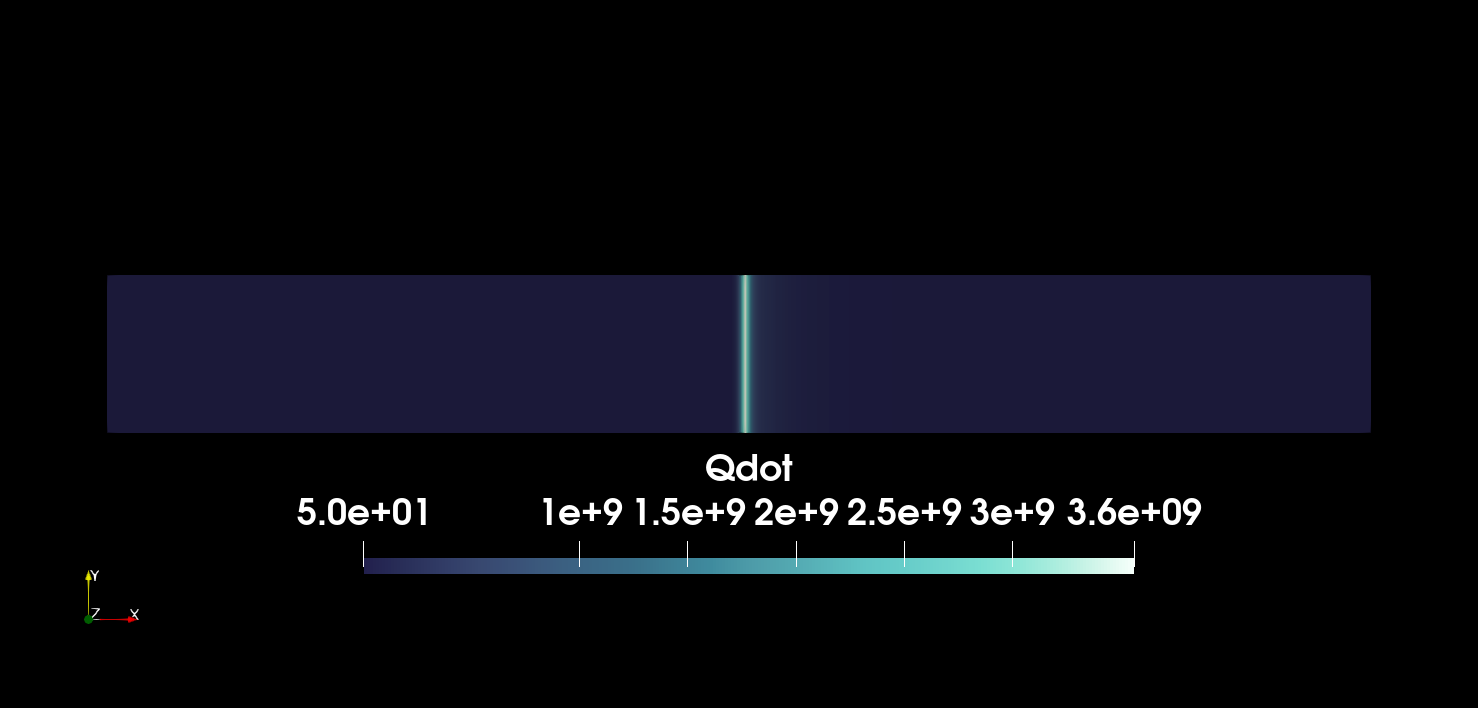} \hfill
\includegraphics[width=0.485\textwidth]{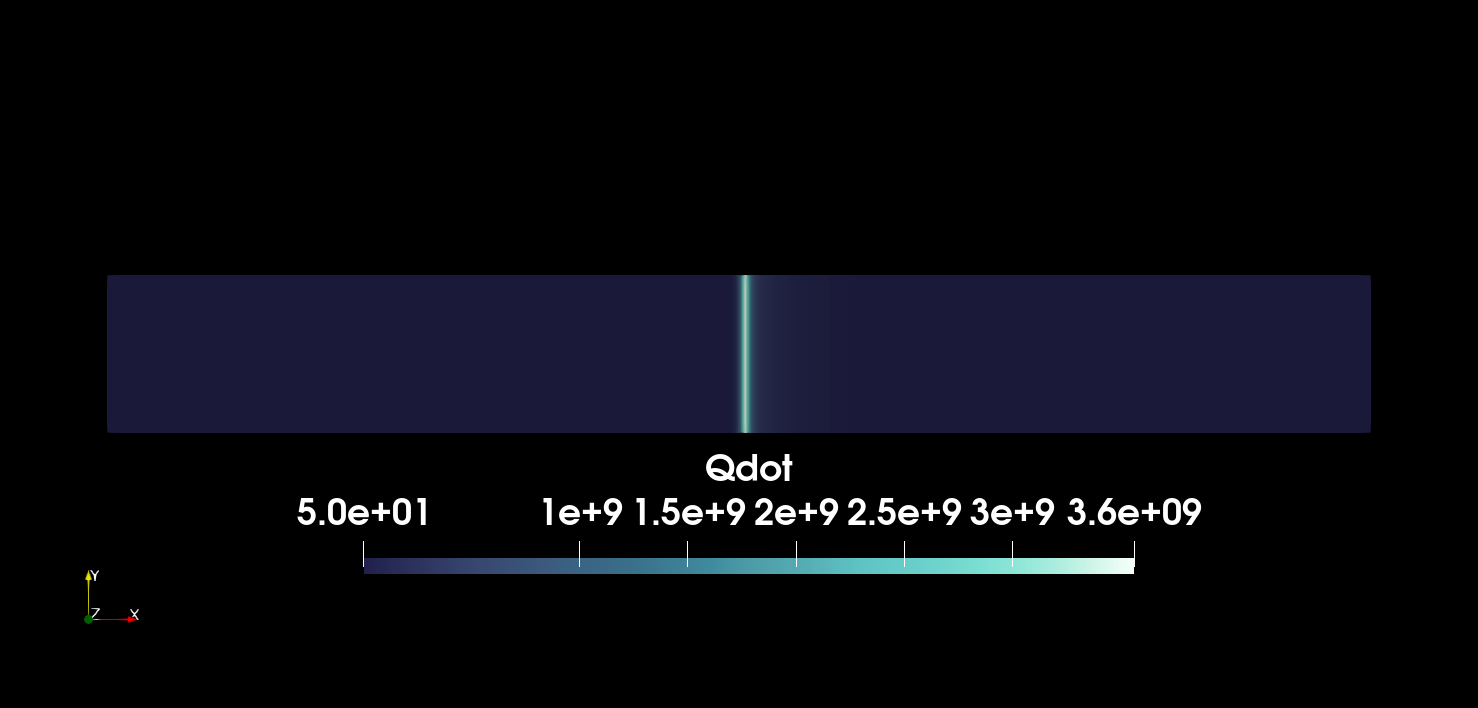}

\vspace{2mm}
\includegraphics[width=0.485\textwidth]{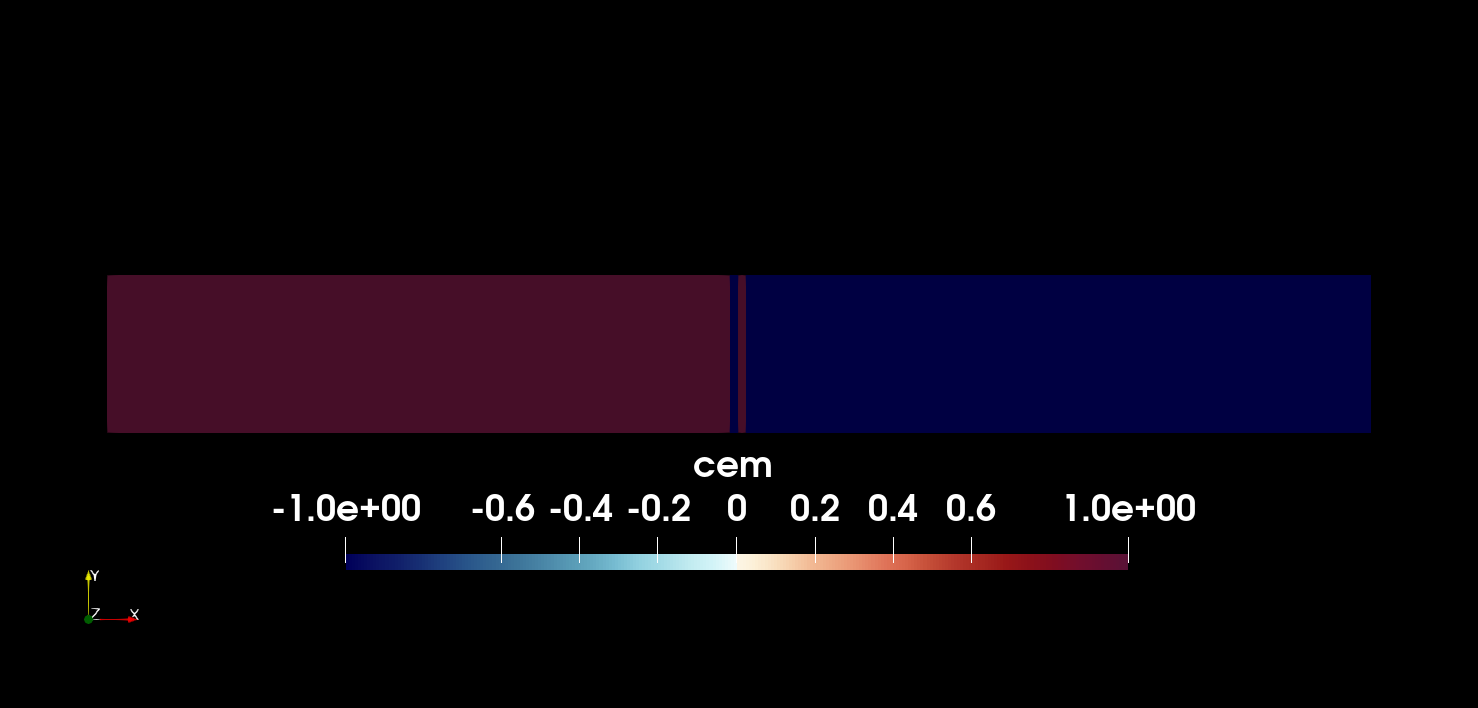} \hfill
\includegraphics[width=0.485\textwidth]{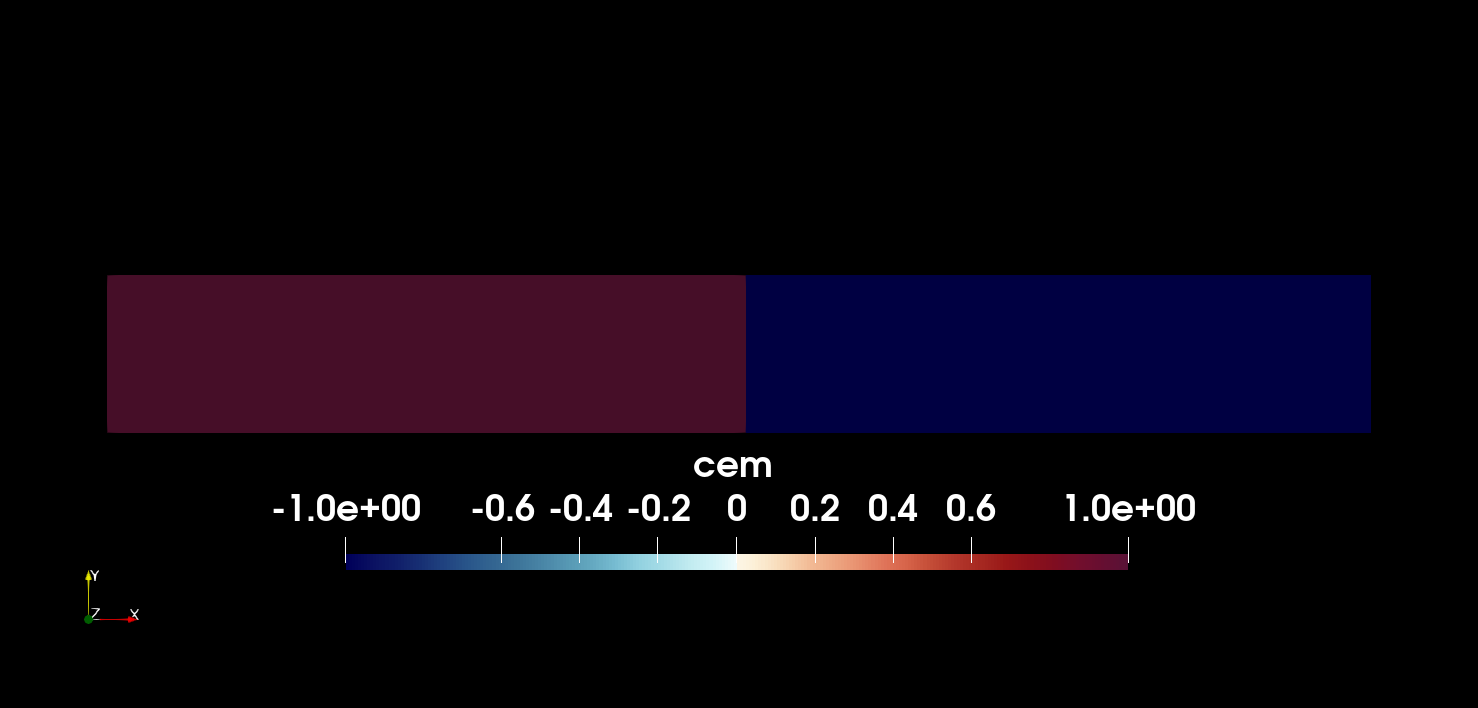}
% \caption{Validation of the developed \texttt{cemaPyjacChemistryModel} to CEMA fields using either semi-analytical concentration based Jacobian (left), and analytical mass fraction based Jacobian (right). As noted, there are discrepancies in the preheat zone for the case with semi-analytical Jacobian. In the case using \texttt{PyJac}, the zero-crossing CEMA field clearly indicates the reaction front.}
\caption{Comparison of semi-analytical (left panel) and analytical (right panel) based development of \texttt{cemaPyjacChemistryModel} for CEMA. Note the discrepancies in the preheat zone for the case of semi-analytical Jacobian due to numerical inaccuracies and insufficient significant digits of the Jacobian matrix as consistently explained in the original CEMA paper~\cite{Lu2010}.}
\label{fig:valid_cem2}
\end{figure}

The final results of this section is a comparison between the analytical and numerical Jacobian based implementation for CEMA computations. In \Cref{fig:valid_cem2}, field plots of temperature, HRR, and \verb|cem| are presented for the numerical based Jacobian (left panel) and analytical based Jacobian (right panel). While both results were almost identical using GRI-3.0 mechanism
% ~\footnote{Note that some tweaking was made on existing GRI-3.0 mechanism in OpenFOAM is manipulated to be  numerical Jacobian using in OpenFOAM}
, the presented results using Yao54 mechanism~\cite{Yao2017} (another long-chain hydrocarbon skeletal mechanism involving $54$ species and $269$ reactions) show discrepancies in the preheat zone of \verb|cem| which is evaluated using the numerical Jacobian. This is possibly due to numerical inaccuracies and insufficient significant digits of the Jacobian matrix resulting by finite-differencing. Such notes are further supported by discussions of the original CEMA developments by Lu et al.~\cite{Lu2010}, hence the necessity of incorporating \verb|pyJac| for analytical Jacobian evaluation in the present developments. Next, we briefly demonstrate the validation of projected CEMA for combustion mode characterization.

% numerical Jacobian was found to yield confusing results for CEMA, which was already noted by Lu et al.~\cite{Lu2010}

% OpenFOAM concentration based semi-analytical Jacobian (dfdPhic) versus PyJac mass fraction based analytical Jacobian (dfdPhi)

\clearpage

\section{Results-III: Projected CEMA demonstration}
\label{sec:cema_results3}

In this section, we demonstrate one application of CEMA which is the projection of diffusion and reaction terms onto CEM to characterize the local combustion mode~\cite{Xu2019}. In brief, by recalling derivations from \Cref{sec:cema}, the projected reaction and diffusion terms onto CEM are described as by the following equations.

\begin{align}
\phi_\omega &= \boldsymbol{b}_\mathrm{exp} \cdot \dot{\boldsymbol \omega}, \\
\phi_s &= \boldsymbol{b}_\mathrm{exp} \cdot \boldsymbol{s}
\end{align}

\begin{figure}[htb!]
\centering
\includegraphics[width=0.95\textwidth]{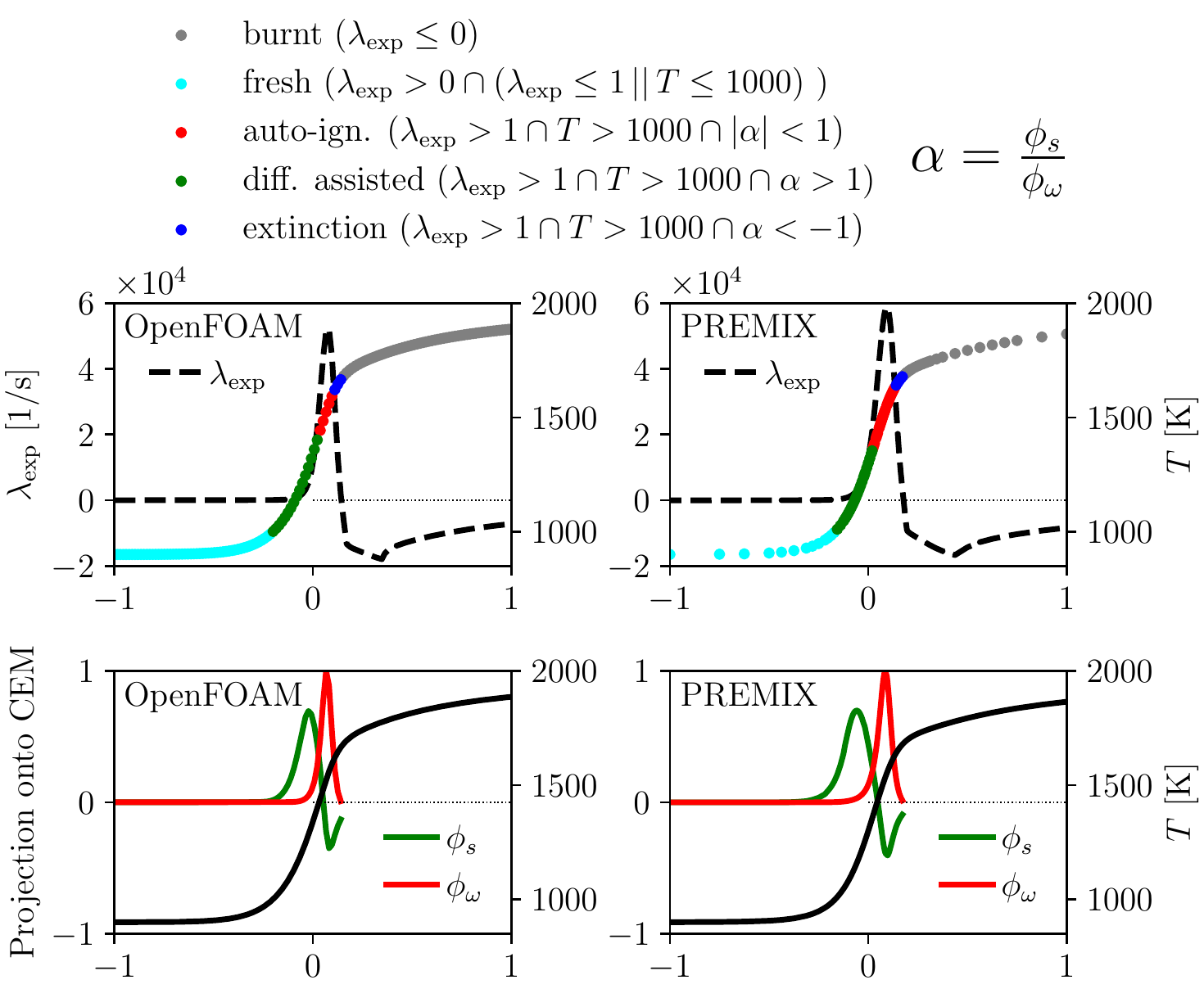}
\caption{Validation of the developed \texttt{cemaPyjacChemistryModel} with implementations of reaction and diffusion terms projections onto CEM for local combustion mode identification. Results are compared against reference case from \texttt{PREMIX} code. Implementation details are not demonstrated in the present report.}
\label{fig:valid_cema_proj}
\end{figure}

% \vspace{-2mm}
The local combustion mode indicator, $\alpha = \phi_s / \phi_\omega$, compares relative alignment of diffusion and chemistry contributions with relevance to fastest CEM. The validation of the developed model for projected CEMA (against reference data set computed using \verb|PREMIX| code) along with definitions for the local combustion modes are depicted in \Cref{fig:valid_cema_proj}. Implementation details of the projected CEMA approach are not shown in the present report.

% The local combustion modes can be categorized as in the following.
% \begin{itemize}
%     \item $\alpha > 1$: {diffusion dominates} chemistry in direction of fastest CEM, {promotes ignition};
%     \item $\alpha < 1$: {diffusion dominates} chemistry in opposite direction of fastest CEM, {inhibits ignition};
%     \item $|\alpha| \leq 1$: {chemistry dominates} diffusion in direction of fastest CEM; % promotes ignition.
% \end{itemize}

% ------------------------------------------------------------ %

\chapter*{Acknowledgment}
The incorporation of the PyJac package into OpenFOAM chemistryModel has been originated in the thesis works of Dr. Heikki Kahila, Wärtsilä Finland Oy, using OpenFOAM-v6. Further information on particular implementations and framework can be found on his doctoral dissertation at Aalto University.

% ----------------------- Back Matter ------------------------ %
%\bibliography{Includes/Bibliography/Bibliography}
% \input{main.bbl}

\chapter*{Study questions}
\begin{enumerate}
% \item How do you ...
% \item What is the purpose of...
\item What is the purpose of direct integration of finite-rate chemistry for chemical source terms?

\item What is the importance of CEMA in the field of computational combustion?

\item Why it is important to incorporate analytical Jacobian for CEMA computations?

\item How to generate analytical Jacobian from given mechanism in Chemkin-II format?

% \item I have a chemical mechanism in Chemkin-II format. How to generate the corresponding files from PyJac?

\item What is the purpose of object orientation in software design?

\item I am not very confident of my programming skills and I would like to use the existing Jacobian formulation from OpenFOAM standard chemictsry model without linking any external third-party libraries. Is that fine?

% \todo{PROVIDED ANSWERS TO BE ADDED TOGETHER AFTER POSSIBLE QUESTIONS ARE POSTED DURING REVIEW PHASE}

\end{enumerate}

\appendix
\chapter{Developed codes}

\section{Tree structure}
\label{ap:tree}

% \dirtree{%
%  .1 \$(WM\_PROJECT\_USER\_DIR)/src/thermophysicalModels.
%  .2 chemistryModel.
%  .3 cemaPyjacChemistryModel.
%  .4 cemaPyjacChemistryModel.C.
%  .4 cemaPyjacChemistryModel.H.
%  .4 cemaPyjacChemistryModelI.H.
%  .3 makeChemistryModel.H.
%  .3 makeChemistryModelTypes.C.
%  .3 makeChemistrySolvers.C.
%  .3 makeChemistrySolverTypes.H.
%  .2 chemistrySolver.
%  .3 odePyjac.
%  .4 odePyjac.C.
%  .4 odePyjac.H.
%  .2 Make.
%  .3 files.
%  .3 options.
%  .2 pyjacInclude.
%  .3 chem\_utils.h.
%  .3 dydt.h.
%  .3 header.h.
%  .3 jacob.h.
%  .3 mass\_mole.h.
%  .3 mechanism.h.
%  .3 rates.h.
%  .3 sparse\_multiplier.h.
% }
\dirtree{%
 % .1 cemaPyjac\_Gadalla\_OSCFD2022.
 .1 CEMAFoam.
 .2 src.
 .3 thermophysicalModels.
 .4 chemistryModel.
 .5 chemistryModel.
 .6 cemaPyjacChemistryModel.
 .7 cemaPyjacChemistryModelI.H.
 .7 cemaPyjacChemistryModel.H.
 .7 cemaPyjacChemistryModel.C.
 .5 chemistrySolver.
 .6 odePyjac.
 .7 odePyjac.H.
 .7 odePyjac.C.
 .5 Make.
 .6 files.
 .6 options.
 .5 makeChemistryModel.H.
 .5 makeChemistryModels.C.
 .5 makeChemistrySolverTypes.H.
 .5 makeChemistrySolvers.C.
 .5 pyjacInclude.
 .6 chem\_utils.h.
 .6 dydt.h.
 .6 header.h.
 .6 jacob.h.
 .6 mass\_mole.h.
 .6 mechanism.h.
 .6 rates.h.
 .6 sparse\_multiplier.h.
}

\clearpage

\dirtree{%
 % .1 cemaPyjac\_Gadalla\_OSCFD2022.
 .1 CEMAFoam.
 .2 tutorials.
 .3 premixedFlame1D.
 .4 Allrun.
 .4 Allclean.
 .4 0.
 .5 U.
 .5 p.
 .5 T.
 .5 T.dat.
 .5 CH4.
 .5 CH4.dat.
 .5 O2.
 .5 O2.dat.
 .5 N2.
 .5 N2.dat.
 .5 CO2.
 .5 CO2.dat.
 .5 H2O.
 .5 H2O.dat.
 .5 Ydefault.
 .4 constant.
 .5 chemistryProperties.
 .5 reactionsGRIPyjac.
 .5 thermo.compressibleGasGRI.
 .5 thermophysicalProperties.
 .5 turbulenceProperties.
 .5 combustionProperties.
 .4 system.
 .5 blockMeshDict.
 .5 decomposeParDict.
 .5 controlDict.
 .5 fvSchemes.
 .5 fvSolution.
 .4 mechanism.
 .5 grimech30.dat.
 .5 thermo30.dat.
 .5 transport.dat.
 .5 out.tgz.
 .5 runCmake.sh.
 .5 CMakeLists.txt.
 .2 utilities.
 .3 compute\_inital\_fields.py.
 .3 interpolate\_inital\_fields.py.
 .3 ndinterpolator.py.
}

\clearpage

% \section{\texttt{icoFoam} solver}
\section{\texttt{cemaPyjacChemistryModel} library}

% If you wish to put your developed codes in the appendix you can use the \verb|listings| package. Here the optional argument \verb|numbers=none| is used to remove the line numbers. It is better if the codes in the appendix do not have line numbers, since people may want to copy-paste them from the output PDF.

\subsection{\texttt{chemistryModel}}

\begin{lstlisting}[style=cpp,title=\texttt{cemaPyjacChemistryModel.H},captionpos=t,numbers=none]
/*---------------------------------------------------------------------------*\
  =========                 |
  \\      /  F ield         | OpenFOAM: The Open Source CFD Toolbox
   \\    /   O peration     |
    \\  /    A nd           | www.openfoam.com
     \\/     M anipulation  |
-------------------------------------------------------------------------------
    Copyright (C) 2011-2017 OpenFOAM Foundation  -- Author: Mahmoud Gadalla
-------------------------------------------------------------------------------
License
    This file is part of CEMAFoam, derived from OpenFOAM.

    https://github.com/Aalto-CFD/CEMAFoam

    OpenFOAM is free software: you can redistribute it and/or modify it
    under the terms of the GNU General Public License as published by
    the Free Software Foundation, either version 3 of the License, or
    (at your option) any later version.

    OpenFOAM is distributed in the hope that it will be useful, but WITHOUT
    ANY WARRANTY; without even the implied warranty of MERCHANTABILITY or
    FITNESS FOR A PARTICULAR PURPOSE.  See the GNU General Public License
    for more details.

    You should have received a copy of the GNU General Public License
    along with OpenFOAM.  If not, see <http://www.gnu.org/licenses/>.

Class
    Foam::cemaPyjacChemistryModel

Description
    Extends base chemistry model by adding a thermo package, and ODE functions.
    Introduces chemistry equation system and evaluation of chemical source
    terms. Replaces approximated jacobian implementation with analytical one.
    Adds new functionality for extended Chemical Explosive Mode Analysis (CEMA)

SourceFiles
    cemaPyjacChemistryModelI.H
    cemaPyjacChemistryModel.C

\*---------------------------------------------------------------------------*/

#ifndef cemaPyjacChemistryModel_H
#define cemaPyjacChemistryModel_H

#include "BasicChemistryModel.H"
#include "Reaction.H"
#include "ODESystem.H"
#include "volFields.H"
#include "simpleMatrix.H"

#include "EigenMatrix.H" // FOR EIGEN DECOMPOSITION

extern "C" {
    #include "chem_utils.h"
    #include "dydt.h"
    #include "jacob.h"
};

// * * * * * * * * * * * * * * * * * * * * * * * * * * * * * * * * * * * * * //

namespace Foam
{

// Forward declaration of classes
class fvMesh;

/*---------------------------------------------------------------------------*\
                      Class cemaPyjacChemistryModel Declaration
\*---------------------------------------------------------------------------*/

template<class ReactionThermo, class ThermoType>
class cemaPyjacChemistryModel
:
    public BasicChemistryModel<ReactionThermo>,
    public ODESystem
{
    // Private Member Functions

        //- Solve the reaction system for the given time step
        //  of given type and return the characteristic time
        template<class DeltaTType>
        scalar solve(const DeltaTType& deltaT);

        //- No copy construct
        cemaPyjacChemistryModel
        (
            const cemaPyjacChemistryModel<ReactionThermo, ThermoType>&
        ) = delete;

        //- No copy assignment
        void operator=
        (
            const cemaPyjacChemistryModel<ReactionThermo, ThermoType>&
        ) = delete;

protected:

    typedef ThermoType thermoType;


    // Protected data

        //- Reference to the field of specie mass fractions
        PtrList<volScalarField>& Y_;

        //- Reactions
        const PtrList<Reaction<ThermoType>>& reactions_;

        //- Thermodynamic data of the species
        const PtrList<ThermoType>& specieThermo_;

        //- Number of species
        label nSpecie_;

        //- Number of reactions
        label nReaction_;

        //- Temperature below which the reaction rates are assumed 0
        scalar Treact_;

        //- List of reaction rate per specie [kg/m3/s]
        PtrList<volScalarField::Internal> RR_;

        //- Temporary concentration field
        mutable scalarField c_;

        //- Temporary rate-of-change of concentration field
        mutable scalarField dcdt_;

        // Enthalpy of formation for every species, from PyJac //###
        scalarList sp_enthalpy_;

    	// Number of elements specified in the reaction file
    	label nElements_;

        // Jacobian from chemistry problem, from pyJac
        mutable scalarSquareMatrix chemJacobian_;

        // CEMA geometric fields
        volScalarField cem_;

    // Protected Member Functions

        //- Write access to chemical source terms
        //  (e.g. for multi-chemistry model)
        inline PtrList<volScalarField::Internal>& RR();


public:

    //- Runtime type information
    TypeName("cemaPyjac");


    // Constructors

        //- Construct from thermo
        cemaPyjacChemistryModel(ReactionThermo& thermo);


    //- Destructor
    virtual ~cemaPyjacChemistryModel();


    // Member Functions

        //- The reactions
        inline const PtrList<Reaction<ThermoType>>& reactions() const;

        //- Thermodynamic data of the species
        inline const PtrList<ThermoType>& specieThermo() const;

        //- The number of species
        virtual inline label nSpecie() const;

        //- The number of reactions
        virtual inline label nReaction() const;

        //- Temperature below which the reaction rates are assumed 0
        inline scalar Treact() const;

        //- Temperature below which the reaction rates are assumed 0
        inline scalar& Treact();

        //- dc/dt = omega, rate of change in concentration, for each species
        virtual void omega
        (
            const scalarField& c,
            const scalar T,
            const scalar p,
            scalarField& dcdt
        ) const;

        //- Return the reaction rate for reaction r and the reference
        //  species and characteristic times
        virtual scalar omega
        (
            const Reaction<ThermoType>& r,
            const scalarField& c,
            const scalar T,
            const scalar p,
            scalar& pf,
            scalar& cf,
            label& lRef,
            scalar& pr,
            scalar& cr,
            label& rRef
        ) const;


        //- Return the reaction rate for iReaction and the reference
        //  species and characteristic times
        virtual scalar omegaI
        (
            label iReaction,
            const scalarField& c,
            const scalar T,
            const scalar p,
            scalar& pf,
            scalar& cf,
            label& lRef,
            scalar& pr,
            scalar& cr,
            label& rRef
        ) const;

        //- Calculates the reaction rates
        virtual void calculate();


        // Chemistry model functions (overriding abstract functions in
        // basicChemistryModel.H)

            //- Return const access to the chemical source terms for specie, i
            inline const volScalarField::Internal& RR
            (
                const label i
            ) const;

            //- Return non const access to chemical source terms [kg/m3/s]
            virtual volScalarField::Internal& RR
            (
                const label i
            );

            //- Return reaction rate of the speciei in reactionI
            virtual tmp<volScalarField::Internal> calculateRR
            (
                const label reactionI,
                const label speciei
            ) const;

            //- Solve the reaction system for the given time step
            //  and return the characteristic time
            virtual scalar solve(const scalar deltaT);

            //- Solve the reaction system for the given time step
            //  and return the characteristic time
            virtual scalar solve(const scalarField& deltaT);

            //- Return the chemical time scale
            virtual tmp<volScalarField> tc() const;

            //- Return the heat release rate [kg/m/s3]
            virtual tmp<volScalarField> Qdot() const;


        // ODE functions (overriding abstract functions in ODE.H)

            //- Number of ODE's to solve
            inline virtual label nEqns() const;

            virtual void derivatives
            (
                const scalar t,
                const scalarField& c,
                scalarField& dcdt
            ) const;

            virtual void jacobian
            (
                const scalar t,
                const scalarField& c,
                scalarField& dcdt,
                scalarSquareMatrix& dfdc
            ) const;

            virtual void solve
            (
                scalarField &c,
                scalar& T,
                scalar& p,
                scalar& deltaT,
                scalar& subDeltaT
            ) const = 0;

            // Perhaps to propagate for other models
            void cema
            (
                scalar& cem
            ) const;

};


// * * * * * * * * * * * * * * * * * * * * * * * * * * * * * * * * * * * * * //

} // End namespace Foam

// * * * * * * * * * * * * * * * * * * * * * * * * * * * * * * * * * * * * * //

#include "cemaPyjacChemistryModelI.H"

// * * * * * * * * * * * * * * * * * * * * * * * * * * * * * * * * * * * * * //

#ifdef NoRepository
    #include "cemaPyjacChemistryModel.C"
#endif

// * * * * * * * * * * * * * * * * * * * * * * * * * * * * * * * * * * * * * //

#endif

// ************************************************************************* //
\end{lstlisting}

\begin{lstlisting}[style=cpp,title=\texttt{cemaPyjacChemistryModelI.H},captionpos=t,numbers=none]
/*---------------------------------------------------------------------------*\
  =========                 |
  \\      /  F ield         | OpenFOAM: The Open Source CFD Toolbox
   \\    /   O peration     |
    \\  /    A nd           | www.openfoam.com
     \\/     M anipulation  |
-------------------------------------------------------------------------------
    Copyright (C) 2011-2017 OpenFOAM Foundation
-------------------------------------------------------------------------------
License
    This file is part of OpenFOAM.

    OpenFOAM is free software: you can redistribute it and/or modify it
    under the terms of the GNU General Public License as published by
    the Free Software Foundation, either version 3 of the License, or
    (at your option) any later version.

    OpenFOAM is distributed in the hope that it will be useful, but WITHOUT
    ANY WARRANTY; without even the implied warranty of MERCHANTABILITY or
    FITNESS FOR A PARTICULAR PURPOSE.  See the GNU General Public License
    for more details.

    You should have received a copy of the GNU General Public License
    along with OpenFOAM.  If not, see <http://www.gnu.org/licenses/>.

\*---------------------------------------------------------------------------*/

#include "volFields.H"
#include "zeroGradientFvPatchFields.H"

// * * * * * * * * * * * * * * * Member Functions  * * * * * * * * * * * * * //

template<class ReactionThermo, class ThermoType>
inline Foam::label
Foam::cemaPyjacChemistryModel<ReactionThermo, ThermoType>::nEqns() const
{
    // nEqns = number of species (N-1) + temperature + pressure
    return nSpecie_ + 1;
}
//!!! REMAINING ARE SAME AS STANDARD

// ************************************************************************* //
\end{lstlisting}

\lstinputlisting[style=cpp,title=\texttt{cemaPyjacChemistryModel.C},captionpos=t,numbers=none]{Codes/lib_cema/chemistryModel/chemistryModel/cemaPyjacChemistryModel/cemaPyjacChemistryModel.C}

% \lstinputlisting[style=cpp,title=\texttt{Makefile},captionpos=t,numbers=none]{Codes/lib_cema/chemistryModel/Make/files}
% \lstinputlisting[style=cpp,title=\texttt{Makeoptions},captionpos=t,numbers=none]{Codes/lib_cema/chemistryModel/Make/options}

% \lstinputlisting[style=cpp,title=\texttt{cemaPyjacChemistryModel.H},captionpos=t,numbers=none]{Codes/lib_cema/chemistryModel/chemistryModel/cemaPyjacChemistryModel/appendix/cemaPyjacChemistryModel.H}

% \lstinputlisting[style=cpp,title=\texttt{cemaPyjacChemistryModelI.H},captionpos=t,numbers=none]{Codes/lib_cema/chemistryModel/chemistryModel/cemaPyjacChemistryModel/appendix/cemaPyjacChemistryModelI.H}

% \lstinputlisting[style=cpp,title=\texttt{cemaPyjacChemistryModel.C},captionpos=t,numbers=none]{Codes/lib_cema/chemistryModel/chemistryModel/cemaPyjacChemistryModel/appendix/cemaPyjacChemistryModel.C}

% \lstinputlisting[style=cpp,title=\texttt{Makefile},captionpos=t,numbers=none]{Codes/lib_cema/chemistryModel/Make/files}
% \lstinputlisting[style=cpp,title=\texttt{Makeoptions},captionpos=t,numbers=none]{Codes/lib_cema/chemistryModel/Make/options}

% \section{Dictionary}
% \section{premixedFlame1D tutorial files}

% You can of course also add any other ascii file, such as dictionaries:

% \lstinputlisting[style=OpenFOAMDict,title=\texttt{controlDict},captionpos=t,numbers=none]{Codes/controlDict}

\subsection{\texttt{chemistrySolver}}

\begin{lstlisting}[style=cpp,title=\texttt{odePyjac.H},captionpos=t,numbers=none]
/*---------------------------------------------------------------------------*\
  =========                 |
  \\      /  F ield         | OpenFOAM: The Open Source CFD Toolbox
   \\    /   O peration     |
    \\  /    A nd           | www.openfoam.com
     \\/     M anipulation  |
-------------------------------------------------------------------------------
    Copyright (C) 2011-2017 OpenFOAM Foundation
-------------------------------------------------------------------------------
License
    This file is part of OpenFOAM.

    OpenFOAM is free software: you can redistribute it and/or modify it
    under the terms of the GNU General Public License as published by
    the Free Software Foundation, either version 3 of the License, or
    (at your option) any later version.

    OpenFOAM is distributed in the hope that it will be useful, but WITHOUT
    ANY WARRANTY; without even the implied warranty of MERCHANTABILITY or
    FITNESS FOR A PARTICULAR PURPOSE.  See the GNU General Public License
    for more details.

    You should have received a copy of the GNU General Public License
    along with OpenFOAM.  If not, see <http://www.gnu.org/licenses/>.

Class
    Foam::odePyjac

Description
    An ODE solver for chemistry

SourceFiles
    odePyjac.C

\*---------------------------------------------------------------------------*/

#ifndef odePyjac_H
#define odePyjac_H

#include "chemistrySolver.H"
#include "ODESolver.H"

// * * * * * * * * * * * * * * * * * * * * * * * * * * * * * * * * * * * * * //

namespace Foam
{

/*---------------------------------------------------------------------------*\
                            Class odePyjac Declaration
\*---------------------------------------------------------------------------*/

template<class ChemistryModel>
class odePyjac
:
    public chemistrySolver<ChemistryModel>
{
    // Private data

        dictionary coeffsDict_;

        mutable autoPtr<ODESolver> odeSolver_;

        // Solver data
        mutable scalarField cTp_;


public:

    //- Runtime type information
    TypeName("odePyjac");


    // Constructors

        //- Construct from thermo
        odePyjac(typename ChemistryModel::reactionThermo& thermo);


    //- Destructor
    virtual ~odePyjac();


    // Member Functions

        //- Update the concentrations and return the chemical time
        virtual void solve
        (
            scalarField& c,
            scalar& T,
            scalar& p,
            scalar& deltaT,
            scalar& subDeltaT
        ) const;
};


// * * * * * * * * * * * * * * * * * * * * * * * * * * * * * * * * * * * * * //

} // End namespace Foam

// * * * * * * * * * * * * * * * * * * * * * * * * * * * * * * * * * * * * * //

#ifdef NoRepository
    #include "odePyjac.C"
#endif

// * * * * * * * * * * * * * * * * * * * * * * * * * * * * * * * * * * * * * //

#endif

// ************************************************************************* //
\end{lstlisting}

\begin{lstlisting}[style=cpp,title=\texttt{odePyjac.C},captionpos=t,numbers=none]
/*---------------------------------------------------------------------------*\
  =========                 |
  \\      /  F ield         | OpenFOAM: The Open Source CFD Toolbox
   \\    /   O peration     |
    \\  /    A nd           | www.openfoam.com
     \\/     M anipulation  |
-------------------------------------------------------------------------------
    Copyright (C) 2011-2017 OpenFOAM Foundation
-------------------------------------------------------------------------------
License
    This file is part of OpenFOAM.

    OpenFOAM is free software: you can redistribute it and/or modify it
    under the terms of the GNU General Public License as published by
    the Free Software Foundation, either version 3 of the License, or
    (at your option) any later version.

    OpenFOAM is distributed in the hope that it will be useful, but WITHOUT
    ANY WARRANTY; without even the implied warranty of MERCHANTABILITY or
    FITNESS FOR A PARTICULAR PURPOSE.  See the GNU General Public License
    for more details.

    You should have received a copy of the GNU General Public License
    along with OpenFOAM.  If not, see <http://www.gnu.org/licenses/>.

\*---------------------------------------------------------------------------*/

#include "odePyjac.H"

// * * * * * * * * * * * * * * * * Constructors  * * * * * * * * * * * * * * //

template<class ChemistryModel>
Foam::odePyjac<ChemistryModel>::odePyjac(typename ChemistryModel::reactionThermo& thermo)
:
    chemistrySolver<ChemistryModel>(thermo),
    coeffsDict_(this->subDict("odeCoeffs")),
    odeSolver_(ODESolver::New(*this, coeffsDict_)),
    cTp_(this->nEqns())
{}


// * * * * * * * * * * * * * * * * Destructor  * * * * * * * * * * * * * * * //

template<class ChemistryModel>
Foam::odePyjac<ChemistryModel>::~odePyjac()
{}


// * * * * * * * * * * * * * * * Member Functions  * * * * * * * * * * * * * //

template<class ChemistryModel>
void Foam::odePyjac<ChemistryModel>::solve
(
    scalarField& c,
    scalar& T,
    scalar& p,
    scalar& deltaT,
    scalar& subDeltaT
) const
{
    // Reset the size of the ODE system to the simplified size when mechanism
    // reduction is active
    if (odeSolver_->resize())
    {
        odeSolver_->resizeField(cTp_);
    }

    const label nSpecie = this->nSpecie();

    // Copy the concentration, T and P to the total solve-vector (N+1)
    cTp_[0] = p;
    cTp_[1] = T;
    // Update for N-1 species
    for (label i=0; i<nSpecie-1; i++)
    {
        cTp_[i+2] = c[i];
    }

    // Here, we call ODE solver...
    odeSolver_->solve(0, deltaT, cTp_, subDeltaT);

    // Back substitute results
    p = cTp_[0];
    T = cTp_[1];
    scalar csum = 0;

    for (label i=0; i<nSpecie-1; i++)
    {
        c[i] = max(0.0, cTp_[i+2]);
        csum += c[i];
    }
    // Last species
    c[nSpecie-1] = 1.0 - csum;
}


// ************************************************************************* //
\end{lstlisting}

% \clearpage
% \section{cemaPyjac Make files}
\subsection{Make \texttt{chemistryModel}}
\label{ap:make}

\begin{lstlisting}[style=cpp, caption=Make (\texttt{files}) for the \texttt{cemaPyjacChemistryModel} library compilation,captionpos=t,numbers=none]
makeChemistryModels.C
makeChemistrySolvers.C

LIB = $(FOAM_USER_LIBBIN)/libcemaPyjacChemistryModel
\end{lstlisting}

\begin{lstlisting}[style=cpp, caption=Make (\texttt{options}) for the \texttt{cemaPyjacChemistryModel} library compilation,captionpos=t,numbers=none]
EXE_INC = \
    -I$(LIB_SRC)/finiteVolume/lnInclude \
    -I$(LIB_SRC)/meshTools/lnInclude \
    -I$(LIB_SRC)/ODE/lnInclude \
    -I$(LIB_SRC)/transportModels/compressible/lnInclude \
    -I$(LIB_SRC)/thermophysicalModels/reactionThermo/lnInclude \
    -I$(LIB_SRC)/thermophysicalModels/basic/lnInclude \
    -I$(LIB_SRC)/thermophysicalModels/specie/lnInclude \
    -I$(LIB_SRC)/thermophysicalModels/functions/Polynomial \
    -I$(LIB_SRC)/thermophysicalModels/thermophysicalFunctions/lnInclude \
    -I$(LIB_SRC)/thermophysicalModels/chemistryModel/lnInclude \
    -IpyjacInclude

LIB_LIBS = \
    -lfiniteVolume \
    -lmeshTools \
    -lODE \
    -lcompressibleTransportModels \
    -lfluidThermophysicalModels \
    -lreactionThermophysicalModels \
    -lspecie \
    -lchemistryModel
\end{lstlisting}

\clearpage
\begin{lstlisting}[style=cpp, caption=Macros for chemistry models based on compressibility and transport types,captionpos=t,numbers=none]
/*---------------------------------------------------------------------------*\
  =========                 |
  \\      /  F ield         | OpenFOAM: The Open Source CFD Toolbox
   \\    /   O peration     |
    \\  /    A nd           | www.openfoam.com
     \\/     M anipulation  |
-------------------------------------------------------------------------------
    Copyright (C) 2011-2017 OpenFOAM Foundation
-------------------------------------------------------------------------------
License
    This file is part of OpenFOAM.

    OpenFOAM is free software: you can redistribute it and/or modify it
    under the terms of the GNU General Public License as published by
    the Free Software Foundation, either version 3 of the License, or
    (at your option) any later version.

    OpenFOAM is distributed in the hope that it will be useful, but WITHOUT
    ANY WARRANTY; without even the implied warranty of MERCHANTABILITY or
    FITNESS FOR A PARTICULAR PURPOSE.  See the GNU General Public License
    for more details.

    You should have received a copy of the GNU General Public License
    along with OpenFOAM.  If not, see <http://www.gnu.org/licenses/>.

Description
    Macros for instantiating chemistry models based on compressibility and
    transport types
    makeChemistryModelType defined: $FOAM_SRC/OpenFOAM/db/typeInfo/className.H

\*---------------------------------------------------------------------------*/

#ifndef makeChemistryModel_H
#define makeChemistryModel_H

#include "addToRunTimeSelectionTable.H"

// * * * * * * * * * * * * * * * * * * * * * * * * * * * * * * * * * * * * * //

namespace Foam
{

// * * * * * * * * * * * * * * * * * * * * * * * * * * * * * * * * * * * * * //

#define makeChemistryModel(Comp)                                               \
                                                                               \
    typedef BasicChemistryModel<Comp> BasicChemistryModel##Comp;               \
                                                                               \
    defineTemplateTypeNameAndDebugWithName                                     \
    (                                                                          \
        BasicChemistryModel##Comp,                                             \
        "BasicChemistryModel<"#Comp">",                                        \
        0                                                                      \
    );                                                                         \
                                                                               \
    defineTemplateRunTimeSelectionTable                                        \
    (                                                                          \
        BasicChemistryModel##Comp,                                             \
        thermo                                                                 \
    );


#define makeChemistryModelType(SS, Comp, Thermo)                               \
                                                                               \
    typedef SS<Comp, Thermo> SS##Comp##Thermo;                                 \
                                                                               \
    defineTemplateTypeNameAndDebugWithName                                     \
    (                                                                          \
        SS##Comp##Thermo,                                                      \
        (#SS"<"#Comp"," + Thermo::typeName() + ">").c_str(),                   \
        0                                                                      \
    );


// * * * * * * * * * * * * * * * * * * * * * * * * * * * * * * * * * * * * * //

} // End namespace Foam

// * * * * * * * * * * * * * * * * * * * * * * * * * * * * * * * * * * * * * //

#endif

// ************************************************************************* //
\end{lstlisting}

\lstinputlisting[style=cpp, caption=Macros for chemistry solvers based on compressibility and transport types,captionpos=t,numbers=none]{Codes/lib_cema/chemistryModel/makeChemistrySolverTypes.H}

\lstinputlisting[style=cpp, caption=Templates to instantiate chemistry models based on compressibility and transport types,captionpos=t,numbers=none]{Codes/lib_cema/chemistryModel/makeChemistryModels.C}

\lstinputlisting[style=cpp, caption=Templates to instantiate chemistry solvers based on compressibility and transport types,captionpos=t,numbers=none]{Codes/lib_cema/chemistryModel/makeChemistrySolvers.C}

% \clearpage
\section{\texttt{pyJac} CMake}
\label{ap:pyjac_cmake}
We assume that \verb|pyJac| was already used to generate a mechanism source code, and subsequently that \verb|out| directory is created.

\begin{lstlisting}[style=cpp, caption=Shell script to build mechanism source code into the shared object \texttt{libc\_pyjac.so},captionpos=t,numbers=none]
rm -rf build

mkdir build

cd build && cmake .. -DCMAKE_C_COMPILER=cc 

make
\end{lstlisting}

\begin{lstlisting}[style=cpp, caption=\texttt{CMakelists.txt} containing set of directives and instructions to compile using \texttt{CMake},captionpos=t,numbers=none]
cmake_minimum_required(VERSION 2.6)

project(pyJac)

set(CMAKE_BUILD_TYPE Release)

enable_language(C)

#GNU on DESKTOP
set(CMAKE_C_FLAGS "-std=c99 -O3 -mtune=native -fPIC")

include_directories(out)
include_directories(out/jacobs)

file(GLOB_RECURSE SOURCES "out/*.c")
add_library(c_pyjac SHARED ${SOURCES})
install(TARGETS c_pyjac DESTINATION .)

\end{lstlisting}

\section{\texttt{pyJac} generic header files}
\label{ap:pyjac_include}
These header files are required for the compilation of the developed library, while definitions if the functions declarations are compiled for specific chemical mechanism and the generated binary is linked to solver during runtime.

\lstinputlisting[style=cpp, caption=\texttt{header.h} initial header file,captionpos=t,numbers=none]{Codes/pyjacInclude/header.h}

\lstinputlisting[style=cpp, caption=\texttt{mechanism.h} for \texttt{GRI-3.0}. Note that file needs to be updated for new mechanism in case enclosed directives are to be used. Otherwise all good,captionpos=t,numbers=none]{Codes/pyjacInclude/mechanism.h}

\lstinputlisting[style=cpp, caption=\texttt{jacob.h} to evaluate Jacobian,captionpos=t,numbers=none]{Codes/pyjacInclude/jacob.h}

\begin{lstlisting}[style=cpp, caption=\texttt{dydt.h} to evaluate derivatives,captionpos=t,numbers=none]
#ifndef DYDT_HEAD
#define DYDT_HEAD

#include "header.h"

void dydt (const double, const double, const double * __restrict__, double * __restrict__);

#endif
\end{lstlisting}

\lstinputlisting[style=cpp, caption=\texttt{rates.h} to evaluate reaction rates,captionpos=t,numbers=none]{Codes/pyjacInclude/rates.h}

\lstinputlisting[style=cpp, caption=\texttt{chem\_utils.h} to evaluate thermodynamic quantities,captionpos=t,numbers=none]{Codes/pyjacInclude/chem_utils.h}

\lstinputlisting[style=cpp, caption=\texttt{mass\_mole.h} to convert between molar and mass basis,captionpos=t,numbers=none]{Codes/pyjacInclude/mass_mole.h}

\lstinputlisting[style=cpp, caption=\texttt{sparse\_multiplier.h},captionpos=t,numbers=none]{Codes/pyjacInclude/sparse_multiplier.h}

% \clearpage
\section{Utilities}
\label{ap:utilities}
We note that the interpolator module is taken from \verb|Bladex| Python package~\cite{Gadalla2019}, available at \url{https://github.com/mathLab/BladeX}.

\lstinputlisting[style=python,title=\texttt{computeInitalFields.py},captionpos=t,numbers=none]{Codes/utilities/computeInitialFields.py}

\clearpage
\lstinputlisting[style=python,title=\texttt{interpolateInitalFields.py},captionpos=t,numbers=none]{Codes/utilities/interpolateInitialFields.py}

% ./premixedFlame1D_ref/cantera/ndinterpolator.py
% ./premixedFlame1D_ref/cantera/plot.py
% ./premixedFlame1D_ref/cantera/pp.py

\printindex
% ------------------------------------------------------------ %

\end{document}